\let\orienddocument\enddocument 
\documentclass[longauth]{aa}
\let\enddocument\orienddocument
\usepackage{lineno}
\usepackage{lastpage}

\usepackage{graphicx}
\usepackage[varg]{txfonts}
\usepackage{verbatim}
\usepackage{amssymb,amsmath,amsfonts}

\usepackage{subcaption}

\usepackage{hyperref}
\hypersetup{
    colorlinks=true,
    linkcolor=blue,
    citecolor=blue,
    filecolor=blue,      
    urlcolor=blue,
    breaklinks=true,
}
\usepackage[normalem]{ulem} 

\usepackage{color}
\definecolor{byzantine}{rgb}{0.74, 0.2, 0.64}

\DeclareUnicodeCharacter{2212}{-}

\usepackage{natbib}
\bibpunct{(}{)}{;}{a}{}{,} 

\newcommand{\trades}[0]{\textsc{TRADES}}
\newcommand{\pyde}[0]{\textsc{PyDE}}
\newcommand{\emcee}[0]{\textsc{emcee}}
\newcommand{\pyorbit}[0]{\textsc{PyORBIT}}

\newcommand{\rmd}[0]{\ensuremath{\mathrm{d}}}
\newcommand{\unif}[2]{\ensuremath{\mathcal{U} (#1,#2)}}
\newcommand{\gauss}[2]{\ensuremath{\mathcal{G}(#1,#2)}}
\newcommand{\halfgauss}[2]{\ensuremath{\mathcal{N}^{+}(#1,#2)}}

\defcitealias{vanderburg_2016}{V16}
\defcitealias{berardo_2019}{B19}
\defcitealias{santerne_2019}{S19}
\defcitealias{kempton_2018}{K18}
\defcitealias{otegi_2022}{O22}

\begin{document} 

   \title{Transit timing variations in HIP\,41378: \\ CHEOPS and TESS confirm a non-transiting sixth planet in the system\thanks{Based on data from CHEOPS Guaranteed Time Observations, collected under Programme ID \texttt{CH\_PR100025.}}}

    \author{
P.~Leonardi\thanks{E-mail: pietro.leonardi.1@studenti.unipd.it}\inst{\ref{inst:1},\ref{inst:2},\ref{inst:3},\ref{inst:4}}\,$^{\href{https://orcid.org/0000-0001-6026-9202}{\protect\includegraphics[height=0.19cm]{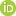}}}$, 
L.~Borsato\inst{\ref{inst:3}}\,$^{\href{https://orcid.org/0000-0003-0066-9268}{\protect\includegraphics[height=0.19cm]{Figure/orcid.jpg}}}$, 
L.~Pagliaro\inst{\ref{inst:2}}, 
D.~Kubyshkina\inst{\ref{inst:4}}, 
J.~A.~Egger\inst{\ref{inst:5}}\,$^{\href{https://orcid.org/0000-0003-1628-4231}{\protect\includegraphics[height=0.19cm]{Figure/orcid.jpg}}}$, 
T.~G.~Wilson\inst{\ref{inst:6}}\,$^{\href{https://orcid.org/0000-0001-8749-1962}{\protect\includegraphics[height=0.19cm]{Figure/orcid.jpg}}}$, 
A.~Heitzmann\inst{\ref{inst:7}}\,$^{\href{https://orcid.org/0000-0002-8091-7526}{\protect\includegraphics[height=0.19cm]{Figure/orcid.jpg}}}$, 
A.~Brandeker\inst{\ref{inst:8}}\,$^{\href{https://orcid.org/0000-0002-7201-7536}{\protect\includegraphics[height=0.19cm]{Figure/orcid.jpg}}}$, 
M.~N.~Günther\inst{\ref{inst:9}}\,$^{\href{https://orcid.org/0000-0002-3164-9086}{\protect\includegraphics[height=0.19cm]{Figure/orcid.jpg}}}$, 
V.~Nascimbeni\inst{\ref{inst:3}}\,$^{\href{https://orcid.org/0000-0001-9770-1214}{\protect\includegraphics[height=0.19cm]{Figure/orcid.jpg}}}$, 
A.~Leleu\inst{\ref{inst:7},\ref{inst:5}}\,$^{\href{https://orcid.org/0000-0003-2051-7974}{\protect\includegraphics[height=0.19cm]{Figure/orcid.jpg}}}$,
A.~Bonfanti\inst{\ref{inst:4}}\,$^{\href{https://orcid.org/0000-0002-1916-5935}{\protect\includegraphics[height=0.19cm]{Figure/orcid.jpg}}}$, S.~G.~Sousa\inst{\ref{inst:10}}\,$^{\href{https://orcid.org/0000-0001-9047-2965}{\protect\includegraphics[height=0.19cm]{Figure/orcid.jpg}}}$,
G.~Mantovan\inst{\ref{inst:2}}, 
G.~Piotto\inst{\ref{inst:3},\ref{inst:1}}\,$^{\href{https://orcid.org/0000-0002-9937-6387}{\protect\includegraphics[height=0.19cm]{Figure/orcid.jpg}}}$, 
L.~Fossati\inst{\ref{inst:4}}\,$^{\href{https://orcid.org/0000-0003-4426-9530}{\protect\includegraphics[height=0.19cm]{Figure/orcid.jpg}}}$, 
D.~Nardiello\inst{\ref{inst:2}}\,$^{\href{https://orcid.org/0000-0003-1149-3659}{\protect\includegraphics[height=0.19cm]{Figure/orcid.jpg}}}$, 
T.~Zingales\inst{\ref{inst:1},\ref{inst:3}}\,$^{\href{https://orcid.org/0000-0001-6880-5356}{\protect\includegraphics[height=0.19cm]{Figure/orcid.jpg}}}$, 
V.~Adibekyan\inst{\ref{inst:10}}\,$^{\href{https://orcid.org/0000-0002-0601-6199}{\protect\includegraphics[height=0.19cm]{Figure/orcid.jpg}}}$, 
C.~Pezzotti\inst{\ref{inst:12}}, 
B.~Akinsanmi\inst{\ref{inst:7}}\,$^{\href{https://orcid.org/0000-0001-6519-1598}{\protect\includegraphics[height=0.19cm]{Figure/orcid.jpg}}}$, 
Y.~Alibert\inst{\ref{inst:13},\ref{inst:5}}\,$^{\href{https://orcid.org/0000-0002-4644-8818}{\protect\includegraphics[height=0.19cm]{Figure/orcid.jpg}}}$, 
R.~Alonso\inst{\ref{inst:14},\ref{inst:15}}\,$^{\href{https://orcid.org/0000-0001-8462-8126}{\protect\includegraphics[height=0.19cm]{Figure/orcid.jpg}}}$, 
T.~Bárczy\inst{\ref{inst:16}}\,$^{\href{https://orcid.org/0000-0002-7822-4413}{\protect\includegraphics[height=0.19cm]{Figure/orcid.jpg}}}$, 
D.~Barrado\inst{\ref{inst:17}}\,$^{\href{https://orcid.org/0000-0002-5971-9242}{\protect\includegraphics[height=0.19cm]{Figure/orcid.jpg}}}$, 
S.~C.~C.~Barros\inst{\ref{inst:10},\ref{inst:18}}\,$^{\href{https://orcid.org/0000-0003-2434-3625}{\protect\includegraphics[height=0.19cm]{Figure/orcid.jpg}}}$, 
W.~Baumjohann\inst{\ref{inst:4}}\,$^{\href{https://orcid.org/0000-0001-6271-0110}{\protect\includegraphics[height=0.19cm]{Figure/orcid.jpg}}}$, 
W.~Benz\inst{\ref{inst:5},\ref{inst:13}}\,$^{\href{https://orcid.org/0000-0001-7896-6479}{\protect\includegraphics[height=0.19cm]{Figure/orcid.jpg}}}$, 
N.~Billot\inst{\ref{inst:7}}\,$^{\href{https://orcid.org/0000-0003-3429-3836}{\protect\includegraphics[height=0.19cm]{Figure/orcid.jpg}}}$, 
C.~Broeg\inst{\ref{inst:5},\ref{inst:13}}\,$^{\href{https://orcid.org/0000-0001-5132-2614}{\protect\includegraphics[height=0.19cm]{Figure/orcid.jpg}}}$, 
M.~Buder\inst{\ref{inst:19}}, 
A.~Collier~Cameron\inst{\ref{inst:20}}\,$^{\href{https://orcid.org/0000-0002-8863-7828}{\protect\includegraphics[height=0.19cm]{Figure/orcid.jpg}}}$, 
C.~Corral~van~Damme\inst{\ref{inst:9}}, 
A.~C.~M.~Correia\inst{\ref{inst:21}}\,$^{\href{https://orcid.org/0000-0002-8946-8579}{\protect\includegraphics[height=0.19cm]{Figure/orcid.jpg}}}$, 
Sz.~Csizmadia\inst{\ref{inst:22}}\,$^{\href{https://orcid.org/0000-0001-6803-9698}{\protect\includegraphics[height=0.19cm]{Figure/orcid.jpg}}}$, 
P.~E.~Cubillos\inst{\ref{inst:4},\ref{inst:23}}, 
M.~B.~Davies\inst{\ref{inst:24}}\,$^{\href{https://orcid.org/0000-0001-6080-1190}{\protect\includegraphics[height=0.19cm]{Figure/orcid.jpg}}}$, 
M.~Deleuil\inst{\ref{inst:25}}\,$^{\href{https://orcid.org/0000-0001-6036-0225}{\protect\includegraphics[height=0.19cm]{Figure/orcid.jpg}}}$, 
A.~Deline\inst{\ref{inst:7}}, 
O.~D.~S.~Demangeon\inst{\ref{inst:10},\ref{inst:18}}\,$^{\href{https://orcid.org/0000-0001-7918-0355}{\protect\includegraphics[height=0.19cm]{Figure/orcid.jpg}}}$, 
B.-O.~Demory\inst{\ref{inst:13},\ref{inst:5}}\,$^{\href{https://orcid.org/0000-0002-9355-5165}{\protect\includegraphics[height=0.19cm]{Figure/orcid.jpg}}}$, 
A.~Derekas\inst{\ref{inst:26}}, 
B.~Edwards\inst{\ref{inst:27}}, 
D.~Ehrenreich\inst{\ref{inst:7},\ref{inst:28}}\,$^{\href{https://orcid.org/0000-0001-9704-5405}{\protect\includegraphics[height=0.19cm]{Figure/orcid.jpg}}}$, 
A.~Erikson\inst{\ref{inst:22}}, 
J.~Farinato\inst{\ref{inst:3}}\,$^{\href{https://orcid.org/0000-0002-5840-8362}{\protect\includegraphics[height=0.19cm]{Figure/orcid.jpg}}}$, 
A.~Fortier\inst{\ref{inst:5},\ref{inst:13}}\,$^{\href{https://orcid.org/0000-0001-8450-3374}{\protect\includegraphics[height=0.19cm]{Figure/orcid.jpg}}}$, 
M.~Fridlund\inst{\ref{inst:29},\ref{inst:30}}\,$^{\href{https://orcid.org/0000-0002-0855-8426}{\protect\includegraphics[height=0.19cm]{Figure/orcid.jpg}}}$, 
D.~Gandolfi\inst{\ref{inst:31}}\,$^{\href{https://orcid.org/0000-0001-8627-9628}{\protect\includegraphics[height=0.19cm]{Figure/orcid.jpg}}}$, 
K.~Gazeas\inst{\ref{inst:32}}\,$^{\href{https://orcid.org/0000-0002-8855-3923}{\protect\includegraphics[height=0.19cm]{Figure/orcid.jpg}}}$, 
M.~Gillon\inst{\ref{inst:33}}\,$^{\href{https://orcid.org/0000-0003-1462-7739}{\protect\includegraphics[height=0.19cm]{Figure/orcid.jpg}}}$, 
M.~Güdel\inst{\ref{inst:34}}, 
Ch.~Helling\inst{\ref{inst:4},\ref{inst:35}}, 
K.~G.~Isaak\inst{\ref{inst:9}}\,$^{\href{https://orcid.org/0000-0001-8585-1717}{\protect\includegraphics[height=0.19cm]{Figure/orcid.jpg}}}$, 
L.~L.~Kiss\inst{\ref{inst:36},\ref{inst:37}}, 
J.~Korth\inst{\ref{inst:38}}\,$^{\href{https://orcid.org/0000-0002-0076-6239}{\protect\includegraphics[height=0.19cm]{Figure/orcid.jpg}}}$, 
K.~W.~F.~Lam\inst{\ref{inst:22}}\,$^{\href{https://orcid.org/0000-0002-9910-6088}{\protect\includegraphics[height=0.19cm]{Figure/orcid.jpg}}}$, 
J.~Laskar\inst{\ref{inst:39}}\,$^{\href{https://orcid.org/0000-0003-2634-789X}{\protect\includegraphics[height=0.19cm]{Figure/orcid.jpg}}}$, 
A.~Lecavelier~des~Etangs\inst{\ref{inst:40}}\,$^{\href{https://orcid.org/0000-0002-5637-5253}{\protect\includegraphics[height=0.19cm]{Figure/orcid.jpg}}}$, 
M.~Lendl\inst{\ref{inst:7}}\,$^{\href{https://orcid.org/0000-0001-9699-1459}{\protect\includegraphics[height=0.19cm]{Figure/orcid.jpg}}}$, 
D.~Magrin\inst{\ref{inst:3}}\,$^{\href{https://orcid.org/0000-0003-0312-313X}{\protect\includegraphics[height=0.19cm]{Figure/orcid.jpg}}}$, 
P.~F.~L.~Maxted\inst{\ref{inst:41}}\,$^{\href{https://orcid.org/0000-0003-3794-1317}{\protect\includegraphics[height=0.19cm]{Figure/orcid.jpg}}}$, 
B.~Merín\inst{\ref{inst:42}}\,$^{\href{https://orcid.org/0000-0002-8555-3012}{\protect\includegraphics[height=0.19cm]{Figure/orcid.jpg}}}$, 
C.~Mordasini\inst{\ref{inst:5},\ref{inst:13}}, 
G.~Olofsson\inst{\ref{inst:8}}\,$^{\href{https://orcid.org/0000-0003-3747-7120}{\protect\includegraphics[height=0.19cm]{Figure/orcid.jpg}}}$, 
R.~Ottensamer\inst{\ref{inst:34}}, 
I.~Pagano\inst{\ref{inst:43}}\,$^{\href{https://orcid.org/0000-0001-9573-4928}{\protect\includegraphics[height=0.19cm]{Figure/orcid.jpg}}}$, 
E.~Pallé\inst{\ref{inst:14},\ref{inst:15}}\,$^{\href{https://orcid.org/0000-0003-0987-1593}{\protect\includegraphics[height=0.19cm]{Figure/orcid.jpg}}}$, 
G.~Peter\inst{\ref{inst:19}}\,$^{\href{https://orcid.org/0000-0001-6101-2513}{\protect\includegraphics[height=0.19cm]{Figure/orcid.jpg}}}$, 
D.~Piazza\inst{\ref{inst:44}}, 
D.~Pollacco\inst{\ref{inst:6}}, 
D.~Queloz\inst{\ref{inst:45},\ref{inst:46}}\,$^{\href{https://orcid.org/0000-0002-3012-0316}{\protect\includegraphics[height=0.19cm]{Figure/orcid.jpg}}}$, 
R.~Ragazzoni\inst{\ref{inst:3},\ref{inst:1}}\,$^{\href{https://orcid.org/0000-0002-7697-5555}{\protect\includegraphics[height=0.19cm]{Figure/orcid.jpg}}}$, 
N.~Rando\inst{\ref{inst:9}}, 
H.~Rauer\inst{\ref{inst:22},\ref{inst:47}}\,$^{\href{https://orcid.org/0000-0002-6510-1828}{\protect\includegraphics[height=0.19cm]{Figure/orcid.jpg}}}$, 
I.~Ribas\inst{\ref{inst:48},\ref{inst:49}}\,$^{\href{https://orcid.org/0000-0002-6689-0312}{\protect\includegraphics[height=0.19cm]{Figure/orcid.jpg}}}$, 
N.~C.~Santos\inst{\ref{inst:10},\ref{inst:18}}\,$^{\href{https://orcid.org/0000-0003-4422-2919}{\protect\includegraphics[height=0.19cm]{Figure/orcid.jpg}}}$, 
G.~Scandariato\inst{\ref{inst:43}}\,$^{\href{https://orcid.org/0000-0003-2029-0626}{\protect\includegraphics[height=0.19cm]{Figure/orcid.jpg}}}$, 
D.~Ségransan\inst{\ref{inst:7}}\,$^{\href{https://orcid.org/0000-0003-2355-8034}{\protect\includegraphics[height=0.19cm]{Figure/orcid.jpg}}}$, 
A.~E.~Simon\inst{\ref{inst:5},\ref{inst:13}}\,$^{\href{https://orcid.org/0000-0001-9773-2600}{\protect\includegraphics[height=0.19cm]{Figure/orcid.jpg}}}$, 
A.~M.~S.~Smith\inst{\ref{inst:22}}\,$^{\href{https://orcid.org/0000-0002-2386-4341}{\protect\includegraphics[height=0.19cm]{Figure/orcid.jpg}}}$, 
M.~Stalport\inst{\ref{inst:12},\ref{inst:33}}, 
S.~Sulis\inst{\ref{inst:25}}\,$^{\href{https://orcid.org/0000-0001-8783-526X}{\protect\includegraphics[height=0.19cm]{Figure/orcid.jpg}}}$, 
Gy.~M.~Szabó\inst{\ref{inst:26},\ref{inst:50}}\,$^{\href{https://orcid.org/0000-0002-0606-7930}{\protect\includegraphics[height=0.19cm]{Figure/orcid.jpg}}}$, 
S.~Udry\inst{\ref{inst:7}}\,$^{\href{https://orcid.org/0000-0001-7576-6236}{\protect\includegraphics[height=0.19cm]{Figure/orcid.jpg}}}$, 
B.~Ulmer\inst{\ref{inst:19}}, 
S.~Ulmer-Moll\inst{\ref{inst:51},\ref{inst:12}}\,$^{\href{https://orcid.org/0000-0003-2417-7006}{\protect\includegraphics[height=0.19cm]{Figure/orcid.jpg}}}$, 
V.~Van~Grootel\inst{\ref{inst:12}}\,$^{\href{https://orcid.org/0000-0003-2144-4316}{\protect\includegraphics[height=0.19cm]{Figure/orcid.jpg}}}$, 
J.~Venturini\inst{\ref{inst:7}}\,$^{\href{https://orcid.org/0000-0001-9527-2903}{\protect\includegraphics[height=0.19cm]{Figure/orcid.jpg}}}$, 
E.~Villaver\inst{\ref{inst:14},\ref{inst:15}}, 
N.~A.~Walton\inst{\ref{inst:52}}\,$^{\href{https://orcid.org/0000-0003-3983-8778}{\protect\includegraphics[height=0.19cm]{Figure/orcid.jpg}}}$, and
S.~Wolf\inst{\ref{inst:5}}
}

\authorrunning{Leonardi et al.}
\titlerunning{HIP\,41378\,b \&\,c with CHEOPS and TESS.}

    \institute{
            \label{inst:1} Dipartimento di Fisica, Università di Trento, Via Sommarive 14, 38123 Povo \and
            \label{inst:2} Dipartimento di Fisica e Astronomia, Università degli Studi di Padova, Vicolo dell’Osservatorio 3, 35122 Padova, Italy \and
            \label{inst:3} INAF, Osservatorio Astronomico di Padova, Vicolo dell'Osservatorio 5, 35122 Padova, Italy \and
            \label{inst:4} Space Research Institute, Austrian Academy of Sciences, Schmiedlstrasse 6, A-8042 Graz, Austria \and
            \label{inst:5} Space Research and Planetary Sciences, Physics Institute, University of Bern, Gesellschaftsstrasse 6, 3012 Bern, Switzerland \and
            \label{inst:6} Department of Physics, University of Warwick, Gibbet Hill Road, Coventry CV4 7AL, United Kingdom \and
            \label{inst:7} Observatoire astronomique de l'Université de Genève, Chemin Pegasi 51, 1290 Versoix, Switzerland \and
            \label{inst:8} Department of Astronomy, Stockholm University, AlbaNova University Center, 10691 Stockholm, Sweden \and
            \label{inst:9} European Space Agency (ESA), European Space Research and Technology Centre (ESTEC), Keplerlaan 1, 2201 AZ Noordwijk, The Netherlands \and
            \label{inst:10} Instituto de Astrofisica e Ciencias do Espaco, Universidade do Porto, CAUP, Rua das Estrelas, 4150-762 Porto, Portugal \and
            \label{inst:11} Dipartimento di Fisica e Astronomia "Galileo Galilei", Università degli Studi di Padova, Vicolo dell'Osservatorio 3, 35122 Padova, Italy \and
            \label{inst:12} Space sciences, Technologies and Astrophysics Research (STAR) Institute, Université de Liège, Allée du 6 Août 19C, 4000 Liège, Belgium \and
            \label{inst:13} Center for Space and Habitability, University of Bern, Gesellschaftsstrasse 6, 3012 Bern, Switzerland \and
            \label{inst:14} Instituto de Astrofísica de Canarias, Vía Láctea s/n, 38200 La Laguna, Tenerife, Spain \and
            \label{inst:15} Departamento de Astrofísica, Universidad de La Laguna, Astrofísico Francisco Sanchez s/n, 38206 La Laguna, Tenerife, Spain \and
            \label{inst:16} Admatis, 5. Kandó Kálmán Street, 3534 Miskolc, Hungary \and
            \label{inst:17} Depto. de Astrofísica, Centro de Astrobiología (CSIC-INTA), ESAC campus, 28692 Villanueva de la Cañada (Madrid), Spain \and
            \label{inst:18} Departamento de Fisica e Astronomia, Faculdade de Ciencias, Universidade do Porto, Rua do Campo Alegre, 4169-007 Porto, Portugal \and
            \label{inst:19} Institute of Optical Sensor Systems, German Aerospace Center (DLR), Rutherfordstrasse 2, 12489 Berlin, Germany \and
            \label{inst:20} Centre for Exoplanet Science, SUPA School of Physics and Astronomy, University of St Andrews, North Haugh, St Andrews KY16 9SS, UK \and
            \label{inst:21} CFisUC, Departamento de Física, Universidade de Coimbra, 3004-516 Coimbra, Portugal \and
            \label{inst:22} Institute of Planetary Research, German Aerospace Center (DLR), Rutherfordstrasse 2, 12489 Berlin, Germany \and
            \label{inst:23} INAF, Osservatorio Astrofisico di Torino, Via Osservatorio, 20, I-10025 Pino Torinese To, Italy \and
            \label{inst:24} Centre for Mathematical Sciences, Lund University, Box 118, 221 00 Lund, Sweden \and
            \label{inst:25} Aix Marseille Univ, CNRS, CNES, LAM, 38 rue Frédéric Joliot-Curie, 13388 Marseille, France \and
            \label{inst:26} ELTE Gothard Astrophysical Observatory, 9700 Szombathely, Szent Imre h. u. 112, Hungary \and
            \label{inst:27} SRON Netherlands Institute for Space Research, Niels Bohrweg 4, 2333 CA Leiden, Netherlands \and
            \label{inst:28} Centre Vie dans l’Univers, Faculté des sciences, Université de Genève, Quai Ernest-Ansermet 30, 1211 Genève 4, Switzerland \and
            \label{inst:29} Leiden Observatory, University of Leiden, PO Box 9513, 2300 RA Leiden, The Netherlands \and
            \label{inst:30} Department of Space, Earth and Environment, Chalmers University of Technology, Onsala Space Observatory, 439 92 Onsala, Sweden \and
            \label{inst:31} Dipartimento di Fisica, Università degli Studi di Torino, via Pietro Giuria 1, I-10125, Torino, Italy \and
            \label{inst:32} National and Kapodistrian University of Athens, Department of Physics, University Campus, Zografos GR-157 84, Athens, Greece \and
            \label{inst:33} Astrobiology Research Unit, Université de Liège, Allée du 6 Août 19C, B-4000 Liège, Belgium \and
            \label{inst:34} Department of Astrophysics, University of Vienna, Türkenschanzstrasse 17, 1180 Vienna, Austria \and
            \label{inst:35} Institute for Theoretical Physics and Computational Physics, Graz University of Technology, Petersgasse 16, 8010 Graz, Austria \and
            \label{inst:36} Konkoly Observatory, Research Centre for Astronomy and Earth Sciences, 1121 Budapest, Konkoly Thege Miklós út 15-17, Hungary \and
            \label{inst:37} ELTE E\"otv\"os Lor\'and University, Institute of Physics, P\'azm\'any P\'eter s\'et\'any 1/A, 1117 Budapest, Hungary \and
            \label{inst:38} Lund Observatory, Division of Astrophysics, Department of Physics, Lund University, Box 118, 22100 Lund, Sweden \and
            \label{inst:39} IMCCE, UMR8028 CNRS, Observatoire de Paris, PSL Univ., Sorbonne Univ., 77 av. Denfert-Rochereau, 75014 Paris, France \and
            \label{inst:40} Institut d'astrophysique de Paris, UMR7095 CNRS, Université Pierre \& Marie Curie, 98bis blvd. Arago, 75014 Paris, France \and
            \label{inst:41} Astrophysics Group, Lennard Jones Building, Keele University, Staffordshire, ST5 5BG, United Kingdom \and
            \label{inst:42} European Space Agency, ESA - European Space Astronomy Centre, Camino Bajo del Castillo s/n, 28692 Villanueva de la Cañada, Madrid, Spain \and
            \label{inst:43} INAF, Osservatorio Astrofisico di Catania, Via S. Sofia 78, 95123 Catania, Italy \and
            \label{inst:44} Weltraumforschung und Planetologie, Physikalisches Institut, University of Bern, Gesellschaftsstrasse 6, 3012 Bern, Switzerland \and
            \label{inst:45} ETH Zurich, Department of Physics, Wolfgang-Pauli-Strasse 2, CH-8093 Zurich, Switzerland \and
            \label{inst:46} Cavendish Laboratory, JJ Thomson Avenue, Cambridge CB3 0HE, UK \and
            \label{inst:47} Institut fuer Geologische Wissenschaften, Freie Universitaet Berlin, Maltheserstrasse 74-100,12249 Berlin, Germany \and
            \label{inst:48} Institut de Ciencies de l'Espai (ICE, CSIC), Campus UAB, Can Magrans s/n, 08193 Bellaterra, Spain \and
            \label{inst:49} Institut d'Estudis Espacials de Catalunya (IEEC), 08860 Castelldefels (Barcelona), Spain \and
            \label{inst:50} HUN-REN-ELTE Exoplanet Research Group, Szent Imre h. u. 112., Szombathely, H-9700, Hungary \and
            \label{inst:51} Leiden Observatory, University of Leiden, Einsteinweg 55, 2333 CA Leiden, The Netherlands \and
            \label{inst:52} Institute of Astronomy, University of Cambridge, Madingley Road, Cambridge, CB3 0HA, United Kingdom
}

\date{Received \today; Accepted September 11, 2025}

  \abstract{
  In multiple-planet systems, gravitational interactions of exoplanets could lead to transit timing variations (TTVs), whose amplitude becomes significantly enhanced when planets are in or near mean-motion resonances (MMRs), making them more easily detectable. In cases where both TTVs and radial velocity (RV) measurements are available, combined analysis can break degeneracies and provide robust planetary and system characterization, even detecting non-transiting planets.

  In this context, HIP\,41378 hosts five confirmed transiting planets with periods ranging from 15 to over 542 days, providing a unique dynamical laboratory  for investigating wide multi-planet systems analogous to the Solar System.
  In this study, we present an intensive space-based photometric follow-up of HIP\,41378, combining 15 new CHEOPS observations with eight TESS sectors, alongside data from K2, Spitzer, HST, and 311 HARPS spectra. We dynamically modeled the TTVs and RV signals  of the two inner sub-Neptunes via N-body integration. These planets, HIP\,41378\,b ($P_{b}$ = 15.57 days, $R_{b} = 2.45\,R_{\oplus}$) and HIP\,41378\,c ($P_{c}$ = 31.71 days, $R_{c} = 2.57\,R_{\oplus}$), are close to ($\Delta\sim1.8$\%) a 2:1 period commensurability. We report a clear detection of TTVs with amplitudes of 20 minutes for planet b and greater than 3 hours for planet c.

  We dynamically confirm the planetary nature of HIP\,41378\,g, a non-transiting planet with a period of about 64 days and a mass of about 7 $M_{\oplus}$, close to a 2:1 commensurability with planet c, suggesting a possible mean-motion resonance chain in the inner system.

  Our precise determination of the masses, eccentricities, and radii of HIP\,41378\, b and c enabled us to investigate their possible volatile-rich compositions. Finally, by leveraging on the last TESS sectors we constrained the period of HIP\,41378\,d to three possible aliases ($P_{d} =$ 278, 371, and 1113~days) suggesting that the system could be placed in a double quasi resonant chain, highlighting its complex dynamical architecture.
  }


   \maketitle

\nolinenumbers

\section{Introduction}
\label{section:intro}
Following the conclusion of the primary \textit{Kepler} mission, NASA’s \textit{K2} mission expanded the search for transiting exoplanets by targeting stars along the ecliptic plane \citep{howel_2014}. Unlike Kepler, which focused on a fixed field, K2 covered a broader sky region, observed a more diverse stellar population, and prioritized brighter stars, making them ideal candidates for radial-velocity (RV) follow-up studies. These extensive long-term surveys led to the unexpected discovery of numerous sub-Neptune-sized planets (1.75~$R_{\oplus}$ $\lesssim R_{p} \lesssim$ 3.5~$R_{\oplus}$, following \citealt{kopparapu_2018}) in compact, coplanar multi-planetary systems \citep[e.g.,][]{borucki_2011, latham_2011, weiss_2018, bean_2021}. The prevalence of such systems has made them key cornerstones of exoplanet research, offering crucial insights into planetary formation and evolution while bridging the gap between the Solar System and the wide diversity of known exoplanetary architectures \citep{bean_2021}. Among this population, sub-Neptunes on long-period orbits are exceptionally valuable. Situated at larger orbital distances, they are shielded from the intense stellar X-ray and ultraviolet (XUV) radiation that drives atmospheric escape via photoevaporation \citep{lopez_fortney_2014, Owen_Wu_2017}. Consequently, these planets are expected to retain their primordial atmospheres. Their present-day atmospheric properties, such as mass fraction and composition, therefore, offer a more direct probe of the conditions within the proto-planetary disk during their formation as well as their evolution. This makes them prime targets for constraining planet formation models through atmospheric characterization studies \citep{Madhusudhan_2019, bean_2021}.
In this context, the multi-planet system around the bright ($m_{V}$ = 8.93) late F-type star HIP\,41378 (K2-93) stands out as one of only five systems, alongside those  around 55 Cnc \citep{butler_1997}, HD 219134 \citep{gillon_2017b}, HD 110067 \citep{luque_2023}, HD 191939 \citep{badenas_agusti_2020}, and Kepler-444 \citep{campante_2015} that host more than four confirmed planets with both mass and radius constraints, while also hosting a star brighter than $m_{V}$ = 9 mag.

The system was first identified by \citet{vanderburg_2016} during Campaign 5 of the \textit{K2} mission (April–July 2015). After its discovery, the system was reobserved three years later during Campaign 18 (May–July 2018) (\citealt{berardo_2019}, hereafter \citetalias{berardo_2019}; \citealt{becker_2019}). The \textit{K2} data revealed a system of five transiting planets (from b to f), with the two inner sub-Neptunes (HIP\,41378\,b and c) exhibiting well-constrained orbital periods of 15.6 and 31.7 days respectively, near a 2:1 period commensurability. However, for the three outer planets, an insufficient number of transits were recovered to determine their orbital periods, leaving only a set of possible period aliases. Based on the 75-day baseline of the \textit{K2} C5 campaign and the long transit durations of the planets, the highest-probability aliases derived suggested long-period orbits for the outer planets ($P > 100$~days).
The first mass measurements of the planets in the system were obtained by \citet[][hereafter \citetalias{santerne_2019}]{santerne_2019} using radial velocity (RV) observations from HARPS, HARPS-N, the Carnegie Planet Finder Spectrograph \citep[PFS;][]{crane_2006, crane_2008, crane_2010} and HIRES (High Resolution Echelle Spectrometer). The authors detected strong RV signals from planets b, c, and f, but were unable to retrieve any signal from planets d and e. Additionally, they identified a periodic signal at $\sim 62~\mathrm{days}$, attributed to a possible non-transiting planet (hereafter HIP\,41378\,g) near a 2:1 period commensurability with planet c raising the possibility of strong dynamical interactions among the inner three planets in a near-resonant chain, which could result in observable transit timing variations (TTVs) \citep{agol_2005, holman_2005}. Their analysis also constrained the orbital period of HIP\,41378\,f to $P_f = 542$ days, making it one of the longest-period planets discovered via transit photometry. A subsequent intensive ground-based follow-up by \citet{bryant_2021} confirmed this period and revealed significant TTVs, indicating strong dynamical interactions among the outer planets.

In this work, we present a global dynamical analysis of the sub-Neptunes HIP\,41378\,b and HIP\,41378\,c. This combines 15 single-visit observations from the \textit{CHaracterising ExOPlanets Satellite} (CHEOPS; \citealt{benz_2021}), eight new sectors from the \textit{Transiting Exoplanet Survey Satellite} (TESS; \citealt{ricker_2015}), archival photometry from \textit{Kepler} (\textit{K2}), \textit{Spitzer}, and the Hubble Space Telescope (HST), and RV measurements from HARPS \citep{Mayor_2003}.
Using measurements from Spitzer, \citetalias{berardo_2019} reports hints of TTV for planet c, with variations exceeding one hour. Motivated by this detection, we analyzed the potential TTVs of HIP\,41378\,c and HIP\,41378\,b. Additionally, we examined the dynamical influence of the non-transiting candidate planet HIP\,41378\,g, assessing its role in shaping the observed TTVs and the overall dynamics of the system.
The paper is organized as follows:
In Section~\ref{section:observations}, we describe the spectroscopic and photometric observations.  
Section~\ref{section:stellar_parameters} presents the newly derived stellar parameters.
Section~\ref{section:data_analysis} details the photometric and dynamical modeling, including transit timing extraction and orbital parameter retrieval. Our results are presented in Section~\ref{section:results}.
In Section~\ref{section:discussion}, we discuss the system's architecture, the implications of the inner planets' "peas-in-a-pod" configuration, and potential planetary compositions. Lastly, Section~\ref{section:conclusions} summarizes the key findings and outlines future observational priorities.

\section{Observations extraction and reduction}
\label{section:observations}
This section presents the observations and data extraction of both proprietary CHEOPS light curves and publicly available spectroscopic and photometric data for HIP\,41378\,b \& c (including K2, TESS, \textit{Spitzer}, HST, and HARPS), spanning over ten years and comprising a total of 47 transit light curves (32 for -b and 15 for -c). For each TESS sector and \textit{K2} campaign, we isolated individual transits of both planets by selecting portions of the light curve that encompassed the transit duration plus three CHEOPS orbits ($\sim98.77$ minutes each). This allowed us to have a consistent out-of-transit baseline across all observations. The center of the transits, determined using a linear ephemeris, and the transit durations were based on the values reported by \citetalias{berardo_2019}.

\paragraph{K2}
The \textit{K2} mission observed the system during campaign 5 (2015 April 27 – 2015 July 10) and campaign 18 (2018 May 12 – 2018 July 02), under the identifier EPIC  211311380 (K2-93), with long- and short-cadence photometry (30 minutes and 1 minute, respectively).
A total of ten transits of planets b \& c were observed during the \textit{K2} mission (seven of b and three of c). 
From the Mikulski Archive for Space Telescopes (MAST\footnote{\url{https://mast.stsci.edu/portal/Mashup/Clients/Mast/Portal.html}}) we extracted the high-level science data products (HLSP) based on the photometric pipeline \texttt{EVEREST} \citep[EPIC Variability Extraction and Removal for Exoplanet Science Targets, version 2.0;][]{luger_2016,luger_2018}.

\paragraph{\textit{Spitzer}}
\label{section:spitzer}
We used photometric data of HIP\,41378 collected with the Infrared Array Camera (IRAC) 4.5 $\mu m$ channel of \textit{Spitzer} telescope \citep{werner_2004}, taken as part of the observing programs 11026 and 13052 (PI: Werner), focused on \textit{K2} follow-ups. The observations, presented by \citetalias{berardo_2019}, cover a transit of planet b and one of planet c.

\paragraph{TESS}
The \textit{TESS} mission \cite{ricker_2015} observed the system (TOI-4304, TIC 366443426) from Cycles 1 to 7 in sectors: 7, 34, 44, 45, 46, 61, 72 and 88 with a cadence of 120~s.
We downloaded the photometric time series processed by the Science Processing Operations Center (SPOC; \citealt{jenkins_2016}) from the MAST archive, and we corrected the simple aperture photometry (SAP) for systematic effects by following the procedure and using the Cotrending Basis Vectors described in \citet{nardiello_2021,nardiello_2025}. As shown in \citet{nardiello_2022}, Pre-search Data Conditioned SAP (PDC-SAP) light curves can suffer from overcorrection problems, that can introduce new systematic errors in the light curves, change the shape of the stellar activity and planetary transits, and also mimic false transit signals. A total of nine transits of planet b and six of planet c were extracted from the light curves. The first transit of planet b during sector 72 fell in a data discontinuity gap; thus, only a partial transit light curve was retrieved.

\paragraph{HST}
We used the publicly available HST/WFC3 \citep{marinelli_2024} transit observations of HIP\,41378\,b from the MAST archive. 
These data, obtained using the G141 grism (1.088 to 1.680 $\mu$m), cover three transits of the planet (January 14, 2018, May 3, 2020, and May 20, 2020) and were taken as part of program GO-15333 (PI: Ian Crossfield).
The data were first published by \cite{Edwards_2023b}, and later reanalyzed by \cite{brande_2024}.
The calibration of the raw WFC3 data, the reduction and the extraction of the white light curves were done using the \textsc{Iraclis} dedicated pipeline \citep{Tsiaras_2016a, Tsiaras_2016b, Tsiaras_2018}, following the methodology of \citet{Edwards_2023b}. Our reduction also includes the frame-splitting method (see \citealt{Edwards_2023a}, for a complete description) that takes into account the persistence effect dependent upon the brightness of the host star, the scanning rate, and the readout scheme employed.

\paragraph{CHEOPS}
\label{Obs_section:CHEOPS}
HIP\,41378 was observed with CHEOPS within the frame of the guaranteed time observation (GTO) as part of two programs: M-R relation in planetary systems\footnote{CH\_PR100025, V. Nascimbeni}, dedicated to the follow-up of TTVs in planetary systems to better constrain masses, orbital parameters, and planetary compositions \citep{nascimbeni_2023, nascimbeni_2024}, and architecture of resonant chains (ARC)\footnote{CH\_PR140080, A. Leleu}, centered on the follow-up of resonant chains \citep{leleu_2021, delrez_2023, leleu_2024}.
We obtained 15 visits acquired between December 23, 2020, and March 11, 2025, of which 11 were centered on transits of planet b and four on those of planet c.
For each observation we used the exposure cadence of 38 seconds, avoiding saturation.
The complete log of the observations is reported in Table \ref{table:log_CHEOPS}.
The CHEOPS raw data were automatically processed by the CHEOPS data reduction pipeline (DRP v13.1.0; \citealt{hoyer_2020}). The DRP corrects for instrumental (e.g bias, flat and dark current) and environmental effects (e.g., cosmic rays, background) \citep{Fortier_2024}. 
The pipeline performs aperture photometry extracting four different light curves. For our study we used the light curve corresponding to the DEFAULT photometric aperture of 25 pixels, which has the lowest rms.
Following the extraction of the light curves we performed a clipping of the outliers with respect to the median flux value of the light curves plus five times the mean absolute deviation (MAD).

\paragraph{HARPS}
\label{section:observations_rv}
We recovered the publicly available HARPS \citep{Mayor_2003} high-precision RV observations from the ESO Science Archive website\footnote{\url{http://archive.eso.org/wdb/wdb/adp/phase3\_main/form}}.
We downloaded 370 spectra obtained under the observing programs 198.C-0169(A) and 0102.C-0171(A) (PI: Santerne) previously published by \cite{santerne_2019}. The system was monitored between January 2017 and April 2019, with a typical integration time of 15 minutes.
The observations yielded a median formal measurement uncertainty of 2 m\,$\mathrm{s^{-1}}$. 
The spectra were extracted using the HARPS online data reduction pipeline (DRS) (\citealt{cosentino_2014}, version 3.8).
We rejected the points taken to monitor granulation and p-modes in the two consecutive nights: March 10, 2018, and March 11, 2018.
We additionally discarded all the data points with an error greater than 5$\sigma$ of the median value. These selections left us with 311 RV measurements.

\section{Stellar parameters}
\label{section:stellar_parameters}

\begin{table}[tb]
\centering\centering\renewcommand{\arraystretch}{1.3}\small
    \caption{Derived stellar parameters of HIP\,41378.} 
    \begin{tabular}{l c}
    \hline\hline
    Parameter & Value\\
    \hline
    T$_\mathrm{eff}$ [K] & $ 6371 \pm 65 $ \\
    $\log g$ [cgs] & $4.32 \pm 0.02 $ \\
    $[\mathrm{Fe}/\mathrm{H}]$\,[dex] & $ 0.046\pm 0.044$ \\
    $v \sin i_{\star}$ [km s$^{-1}$] & $ 7.5 \pm 0.5$ \\
    $M_{\star}$ [$M_{\odot}$] & $1.245^{+0.037}_{-0.043}$ \\ 
    $R_{\star}$ [$R_{\odot}$] & $1.306 \pm 0.010 $ \\
    $\rho_{\star}$ [$\rho_{\odot}$] & $0.557 \pm 0.016$ \\
    Age [Gyr] & $1.8_{-0.6}^{+0.7}$ \\
    $[\mathrm{Mg}/\mathrm{H}]$\,[dex] &  $0.04 \pm 0.07$ \\
    $[\mathrm{Si}/\mathrm{H}]$\,[dex] & $0.04 \pm 0.04$ \\
    \hline
    \end{tabular}
    \label{table:derived_stellar_parameters}
\end{table}
 
The stellar spectroscopic parameters ($T_{\mathrm{eff}}$, $\log g$, microturbulence $v_{\rm tur}$, and [Fe/H]) were derived using the ARES+MOOG methodology as described in \citet{Santos2013,Sousa2014,Sousa2021}. For this we used the latest version of ARES\footnote{The latest version, ARES v2, can be downloaded at \url{https://github.com/sousasag/ARES}} \citep{Sousa2007, Sousa2015} to consistently measure the equivalent widths (EW) of selected iron lines in the combined HARPS spectrum of HIP\ 41378. For this, we used the iron line list presented in \citet{Sousa2008}. The best spectroscopic parameters are found by converging into ionization and excitation equilibrium. In this process, a grid of Kurucz model atmospheres \citep{Kurucz1993} and the radiative transfer code MOOG \citep{Sneden1973} are used. We also derived a more accurate trigonometric surface gravity using the {\it Gaia} DR3 data following the same procedure as described in \citet{Sousa2021}. Stellar abundances of Si and Mg were then derived using the classical curve-of-growth analysis method assuming local thermodynamic equilibrium. The same codes and models were used for abundance determinations. For the derivation of chemical abundances of refractory elements, we closely followed the methods described in \citep[e.g.,][]{Adibekyan2012, Adibekyan2015}. All of the [X/H] ratios were obtained by performing a differential analysis with respect to a high S/N solar (Vesta) spectrum from HARPS. We determined the HIP\,41378 stellar radius using a Markov chain Monte Carlo (MCMC) modified infrared flux method \citep[IRFM --][]{blackwell_1977,schanche_2020}. Within this MCMC framework, we produced synthetic photometry from a constructed spectral energy distribution (SED) based on stellar atmosphere models \citep{castelli_2003}, using our spectroscopically derived stellar parameters as priors. To compute the stellar bolometric flux, we compared these simulated data to broadband fluxes in the following bandpasses:  2MASS $J$, $H$, and $K$, WISE $W1$ and $W2$, and \textit{Gaia} $G$, $G_\mathrm{BP}$, and $G_\mathrm{RP}$ \citep{skrutskie_2006,wright_2010,gaiacollaboration_2022}. Lastly, this was converted into the effective temperature and angular diameter, from which we derived the stellar radius via combination with the offset-corrected \textit{Gaia} parallax \citep{lindegren_2021}. Assuming $T_{\mathrm{eff}}$, [Fe/H], and $R_{\star}$ along with their uncertainties as input parameters, we derived the isochronal mass and age using two different stellar evolutionary models. The first set of mass and age estimates was computed using the isochrone placement routine \citep{bonfanti_2015,bonfanti_2016}, which interpolates the input parameters within precomputed grids of PARSEC\footnote{\textsl{PA}dova and T\textsl{R}ieste \textsl{S}tellar \textsl{E}volutionary \textsl{C}ode: \url{https://stev.oapd.inaf.it/cgi-bin/cmd}} v1.2S \citep{marigo_2017} isochrones and evolutionary tracks. The second set of mass and age values, instead, was estimated via the CLES \citep[Code Liègeois d'Évolution Stellaire;][]{scuflaire_2008} code, which builds up the best-fit evolutionary track following a Levenberg-Marquadt minimization scheme \citep[see, e.g.,][]{salmon_2021}.
After checking the mutual consistency of the two respective pairs of outcomes via the $\chi^2$ criterion outlined in \citet{bonfanti_2021}, we finally computed our final estimates for the mass and age that turned out to be $M_{\star}=1.245_{-0.043}^{+0.037}\,M_{\odot}$ and $t_{\star}=1.8_{-0.6}^{+0.7}$ Gyr; see \citet{bonfanti_2021} for further details about the statistical treatment. The stellar parameters are summarized in Table~\ref{table:derived_stellar_parameters}. 
Our derived stellar parameters were found to be consistent with asteroseismic values from \cite{Lund_2019}. Given the higher precision offered by asteroseismology, we adopted the asteroseismic constraints from \cite{Lund_2019} as priors for all our subsequent analysis.

\section{Data analysis}
\label{section:data_analysis}

\subsection{CHEOPS-only analysis}
\label{section:cheops_detrending}
Upon visually examining the CHEOPS data, each light curve presented significant systematics.
We used the \pyorbit\footnote{\url{https://github.com/LucaMalavolta/PyORBIT}} software \citep{malavolta_2016, malavolta_2018} to simultaneously fit a transit model and detrend each individual CHEOPS light curve. 
All visits corresponding to the same planet were modeled using a common transit model.  A set of 14 instrumental and environmental detrending vectors\footnote{Here, $f$ refers to the measured flux.} was applied globally across all light curves. These included: spacecraft roll-angle($\phi$) ($\rmd f/\rmd \cos\phi$,  $\rmd f/\rmd \sin\phi$, 
$\rmd f/\rmd \cos2\phi$, $\rmd f/\rmd \sin2\phi$,
$\rmd f/\rmd \cos3\phi$, $\rmd f/\rmd \sin3\phi$), background level ($\rmd f/\rmd bg$), photometric contamination estimate ($\mathrm{d}f/\mathrm{d}\textrm{contam}$), smear estimate ($\mathrm{d}f/\mathrm{d}\textrm{smear}$), thermal variation $\Delta T$ of CHEOPS sensors, x centroid and y centroid ($\mathrm{d}f/\mathrm{d}x$, $\mathrm{d}^2f/\mathrm{d}x^2$, $\mathrm{d}f/\mathrm{d}y$,  $\mathrm{d}^2f/\mathrm{d}y^2$).
The detrended transit light curves, where then used for the global fitting.

\subsection{Global transit light curve modeling}
\label{section:pyorbit_analysis}
We homogeneously analyzed all the transit light curves to retrieve the individual planetary parameters from the photometric datasets (i.e., CHEOPS, TESS, \textit{K2}, HST and \textit{Spitzer}) using \pyorbit{}.
The transits were modeled with the \texttt{batman} package \citep{kreidberg_2015}, and for the \textit{K2} transits we used a super-sampling factor of 30. We assumed circular orbits, by fixing the eccentricity value to zero, for the two planets. We fixed the orbital periods (P) to the values we inferred from the CHEOPS pre-modeling i.e., $P_\mathrm{b}=15.571893 \pm 0.000068$, $P_\mathrm{c}=31.70838 \pm 0.00041$ (see Sec.~\ref{section:cheops_detrending}), in order to retrieve the transit timings of each individual event. We included a third-order polynomial temporal trend (4 free parameters) for each of the \textit{K2}, Spitzer, and HST light curves. For all five instrument passbands, the stellar parameters (see Table~\ref{table:derived_stellar_parameters}) were used as Gaussian priors to compute the quadratic limb darkening (LD) coefficients with \textsc{PyLDTk} \citep{Husser_2013, Parviainen_2015}. We used these computed values, with a conservative uncertainty of 0.05, as Gaussian priors in the global analysis, using the LD parameterization ($q_{1}$, $q_{2}$) introduced by \cite{kipping_2013}. Gaussian priors were imposed on the stellar radius and mass based on the stellar spectroscopic analysis conducted in Section~\ref{section:stellar_parameters}. Uninformative uniform priors were imposed on all other free parameters (see Table~\ref{table:fit_orbital_parameters}. The analysis had a total of 134 fitting parameters: 10 LD coefficients (two for CHEOPS, TESS, \textit{K2}, \textit{Spitzer}, and HST), four planetary parameters ($b$, $R_p/R_\star$ for -b and -c), 47 transit times ($T_{0}$), stellar density ($\rho_{\star}$), 12 jitter parameters (one for each telescope, and one for each TESS sector), and 60 (15 $\times$ 4) polynomial trend coefficients. Global parameter optimization was carried out using the \pyde\footnote{\url{https://github.com/hpparvi/PyDE}} differential evolution algorithm \citep{Storn_97, Parviainen_2016}, using 100\,000 generations with a population of 10 $\times N_\mathrm{para}$, where $N_\mathrm{para}$ is the number of free parameters. The output parameters were used as the initial values for the Bayesian analysis, performed using the \texttt{emcee} package \citep{foremanmackey_2013}, which implements the affine invariant MCMC ensemble sampler \citep{goodman_weare_2010}.
We performed an autocorrelation analysis on the chains and the chains were considered converged if they were longer than 100 times the estimated autocorrelation time and this estimate varied by less than 1\%. We ran the sampler with $10\times n_\mathrm{dim}$ walkers (where $n_\mathrm{dim}$ is the number of dimensions of the model) for 450\,000 steps. We discarded the first 100\,000 steps, assuring the convergence of the chains, and set a thinning factor of 100. All the fitted parameters and their corresponding priors as well as the derived posteriors are shown in Table~\ref{table:fit_orbital_parameters}.
All the inferred central times of transit ($T_{0}$) are displayed in Tables~\ref{table:T0_b} and \ref{table:T0_c}.
We show the phase-folded light curves with the best fit transit model in Figure.~\ref{Figure:LC_phase_8}.

\begin{table}[tb]
    \small\centering\renewcommand{\arraystretch}{1.2}
    \caption{Transit times of HIP\,41378\,b from the global photometric analysis with \pyorbit.}
    \begin{tabular}{cccr}
    \hline\hline
    $T_0$ ($\mathrm{BJD_{TDB}}$) & $\sigma_{T0}$ (days) & O-C (minutes) & Telescope \\
    \hline
$2\,457\,152.2845$ & $0.0027$ & $-7.66$ & K2 \\
$2\,457\,167.8510$ & $0.0048$ & $-15.31$ & K2 \\
$2\,457\,183.4250$ & $0.0028$ & $-12.16$ & K2 \\
$2\,457\,199.0000$ & $0.0042$ & $-7.57$ & K2 \\
$2\,457\,790.7290$ & $0.0065$ & $-7.43$ & Spitzer \\
$2\,458\,133.3144$ & $0.01$ & $0.51$ & HST \\
$2\,458\,242.3269$ & $0.001$ & $14.63$ & HST \\
$2\,458\,257.8973$ & $0.0026$ & $12.60$ & K2 \\
$2\,458\,273.4700$ & $0.0028$ & $13.88$ & K2 \\
$2\,458\,289.0427$ & $0.0027$ & $15.15$ & K2 \\
$2\,458\,507.0402$ & $0.011$ & $3.80$ & TESS \\
$2\,458\,989.7603$ & $0.0011$ & $-4.99$ & HST \\
$2\,459\,207.7791$ & $0.0025$ & $14.33$ & CHEOPS \\
$2\,459\,223.3556$ & $0.013$ & $21.07$ & CHEOPS \\
$2\,459\,238.9310$ & $0.018$ & $26.24$ & TESS \\
$2\,459\,254.4889$ & $0.0028$ & $6.20$ & CHEOPS \\
$2\,459\,285.6473$ & $0.0085$ & $27.48$ & CHEOPS \\
$2\,459\,503.6172$ & $0.015$ & $-23.62$ & TESS \\
$2\,459\,519.1950$ & $0.0078$ & $-15.00$ & TESS \\
$2\,459\,534.7625$ & $0.012$ & $-21.21$ & TESS \\
$2\,459\,550.3300$ & $0.012$ & $-27.42$ & TESS \\
$2\,459\,581.4911$ & $0.0067$ & $-2.26$ & CHEOPS \\
$2\,459\,597.0476$ & $0.0053$ & $-24.31$ & CHEOPS \\
$2\,459\,628.1953$ & $0.0014$ & $-18.44$ & CHEOPS \\
$2\,459\,659.3496$ & $0.0072$ & $-3.07$ & CHEOPS \\
$2\,459\,955.2217$ & $0.0018$ & $7.95$ & CHEOPS \\
$2\,459\,970.7952$ & $0.005$ & $10.38$ & TESS \\
$2\,459\,970.7946$ & $0.0017$ & $9.51$ & CHEOPS \\
$2\,459\,986.3647$ & $0.0084$ & $7.04$ & TESS \\
$2\,460\,017.5165$ & $0.0034$ & $18.81$ & CHEOPS \\
$2\,460\,266.6507$ & $0.0069$ & $-2.51$ & TESS \\
$2\,460\,282.2123$ & $0.0035$ & $-17.22$ & TESS \\
$2\,460\,702.6520$ & $0.0045$ & $-16.14$ & TESS \\
    \hline
    \end{tabular}
    \tablefoot{The O-C values (third column) are computed with respect to the linear ephemeris: \\ $T_\mathrm{ref} = 2\,457\,152.2898 \pm0.0039\, \mathrm{BJD_{TDB}}$\\ $P_\mathrm{lin} = 15.571831\pm 0.000029$~days. \\The transit times are given in the $\mathrm{BJD_{TDB}}$ standard \citep{eastman_2010}; the second column reports the associated 1-$\sigma$ error.}
    \label{table:T0_b}
\end{table}

\begin{table}[tb]
    \small\centering\renewcommand{\arraystretch}{1.2}
    \caption{Transit times of HIP\,41378\,c from the global photometric analysis with \pyorbit{}.}
    \begin{tabular}{c c c r}
    \hline\hline
    $T_0$ ($\mathrm{BJD_{TDB}}$) & $\sigma_{T0}$ (days) & O-C (minutes) & Telescope \\
    \hline
$2\,457\,163.1671$ & $0.0035$ & $112.25$ & K2 \\
$2\,457\,194.8661$ & $0.004$ & $92.56$ & K2 \\
$2\,457\,606.9849$ & $0.004$ & $-117.49$ & Spitzer \\
$2\,458\,272.8811$ & $0.003$ & $-218.00$ & K2 \\
$2\,458\,494.8820$ & $0.015$ & $-200.39$ & TESS \\
$2\,459\,509.8919$ & $0.0094$ & $94.18$ & TESS \\
$2\,459\,541.5800$ & $0.017$ & $58.80$ & TESS \\
$2\,459\,573.3330$ & $0.014$ & $116.88$ & TESS \\
$2\,459\,636.7637$ & $0.0043$ & $124.61$ & CHEOPS \\
$2\,459\,922.1468$ & $0.004$ & $80.11$ & CHEOPS \\
$2\,459\,985.5543$ & $0.0051$ & $54.43$ & CHEOPS \\
$2\,459\,985.5541$ & $0.0089$ & $54.14$ & TESS \\
$2\,460\,270.9410$ & $0.015$ & $15.11$ & TESS \\
$2\,460\,714.8127$ & $0.0069$ & $-137.00$ & TESS \\
$2\,460\,746.5209$ & $0.0028$ & $-143.43$ & CHEOPS \\

    \hline
    \end{tabular}
    \tablefoot{The O-C values (third column) are computed with respect to the linear ephemeris: \\ $T_\mathrm{ref} = 2\,457\,163.089 \pm0.047\, \mathrm{BJD_{TDB}}$\\ $P_\mathrm{lin} = 31.71266\pm 0.00063 $~days. \\ The transit times are given in the $\mathrm{BJD_{TDB}}$ standard \citep{eastman_2010}; the second column reports the associated 1-$\sigma$ error.}
    \label{table:T0_c}
\end{table}

\subsection{Dynamical modeling with \trades{}}
\label{section:trades_analysis}
When two neighboring planets are close to a mean motion resonance (MMR), their orbital periods approach a ratio of small integers ($p/q$). To quantify the proximity to resonance, we used the fractional deviation $\Delta$, as introduced by \citet{lithwick_2012}, defined as:
\begin{equation}
    \Delta = \frac{P_{out}/P_{in}}{p/q} - 1,
\end{equation}
where $P_{out}$ and $P_{in}$ are the orbital periods of the outer and inner planets, respectively. As discussed in Section~\ref{section:intro}, the periods of the two inner planets are near the 2:1 period commensurability (p=2, q=1), with $\Delta$ $\approx$ 0.018, which suggests we might observe large TTV signals due to strong mutual gravitational interactions (\citealt{agol_2005, lithwick_2012,steffen_2012}).
We dynamically simulated the TTV signals and integrated the system parameters of HIP\,41378\,b \& c simultaneously fitting the retrieved $\mathrm{T_{0}s}$, (see Section~\ref{section:pyorbit_analysis} and Tables~\ref{table:T0_b}-\ref{table:T0_c}), and RVs (see Section~\ref{section:observations_rv}) using the N-body dynamical integrator \trades\footnote{\url{https://github.com/lucaborsato/trades}} \citep{borsato_2014, borsato_2019, borsato_2021, borsato_2022, nascimbeni_2023, borsato_2024, nascimbeni_2024}.
We selected as the start of the integration and reference time $T_{ref, dyn}$= 2\,457\,137 ($\mathrm{BJD_{TDB}}$), with the integration time ($T_\mathrm{int} = 3620$ days) chosen to cover the entire time span of all observations. 
Following the detection of an additional planetary RV signal by \citetalias{santerne_2019}, with an estimated period of approximately 62 days, near the 2:1 period commensurability with planet c, we decided to investigate this possibility further by testing two models in our analysis.

\begin{figure}[ht]
    \centering
    \includegraphics[width=1\columnwidth,keepaspectratio] {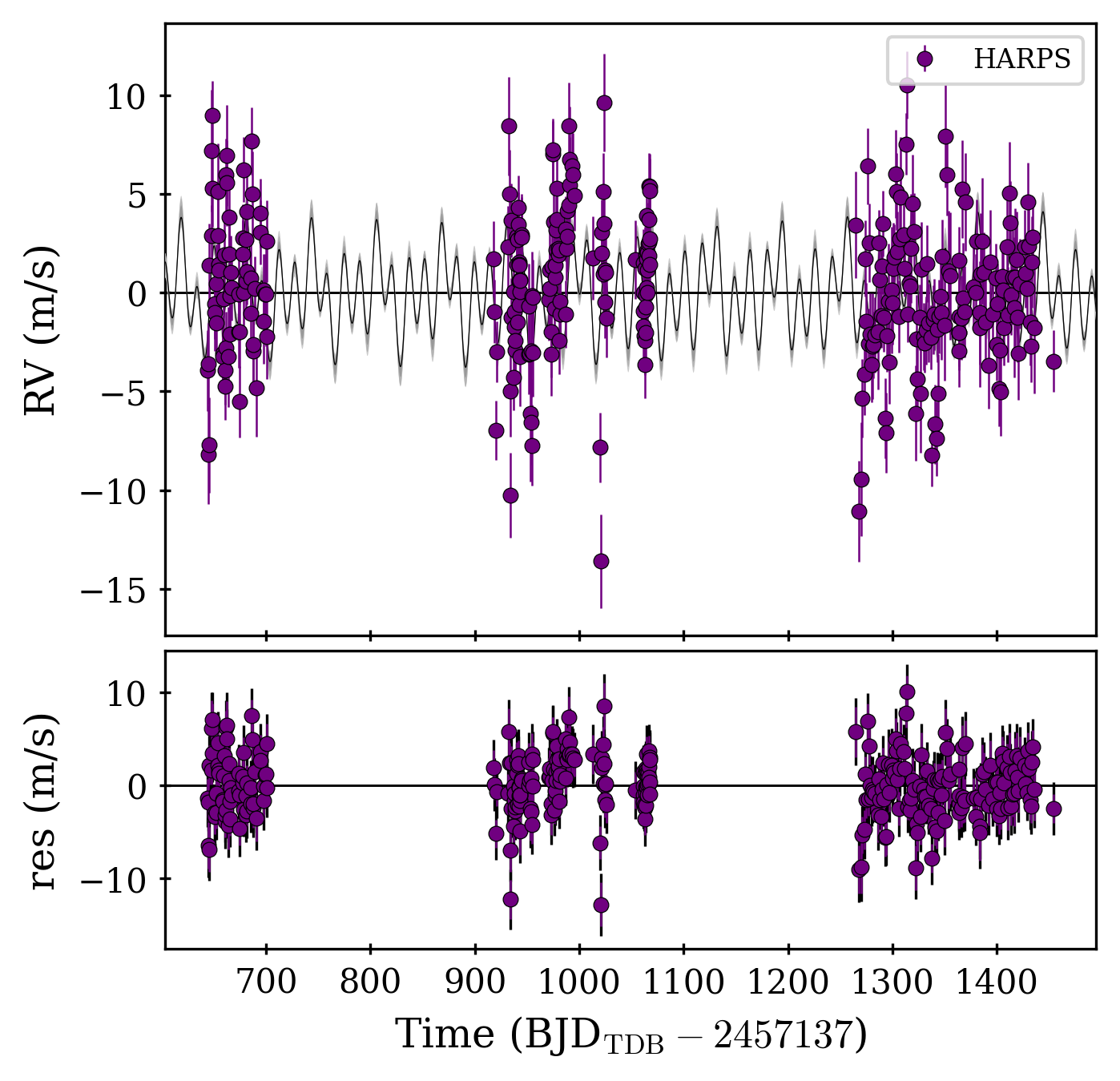}
    \caption{Radial velocities of HIP\,41378. Upper panel: RV plot minus the offset $\mathrm{\gamma}$. \trades{} MAP model is shown as a black line with the shaded gray areas indicating the 1$\sigma$, 2$\sigma$, and 3$\sigma$ confidence intervals. HARPS observations are depicted as purple circles. Lower Panel: RV residuals after subtracting the model.}
    \label{Figure:RV}
\end{figure}

\subsubsection{Two-planet model}
In the first configuration we tested a two-planet model formed by HIP\,41378\,b\, and HIP\,41378\,c.
We used as fitting parameters the planetary-to-star mass ratio $M_\mathrm{p}/M_\star$, the period $P$, the eccentricity $e$, and the mean longitude $\lambda$\footnote{$\lambda = \mathcal{M} + \omega + \Omega$, where $\mathcal{M}$ is the mean anomaly, $\omega$ is the argument of periastron (or pericenter), and $\Omega$ is the longitude of the ascending node.}.
Rather than fitting eccentricity $e$ and argument of periastron $\omega$ individually, we used the parametrization $(\sqrt{e} \cos \omega, \sqrt{e} \sin \omega)$. The mass ratios are used as fitting parameters instead of absolute masses, since TTVs only provide constraints on the relative masses of the planets and the host star. We set the longitude of ascending node $\Omega = 180^\circ$ for -b, and fit it for -c (following \citealt{winn_2010, borsato_2014}). We fixed the planetary and stellar radii ($R_\mathrm{p}$, $R_\mathrm{\star}$), stellar mass ($M_\mathrm{\star}$) and the inclination $i$, according to the values from Table~\ref{table:derived_stellar_parameters} and \ref{table:fit_orbital_parameters}.
For the RV dataset we fit a jitter term ($\sigma_\mathrm{j}$) in $\log_{2}$ and an offset (RV$_\gamma$).
We imposed half-Gaussian priors on the eccentricities following \cite{vaneylen_2019}. To assess potential biases introduced by this choice, we performed an additional fit using uninformative-uniform priors on $e$. This alternative analysis yielded posteriors consistent within 1$\sigma$ but resulted in a higher BIC\footnote{Bayesian information criterion (BIC; \citealt{schwarz_1978})} value. When comparing models using the BIC, a lower value indicates a better fit to the data, accounting for the complexity of the model. Therefore, we chose to adopt the results from the model with the half-Gaussian prior as our reference solution.
For each remaining parameter we imposed uniform-uninformative priors (see Table~\ref{table:fit_TRADES_orbital_parameters}).

\subsubsection{Three-planet model}
Building on the two-planet configuration, we performed a dynamical analysis by including an additional planet, HIP\,41378\,g. The setup for the three-planet model was identical to that of the two-planet model, with a few additional parameters introduced for the third planet. In contrast to the other planets, for planet g we additionally fit the inclination ($i$). The prior ranges for these parameters were chosen to allow for both transiting and non-transiting orbital configurations. Half-Gaussian priors were applied to the eccentricities, and uninformative priors were used for the remaining parameters (see Table~\ref{table:fit_TRADES_orbital_parameters}).
\subsection{Analysis}
For both two- and three-planet models, we first ran \pyde{} with a population size of 120  (i.e. the number of different initial parameter sets) for 70\,000 steps. The best-fit outputs from \pyde{} were then used as initial conditions for the \emcee{} package, which we ran for 600\,000 steps using 120 walkers, corresponding to 9 and 6 times the dimensionality (number of free parameters) in the respective models.
Following the methodology described in \cite{nascimbeni_2024}, we employed a combination of the differential evolution proposal (80\% of the walkers; \citealt{nelson_2014}) and the snooker differential evolution proposal (20\% of the walkers; \citealt{terbraak_2008}) as the sampler within \emcee.

We used a thinning factor of 100 and discarded 450\,000 steps as burn-in, long after the chains converged according to the Geweke \citep{geweke_1991}, Gelman-Rubin \citep{gelmanrubin_1992}, autocorrelation function \citep{goodman_weare_2010}, and visual inspection criteria. The uncertainty associated with each parameter was computed as the highest density interval (HDI) at the 68.27\% credibility level from the marginalized posterior distribution, representing the most probable region of the posterior.
Best-fit values were defined as the maximum a posteriori (MAP) estimates, computed from the posterior distributions and constrained to lie within the HDIs of the fitting parameters.

\begin{figure*}
    \centering
    \includegraphics[width=0.95\textwidth]{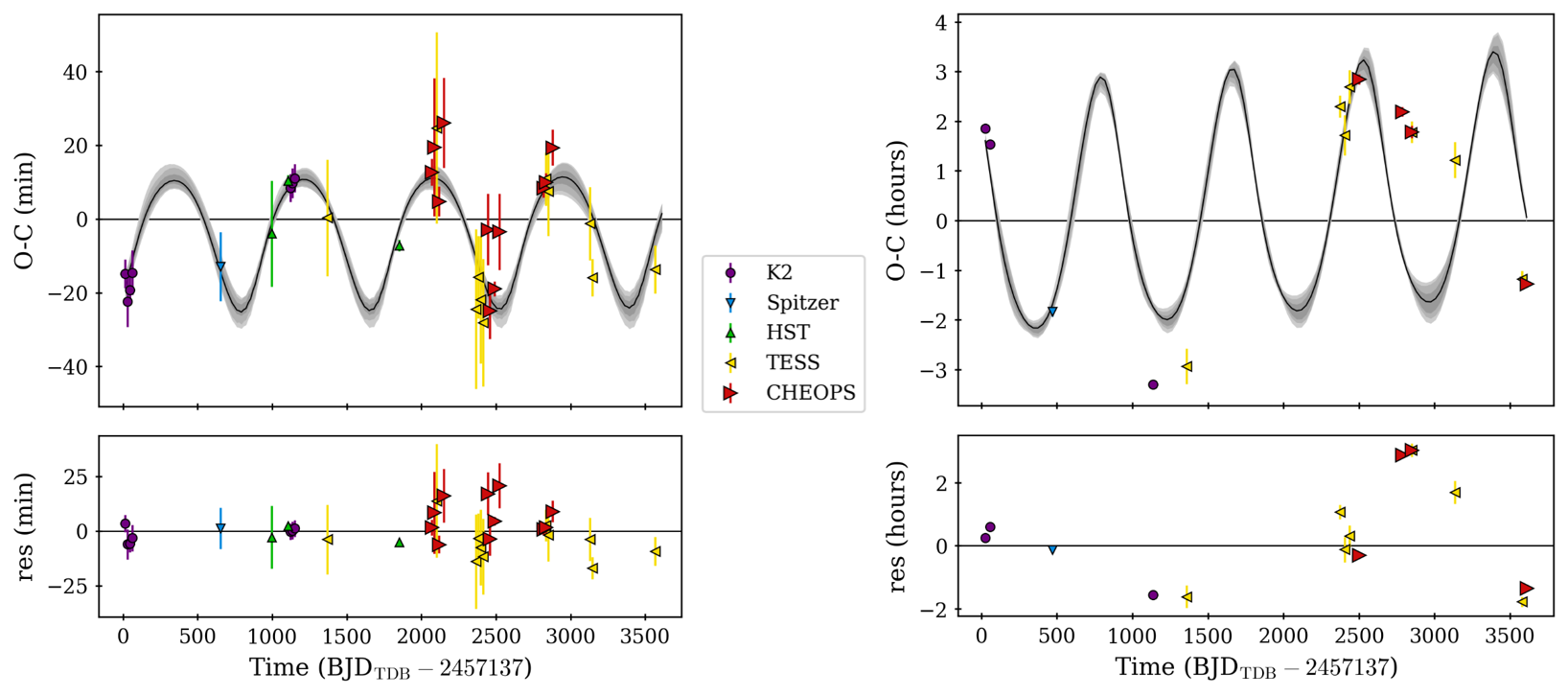}\\[1em]
    \includegraphics[width=0.95\textwidth]{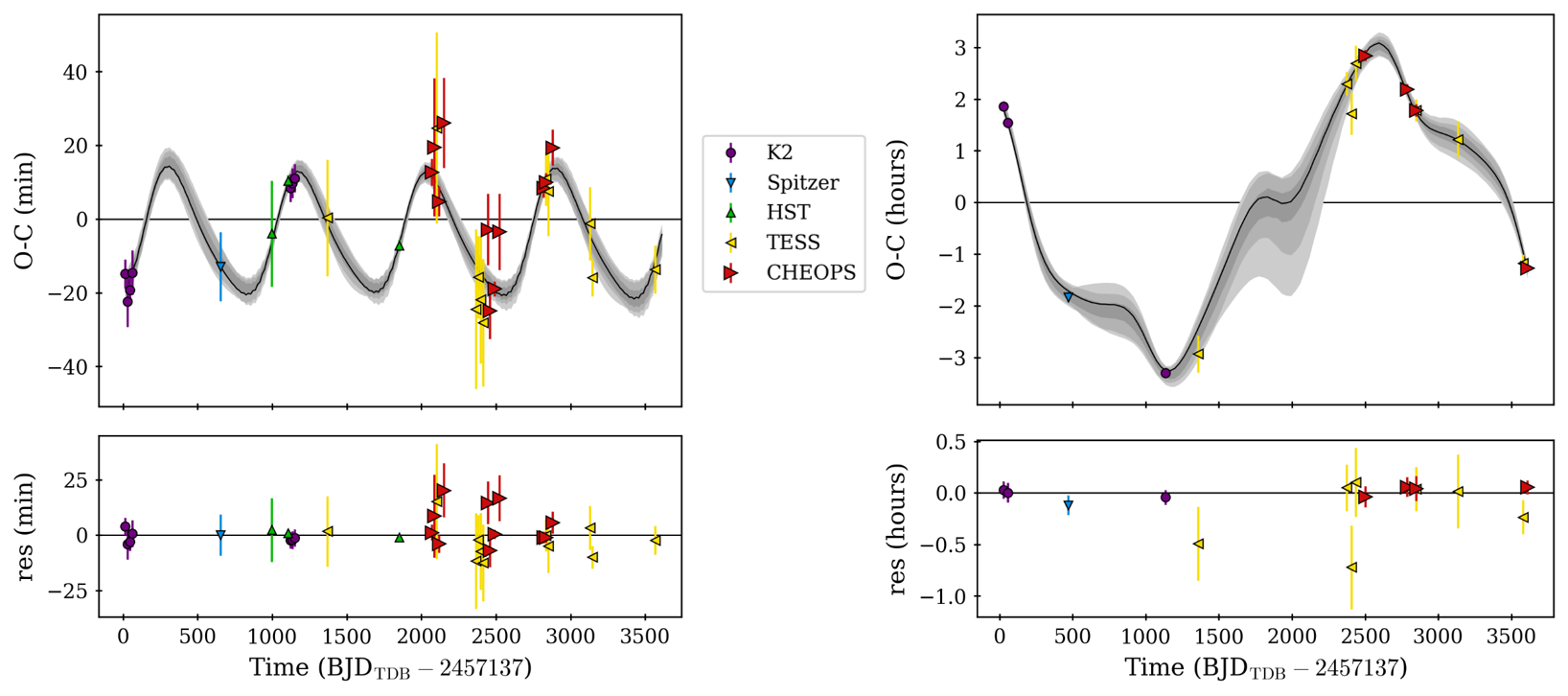}

    \caption{
        O-C diagrams of HIP\,41378\,b (left) and c (right). 
        Each dataset is plotted with a different marker and color. 
        Top: Best-fit two-planet \trades{} model (black line with gray-shaded $1\sigma$, $2\sigma$, and $3\sigma$ confidence intervals). 
        Bottom: Best-fit three-planet \trades{} model. 
        Lower panels in each plot show the residuals with respect to the corresponding model.
    }
    \label{Figure:O-C_combined}
\end{figure*}

\section{Results}
\label{section:results}

Using the N-body dynamical integrator within \trades{}, we performed joint modeling of the TTV and RV signals, and successfully retrieved the orbital configuration of HIP\,41378\,b, HIP\,41378\,c and HIP\,41378\,g (see Figures~\ref{Figure:RV}-\ref{Figure:O-C_combined}).
The best-fit model was selected using both the BIC and the log Bayes factor, which provide complementary model comparison metrics. The comparison between the two-planet and three-planet models ($\mathrm{BIC_{2p}} = 4170.08$ and $\mathrm{BIC_{3p}} = 1353.72$) yields a difference of $\Delta \mathrm{BIC} = 2816.36$, strongly favoring the inclusion of the third planet, HIP\,41378\,g. To strengthen this conclusion, we also computed the logarithm of the Bayes factor \citep{kass_95} between the two models, by using the the approximation found on page seven of \cite{Shen_Gonzalez_2021}, obtaining $\log \mathcal{B}_{3p,2p} = 1400$, which provides decisive evidence in favor of the three-planet model. This confirms that the addition of HIP\,41378\,g significantly improves the fit and is statistically justified. Thereafter, we decided to adopt the posteriors of the three-planet model as reference hereafter in the paper. We were able to determine the planetary mass of HIP\,41378\,g with a $\sim 6\sigma$ level of significance, enabled by the dynamical simultaneous modeling of both TTVs and RVs data. Our analysis places the orbital period of HIP\,41378\,g at approximately $\sim 64$ days, close to the 2:1 period commensurability with HIP\,41378\,c ($P_g/P_c \sim 2.04$), in agreement with the RV signal found by \citetalias{santerne_2019}. The orbital solution shows a compact inner system comprising three sub-Neptunes, which are near a 1:2:4 period ratio.
Our final posterior values along with the priors and the uncertainty intervals are presented in Table~\ref{table:fit_TRADES_orbital_parameters}. Plots of the Observed-minus-Calculated (O-C) diagrams of both planets and the RV plots are shown in Figs.~\ref{Figure:RV}-\ref{Figure:O-C_combined}.

As a complementary result of our dynamical analysis, \trades{} evaluates the Hill stability of the system using the AMD-Hill criterion (Eq. 26 \citealt{petit_2018}), which is based on the angular momentum deficit (AMD; \citealt{laskar_1997}, \citealt{laskar_2000}; \citealt{laskar_petit_2017}). We find that the entire posterior distribution satisfies the AMD-Hill stability criterion, indicating the long-term dynamical stability of the posterior. However, to further assess the stability and chaotic behavior of the posterior solutions, considering effects of planet-planet interactions, mean-motion resonances and planetary ejections, we used the N-body integrator \texttt{rebound} \citep{rein_liu_2012, rein_tamayo_2016}. Specifically, we employed the Mean Exponential Growth factor of Nearby Orbits (MEGNO; or $\langle \mathrm{Y} \rangle$) chaos indicator \citep{cincotta_simo_2000, cincotta_2003}.
A planetary system is considered to be in a stable configuration if it satisfies the condition $\langle \mathrm{Y} \rangle \lesssim 2$, while a planet is considered as ejected if its semi-major axis exceeded 100 times that of planet c. We computed the orbits for a total of 100 Kyr using the symplectic integrator \texttt{WHFast}, with a step size corresponding to 10\% of the orbital period of planet b. We obtained $\langle \mathrm{Y} \rangle \lesssim 2$ for the best-fit (MAP within HDI) solution indicating that the configuration is stable in the integrated time. We then checked a family of solutions randomly selected from our posterior distribution. After running the same integration for 200 solutions, we find that 87.5\% of the simulations exhibited strong stable dynamics.

\section{Discussion}
\label{section:discussion}
Our discussion is divided into six parts. In Section~\ref{section:planets_in_context}, we contextualize the derived planetary parameters with literature values and the broader sub-Neptune population using the mass-radius diagram. The system’s potential architecture is discussed in~\ref{section:architecture}. The interior composition of the planets is analyzed in~\ref{section:interior_comp}.

\subsection{Planets in context}
\label{section:planets_in_context}
\begin{figure}[htb]
    \includegraphics[width=1\columnwidth, keepaspectratio]{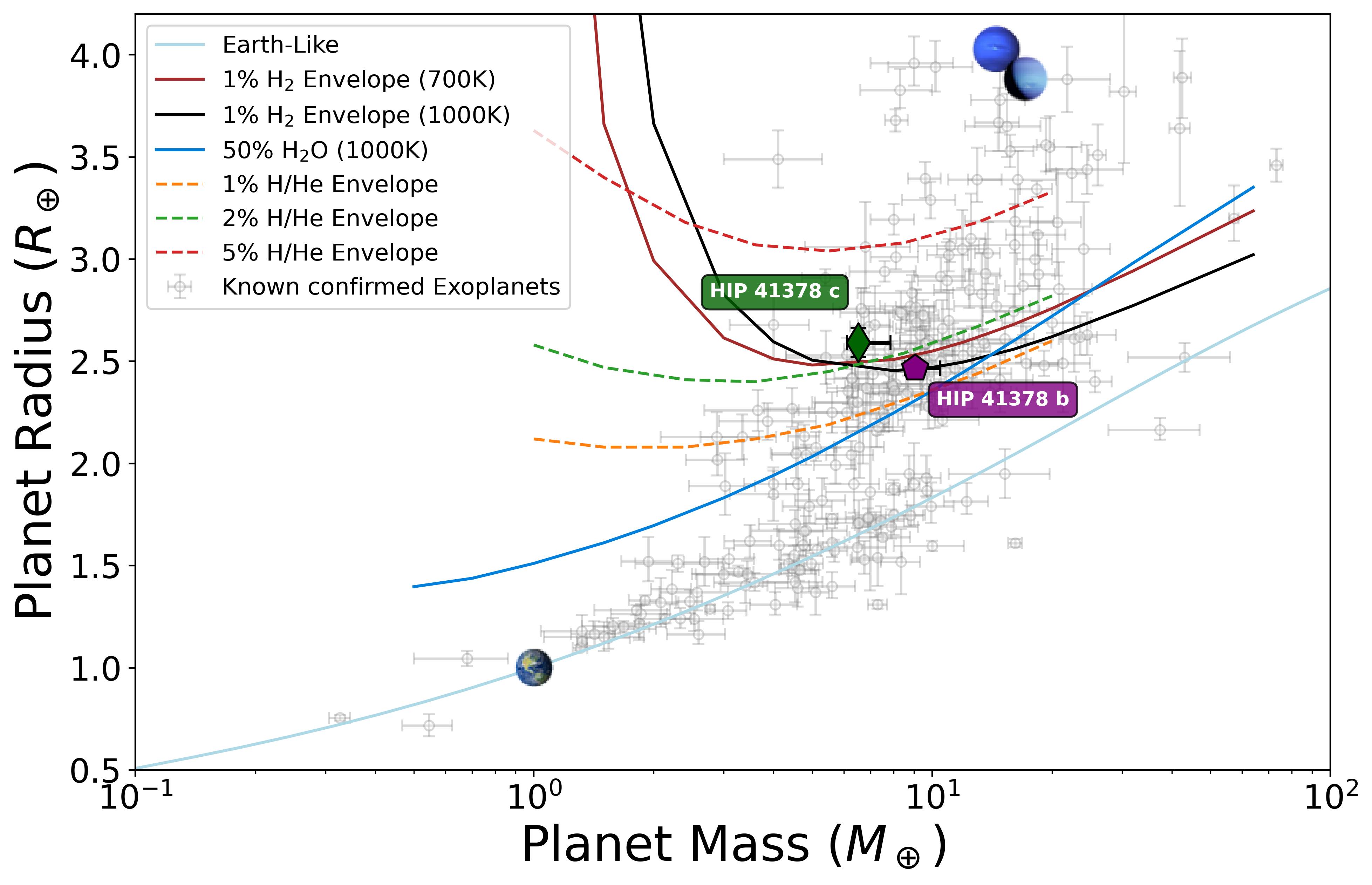}
    \caption{Mass-radius diagram including all confirmed planets with radii below 4~$R_{\oplus}$ with masses and radii measured to better than 30\% precision (Gray points). Planet parameters are taken from the NASA Exoplanet Archive's confirmed planets table, as queried on January 20, 2025.
    The purple pentagon represents the position of HIP\,41378\,b, while the green diamond represents the postion of HIP\,41378\,c. The color-coded lines show different theoretical mass–radius relations corresponding to the planet compositions taken from \cite{zeng_2019} (solid lines) and \cite{lopez_fortney_2014} (dashed lines; 1 Gyr, solar metallicity and incident flux of 10 $S_{\oplus}$). Also shown are Earth, Uranus and Neptune, for reference.}
    \label{Figure:M-R}
\end{figure}

When compared with the values reported by \citetalias{santerne_2019}, we find some statistical discrepancies in the masses of planets b and c. \citetalias{santerne_2019} derived a mass for planet c of $\mathrm{M_{c} = 4.4 \pm 1.1}~\mathrm{M_{\oplus}}$, suggesting a low bulk density ($\mathrm{\rho_{c} = 1.19 \pm 0.30}~\mathrm{g~cm^{-3}}$), which placed it among the sub-Neptune puffy population. This made it the second lowest-density planet in the system, alongside the super-puff HIP\,41378\,f ($\mathrm{\rho_{f}} = 0.09 \pm 0.02~\mathrm{g~cm^{-3}}$), whose unusually low density has been proposed to result from the presence of opaque, oblique planetary rings \citep{piro_vissapragada_2020, akinsanmi_2020}.
In contrast, our updated mass for planet c ($\mathrm{M_{c} = 6.53_{-0.42}^{+1.33}}$ $M_{\oplus}$) places it within the typical range for sub-Neptunes ($\mathrm{\rho_{c} = 1.854_{-0.031}^{+0.572}}~\mathrm{g~cm^{-3}}$), ruling out the puffy scenario (see Figure~\ref{Figure:M-R}). This reassessment of planet c highlights the potential for a similar reevaluation for HIP\,41378\,f as additional RV and photometry data become available, potentially refining its mass and density estimates.

Although \citetalias{santerne_2019} identified the RV signal of HIP\,41378\,g and provided a minimum mass estimate, they were unable to detect any transits in their available photometry. This is consistent with the duration of the \textit{K2} C5 campaign, which lasted 75 days, slightly longer than our inferred period of $\sim 64$ days. In our dynamical analysis (see Sect.~\ref{section:trades_analysis}), we explored both transiting and non-transiting configurations for the planet. The resulting inclination of ${95}_{-10}^{+1}~\mathrm{deg}$ suggests a non-transiting orbit. From our line of sight, any planet farther out than planet c (i.e., with a semimajor axis exceeding $a_c=34.60\,R_{\star}$) will not transit its host star unless its orbital inclination lies within a narrow range, $88.32^\circ < i_{g}  < 91.68^\circ$, as derived from $\arccos(\pm \mathrm{b}/34.60)$, with $b = (R_{\star} + R_{g}) / R_{\star}$\footnote{The radius of planet g was estimated using the Bayesian radius-density-mass relation for small planets implemented in \texttt{spright} \citep{parviainen_2024}, obtaining a value of $2.42_{-0.5}^{+0.5} R_{\oplus}$}. Supporting the results of the dynamical analysis, we detected no transits of HIP\,41378\,g in the newly available TESS and CHEOPS photometric data.
These results differ from statistical studies on \textit{Kepler} multi-planet systems, which have been shown to be largely coplanar with a typical scatter of $\pm\,3^{\circ}$\citep{fang_margot_2012, weiss_2023}. 

To compare the planets with the current sub-Neptune population and to provide an initial constraint on their possible bulk compositions, we placed the two planets in the mass-radius diagram (see Figure~\ref{Figure:M-R}).
Since HIP\,41378\,g does not transit its radius remains undetermined; hence, we did not include it in the diagram.
By placing the planets in the mass-radius diagram, we can see that they are consistent with the current sub-Neptune population. However the position of the planets does not allow us to uniquely determine their composition. We displayed the mass-radius models from \cite{zeng_2019} and \cite{lopez_fortney_2014}, which account for a $50\%$ water-world composition and an Earth-like core with different envelope compositions. The planets fall at the intersections of multiple composition tracks, making it challenging to break the degeneracy between H/He envelope and water-world compositions, based on mass and radius alone. For a more detailed analysis of their internal composition, see Section~\ref{section:interior_comp}.

\subsection{On the architecture of the planetary system}\label{section:architecture}
The HIP\,41378 system hosts six planets, five of which transit the star along our line of sight. As of now, the periods of two outer planets, HIP\,41378\,d and HIP\,41378\,e, remain unresolved. A single transit of HIP\,41378\,e was observed during \textit{K2} Campaign 5, preventing an estimation of its orbital period. On the other hand, HIP\,41378\,d was observed to transit during both \textit{K2} campaigns (5 and 18), enabling the identification of a set of 23 orbital period aliases. The new TESS sector observations, combined with the findings of \citet{sulis_2024}, effectively ruled out the shortest orbital periods for HIP\,41378\,d, narrowing the set of aliases to three values: $\sim278$,  $\sim371$, and $\sim1113$ days. Each of these periods places the planet near a first-order period commensurability with HIP\,41378\,f (i.e. 2:1, 3:2 and 1:2), which could explain the strong TTVs ($>$ 4 hours) observed for the planet \citep{bryant_2021, alam_2022}.
This could suggest the potential existence of a second quasi-resonance chain involving the three outer planets.

Based on the possible orbital periods of the two unresolved outer planets, and the confirmed planetary signal of planet g, we suggest two dynamical configurations for the system: i) a system-wide quasi-resonant chain or ii) hierarchical multi-planet architecture. In the first scenario, the planets follow a continuous quasi-resonant chain, with period ratios that closely align with small integer values. This would suggest a dynamically structured system, where all the planets may have experienced convergent migration and undergone resonant capture \citep{wong_2024}. This would imply the potential presence of additional, yet-undetected planets between the inner and outer regions, completing the resonance chain.
In the second configuration, the system possesses a middle-gap, separating the three inner sub-Neptunes from the three outer Neptunes. As a result, the system is assumed to be hierarchical, meaning that it can be divided into two independently stable subsystems \citep{laskar_petit_2017}. In this scenario, the outer planets would be dynamically decoupled from the inner trio, suggesting that they should not significantly influence the observed TTV signals of the inner planets. Additionally, the three outer planets could be close to a resonant chain, potentially near a low-order period commensurability. This would suggest that, while the system remains hierarchical, the outer planets could still be dynamically linked through resonant interactions. Similar architectures have been observed in systems such as Kepler-90 \citep{cabrera_2014}, which hosts both 2:3:4 and 5:4 quasi-resonant chains, and HD 191939 \citep{orell_miquel_2023}, exhibiting coupled 1:3:4 and 3:1 configurations. Further observations of the outer planets will be necessary to precisely constrain their orbital periods and determine the system’s unique dynamical configuration. The outer-system architecture will be further explored in Grouffal et al. (in prep.). To investigate the two hypotheses and put further constraints on the known outer planets (HIP\,41378\,d, HIP\,41378\,e, and HIP\,41378\,f), we conducted a four- and six-planet dynamical analysis with \trades. This included testing for a hypothetical planet "h" situated between planets g and d. The setup mirrored the one used for HIP\,41378\,g, employing broad priors to explore a wide parameter space for radius, mass, and period, based on the findings of \citetalias{santerne_2019} and \citetalias{berardo_2019}. However, we were unable to constrain the masses and orbital periods of the additional planets, resulting in poor fits for all planets, with $\mathrm{BIC_{3p}} \ll \mathrm{BIC_{6p}},  \mathrm{BIC_{4p}} $.

For now, the planetary architecture for HIP\,41378 remains an unresolved puzzle. However, the dynamical evidence of HIP\,41378\,g enables us to further investigate this peculiar multi-planet configuration. Given the limited constraints on the orbital architecture of the outer planets, we focused on the configuration of the inner planets. Following the classification scheme proposed by \cite{howe_2025}, the inner planets display a closely spaced "peas-in-a-pod" configuration, indicating a high degree of uniformity in their orbital and physical properties.
To measure the degree of this similarity, we applied the approach of \cite{otegi_2022}. Specifically, we evaluated the distances in logarithmic space for the mass ($D_M$), radius ($D_R$), and their combined global distance ($D$). In this metric, lower values correspond to greater similarity. According to our calculated values of $D_{R} =  0.007$, $D_{M} = 0.08$, and $D = 0.08$, the three planets exhibit strong similarity.
These results classify the inner system of HIP\,41378 as the fifth\footnote{The four above are: Kepler-60, Kepler-29, TOI\,763 and L\,98-59} most uniform in the sample of 48 systems analyzed by \cite{otegi_2022}. This reinforces the "closed-spaced peas-in-a-pod" scenario \citep{weiss_2018, howe_2025} and potentially hints at a common formation pathway. Furthermore, consistent with the trend reported by \cite{otegi_2022}, the inner planetary system shows greater similarity in radius than in mass. As pointed out by the authors, this could be attributed to the similarity in planetary density within a system. Given that the density is three times more sensitive to radius variations than to mass variations, it is expected to have a stronger uniformity in radius than in mass.

\subsection{Investigation on the mean-motion resonant state}
\begin{figure}[!t]
    \centering
    \includegraphics[width=1\columnwidth,keepaspectratio]{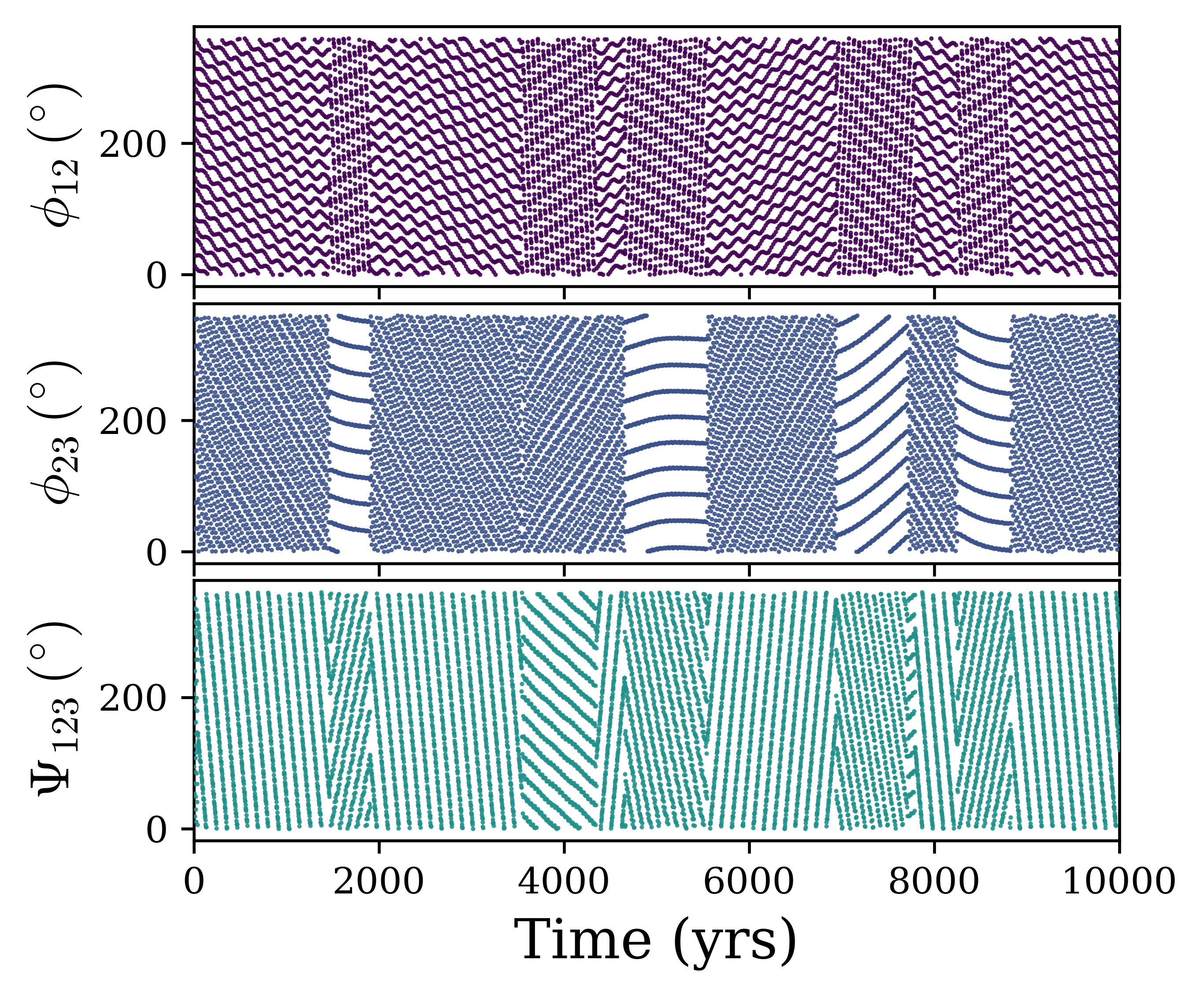}
    \caption{Three-body Laplace resonant angles evolution for HIP\,41378\,b, c and g, for the $\mathrm{MAP_{HDI}}$ solutions. The top two panels show the evolution of the critical angles $\phi_{12}$ and $\phi_{23}$. The bottom panel shows the three-body angle, which is a linear combination of the mean longitudes of the planets.}
    \label{Figure: Critical_angles}
\end{figure}

The inferred orbital period of HIP\,41378\,g, near a 2:1 period commensurability with planet c, positions the inner HIP\,41378 system close to a three-body 1:2:4 resonant chain.
Dynamical configurations in or near MMRs are expected to arise during the early stages of planetary system formation, within the gas-rich protoplanetary disk. For low-mass planets these configurations are thought to result from Type-I convergent migration in gas-rich disks, during which the forming planets are captured in resonance \citep[e.g.,][]{malholtra_1993, kley_nelson_2012, delisle_2017, izidoro_2017, Macdonald_Dawson_2018}.
To investigate whether the three-body chain is in a true mean-motion resonance, we studied the evolution of the three-body angle $\Psi_{123}$.  This angle is defined by the difference between two critical resonant angles:
\begin{equation}
\begin{split}
    \phi_{12} &= 2\lambda_c - \lambda_b - \varpi_c, \\ 
    \phi_{23} &= 2\lambda_g - \lambda_c + \varpi_c, \\ 
    \Psi_{123} &= \phi_{12} -  \phi_{23} = 3\lambda_c - \lambda_b - 2\lambda_g. 
\end{split}
\end{equation}
For a (2:1, 2:1) resonant configuration, \citet{Siegel_Fabrycky_2021} predict that the three-body angle should librate (oscillate around a fixed value) around $180^\circ$. We integrated the $\mathrm{MAP_{HDI}}$ solutions using the N-body code \texttt{rebound} \citep{rein_liu_2012} and the symplectic integrator \texttt{WHFast} \citep{rein_tamayo_2016}, for a total duration of 10\,000 years. Our results, depicted in Figure~\ref{Figure: Critical_angles}, indicate that the three planets are out of resonance, with the three-body angle circulating from 0 to $360^\circ$.

\subsection{Interior bulk composition}
\label{section:interior_comp}
Based on the derived planetary parameters, we used the internal structure modeling framework \texttt{plaNETic}\footnote{\url{https://github.com/joannegger/plaNETic}} \citep{egger_2024} to infer the interior compositions of HIP\,41378\,b and HIP\,41378\,c. \texttt{plaNETic} uses a neural network as a surrogate model for the planetary structure forward model BICEPS \citep{haldemann_2024} to speed up the inference process, which allows for a fast, but still reliable analysis. The modeled planets are assumed to consist of an envelope of uniformly mixed H/He and water, a mantle layer composed of silicon, magnesium, and iron oxides, and an inner iron core, diluted with up to 19\% sulfur as a placeholder for any lighter elements. Both observed planets are modeled simultaneously.

Since this remains a highly degenerate problem with many possible interior structures compatible with the observed planetary bulk properties, the outcome of the inference process depends to a certain extent on the chosen priors. We therefore ran six models assuming different combinations of priors, compatible with different assumptions on the system's formation and evolution history. First, we consider two distinct priors for the planet's water content: one based on a formation scenario beyond the ice-line (Case A, water-rich) and another corresponding to formation within the ice-line (Case B, water-poor). For each of these water priors, we consider three different assumptions for the planetary Si/Mg/Fe ratios. In the first case, we assume the planet's composition directly reflects the stellar Si/Mg/Fe ratios \citep{thiabaud_2015}. In the second case, we account for iron enrichment relative to the host star using the empirical fit from \citep{adibekyan_2021}. In the third case, we model the planet's composition independently of the stellar ratios, sampling the Si, Mg, and Fe molar fractions uniformly from the simplex where their sum is 1, with an upper limit of 0.75 for Fe. A detailed description of these priors can be found in \citep{egger_2024}.

The resulting posterior distributions for the most important internal structure parameters are visualized in Figures~\ref{fig:int_struct_b} and \ref{fig:int_struct_c}, with the results for the full list of internal structure parameters summarized in Tables~\ref{tab:internal_structure_results_b} and \ref{tab:internal_structure_results_c}. We do not constrain the core and mantle layer mass fractions for either planet; the inferred posteriors mostly match the priors. In the water-rich case, we infer envelope mass fractions between 38 and 42\% for planet~b and of 31\% for planet~c, with envelope water mass fractions of almost 100\% for planet~b and around 90\% for planet~c. In the water-poor case, on the other hand, we constrain the envelope mass fractions quite well with values of the order on 0.3 to 0.6\% for planet~b and 0.8 to 1.2\% for planet~c. A future JWST transmission spectrum of HIP\,41378\,b and HIP\,41378\,c could help resolve this degeneracy by directly probing the atmospheric composition of their upper envelope layers.

\section{Summary and conclusions}
\label{section:conclusions}
Using CHEOPS observations, TESS sectors, and archival data, we conducted a dynamical analysis of the two inner sub-Neptune planets transiting the bright star HIP\,41378 ($m_{V}$ = 8.9 mag).
We report the detection of large TTVs in the multi-planet HIP\,41378 system, with amplitudes of 30 minutes and $\gtrsim3$ hours for planets b and c, respectively (see Figure~\ref{Figure:O-C_combined}). Combining these TTVs with RV data, we significantly refined the planetary parameters (see Tables~\ref{table:fit_orbital_parameters}-~\ref{table:fit_TRADES_orbital_parameters}) and dynamically confirmed the additional non-transiting planet, HIP\,41378\,g, which lies close to the 2:1 period commensurability with planet c ($P_{g}/P_{c} \sim 2.04$), suggesting a near 1:2:4 resonant chain for the inner planets.
This detection raises further questions regarding the overall architecture which remains only partially resolved, with the orbital properties of two long-period planets (HIP\,41378\,d and e) yet to be constrained.

Our interior structure analysis revealed significant degeneracy in the interior structures of HIP\,41378\,b and c, with solutions heavily dependent on formation assumptions, highlighting the compositional degeneracy inherent to sub-Neptunes (Figures~\ref{fig:int_struct_b}--\ref{fig:int_struct_c}). Water-rich scenarios suggest $\sim$31--42\% water-dominated envelopes, while water-poor cases yield compact H/He envelopes ($\sim$0.3--1\%). These ambiguities highlight the need for \textit{JWST} atmospheric spectroscopy to distinguish between competing models. For sub-Neptunes such as HIP\,41378\,b and HIP\,41378\,c, mass measurements paired with atmospheric studies will help resolve the compositional degeneracy between gas dwarfs (rocky planets with H/He envelopes; \citealt{lopez_fortney_2014}; \citealt{rogers_2023}) and water worlds (rocky planets with water-rich compositions; \citealt{leger_2004}; \citealt{dorn_2021}; \citealt{aguichine_2021}; \citealt{luque_2022}).

Finally, HIP\,41378 will not be reobserved by \textit{TESS} in year 8. This calls for new observations that could shed light on the possible quasi-resonant chain of the outer system, by exploring the possible period alias of the transiting HIP\,41378\,d or even detecting another transit of HIP\,41378\,e.

\begin{acknowledgements}
This publication was produced while attending the PhD program in Space Science and Technology at the University of Trento, Cycle XXXVIII, with the support of a scholarship cofinanced by the Ministerial Decree no. 351 of 9th April 2022, based on the NRRP - funded by the European Union - NextGenerationEU - Mission 4 "Education and Research", Component 2 "From Research to Business", Investment 3.3 -- CUP E63C22001340001.
CHEOPS is an ESA mission in partnership with Switzerland with important contributions to the payload and the ground segment from Austria, Belgium, France, Germany, Hungary, Italy, Portugal, Spain, Sweden, and the United Kingdom. The CHEOPS Consortium would like to gratefully acknowledge the support received by all the agencies, offices, universities, and industries involved. Their flexibility and willingness to explore new approaches were essential to the success of this mission. CHEOPS data analyzed in this article will be made available in the CHEOPS mission archive (\url{https://cheops.unige.ch/archive_browser/}). We thank Elena Gol for her artistic advice.
LBo, VNa, GPi, TZi, IPa, RRa, and GSc acknowledge support from CHEOPS ASI-INAF agreement n. 2019-29-HH.0.
LBo acknowledges financial support from the Bando Ricerca Fondamentale INAF 2023, Mini-Grant:
``Decoding the dynamical properties of planetary systems observed by TESS and CHEOPS through TTV analysis with parallel computing''.
LPa acknowledge support from a scholarship cofinanced by the Ministerial Decree no. 118 of 2nd March 2023, based on the NRRP - funded by the European Union - NextGenerationEU - Mission 4 Component 1 -- CUP C96E23000340001.
This work has been carried out within the framework of the NCCR PlanetS supported by the Swiss National Science Foundation under grants 51NF40\_182901 and 51NF40\_205606. 
D.K. was supported by a Schr\"odinger Fellowship supported by the Austrian Science Fund (FWF) project number J4792 (FEPLowS). 
TWi acknowledges support from the UKSA and the University of Warwick. 
ABr was supported by the SNSA. 
MNG is the ESA CHEOPS Project Scientist and Mission Representative. BMM is the ESA CHEOPS Project Scientist. KGI was the ESA CHEOPS Project Scientist until the end of December 2022 and Mission Representative until the end of January 2023. All of them are/were responsible for the Guest Observers (GO) Programme. None of them relay/relayed proprietary information between the GO and Guaranteed Time Observation (GTO) Programmes, nor do/did they decide on the definition and target selection of the GTO Programme. 
This work has been carried out within the framework of the NCCR PlanetS supported by the Swiss National Science Foundation under grants 51NF40\_182901 and 51NF40\_205606. AL acknowledges support of the Swiss National Science Foundation under grant number  TMSGI2\_211697. 
YAl acknowledges support from the Swiss National Science Foundation (SNSF) under grant 200020\_192038. 
RAl, DBa, EPa, IRi, and EVi acknowledge financial support from the Agencia Estatal de Investigación of the Ministerio de Ciencia e Innovación MCIN/AEI/10.13039/501100011033 and the ERDF “A way of making Europe” through projects PID2021-125627OB-C31, PID2021-125627OB-C32, PID2021-127289NB-I00, PID2023-150468NB-I00 and PID2023-149439NB-C41. 
SCCB acknowledges the support from Fundação para a Ciência e Tecnologia (FCT) in the form of work contract through the Scientific Employment Incentive program with reference 2023.06687.CEECIND. 
CBr and ASi acknowledge support from the Swiss Space Office through the ESA PRODEX program. 
ACC acknowledges support from STFC consolidated grant number ST/V000861/1, and UKSA grant number ST/X002217/1. 
ACMC acknowledges support from the FCT, Portugal, through the CFisUC projects UIDB/04564/2020 and UIDP/04564/2020, with DOI identifiers 10.54499/UIDB/04564/2020 and 10.54499/UIDP/04564/2020, respectively. 
A.C., A.D., B.E., K.G., and J.K. acknowledge their role as ESA-appointed CHEOPS Science Team Members. 
P.E.C. is funded by the Austrian Science Fund (FWF) Erwin Schroedinger Fellowship, program J4595-N. 
This project was supported by the CNES. 
A.De. 
This work was supported by FCT - Funda\c{c}\~{a}o para a Ci\^{e}ncia e a Tecnologia through national funds and by FEDER through COMPETE2020 through the research grants UIDB/04434/2020, UIDP/04434/2020, 2022.06962.PTDC. 
O.D.S.D. is supported in the form of work contract (DL 57/2016/CP1364/CT0004) funded by national funds through FCT. 
B.-O. D. acknowledges support from the Swiss State Secretariat for Education, Research and Innovation (SERI) under contract number MB22.00046. 
ADe, BEd, KGa, and JKo acknowledge their role as ESA-appointed CHEOPS Science Team Members. 
This project has received funding from the Swiss National Science Foundation for project 200021\_200726. It has also been carried out within the framework of the National Centre of Competence in Research PlanetS supported by the Swiss National Science Foundation under grant 51NF40\_205606. The authors acknowledge the financial support of the SNSF. 
MF and CMP gratefully acknowledge the support of the Swedish National Space Agency (DNR 65/19, 174/18). 
DG gratefully acknowledges financial support from the CRT foundation under Grant No. 2018.2323 “Gaseousor rocky? Unveiling the nature of small worlds”. 
M.G. is an F.R.S.-FNRS Senior Research Associate. 
CHe acknowledges financial support from the Österreichische Akademie 1158 der Wissenschaften and from the European Union H2020-MSCA-ITN-2019 1159 under Grant Agreement no. 860470 (CHAMELEON). Calculations were performed using supercomputer resources provided by the Vienna Scientific Cluster (VSC). 
K.W.F.L. was supported by Deutsche Forschungsgemeinschaft grants RA714/14-1 within the DFG Schwerpunkt SPP 1992, Exploring the Diversity of Extrasolar Planets. 
This work was granted access to the HPC resources of MesoPSL financed by the Region Ile de France and the project Equip@Meso (reference ANR-10-EQPX-29-01) of the programme Investissements d'Avenir supervised by the Agence Nationale pour la Recherche. 
ML acknowledges support of the Swiss National Science Foundation under grant number PCEFP2\_194576. 
PM acknowledges support from STFC research grant number ST/R000638/1. 
This work was also partially supported by a grant from the Simons Foundation (PI Queloz, grant number 327127). 
NCSa acknowledges funding by the European Union (ERC, FIERCE, 101052347). Views and opinions expressed are, however, those of the author(s) only and do not necessarily reflect those of the European Union or the European Research Council. Neither the European Union nor the granting authority can be held responsible for them. 
S.G.S. acknowledge support from FCT through FCT contract nr. CEECIND/00826/2018 and POPH/FSE (EC). 
The Portuguese team thanks the Portuguese Space Agency for the provision of financial support in the framework of the PRODEX Programme of the European Space Agency (ESA) under contract number 4000142255. 
GyMSz acknowledges the support of the Hungarian National Research, Development and Innovation Office (NKFIH) grant K-125015, a a PRODEX Experiment Agreement No. 4000137122, the Lend\"ulet LP2018-7/2021 grant of the Hungarian Academy of Science and the support of the city of Szombathely. 
V.V.G. is an F.R.S-FNRS Research Associate. 
JV acknowledges support from the Swiss National Science Foundation (SNSF) under grant PZ00P2\_208945. 
NAW acknowledges UKSA grant ST/R004838/1.
\end{acknowledgements}

\bibliographystyle{aa}
\bibliography{references}

\begin{thebibliography}{136}
\expandafter\ifx\csname natexlab\endcsname\relax\def\natexlab#1{#1}\fi

\bibitem[{{Adibekyan} {et~al.}(2021){Adibekyan}, {Dorn}, {Sousa}, {Santos}, {Bitsch}, {Israelian}, {Mordasini}, {Barros}, {Delgado Mena}, {Demangeon}, {Faria}, {Figueira}, {Hakobyan}, {Oshagh}, {Soares}, {Kunitomo}, {Takeda}, {Jofr{\'e}}, {Petrucci}, \& {Martioli}}]{adibekyan_2021}
{Adibekyan}, V., {Dorn}, C., {Sousa}, S.~G., {et~al.} 2021, Science, 374, 330

\bibitem[{{Adibekyan} {et~al.}(2015){Adibekyan}, {Figueira}, {Santos}, {Sousa}, {Faria}, {Delgado-Mena}, {Oshagh}, {Tsantaki}, {Hakobyan}, {Gonz{\'a}lez Hern{\'a}ndez}, {Su{\'a}rez-Andr{\'e}s}, \& {Israelian}}]{Adibekyan2015}
{Adibekyan}, V., {Figueira}, P., {Santos}, N.~C., {et~al.} 2015, \aap, 583, A94

\bibitem[{{Adibekyan} {et~al.}(2012){Adibekyan}, {Sousa}, {Santos}, {Delgado Mena}, {Gonz{\'a}lez Hern{\'a}ndez}, {Israelian}, {Mayor}, \& {Khachatryan}}]{Adibekyan2012}
{Adibekyan}, V.~Z., {Sousa}, S.~G., {Santos}, N.~C., {et~al.} 2012, \aap, 545, A32

\bibitem[{{Agol} {et~al.}(2005){Agol}, {Steffen}, {Sari}, \& {Clarkson}}]{agol_2005}
{Agol}, E., {Steffen}, J., {Sari}, R., \& {Clarkson}, W. 2005, \mnras, 359, 567

\bibitem[{{Aguichine} {et~al.}(2021){Aguichine}, {Mousis}, {Deleuil}, \& {Marcq}}]{aguichine_2021}
{Aguichine}, A., {Mousis}, O., {Deleuil}, M., \& {Marcq}, E. 2021, \apj, 914, 84

\bibitem[{{Akinsanmi} {et~al.}(2020){Akinsanmi}, {Santos}, {Faria}, {Oshagh}, {Barros}, {Santerne}, \& {Charnoz}}]{akinsanmi_2020}
{Akinsanmi}, B., {Santos}, N.~C., {Faria}, J.~P., {et~al.} 2020, \aap, 635, L8

\bibitem[{{Alam} {et~al.}(2022){Alam}, {Kirk}, {Dressing}, {L{\'o}pez-Morales}, {Ohno}, {Gao}, {Akinsanmi}, {Santerne}, {Grouffal}, {Adibekyan}, {Barros}, {Buchhave}, {Crossfield}, {Dai}, {Deleuil}, {Giacalone}, {Lillo-Box}, {Marley}, {Mayo}, {Mortier}, {Santos}, {Sousa}, {Turtelboom}, {Wheatley}, \& {Vanderburg}}]{alam_2022}
{Alam}, M.~K., {Kirk}, J., {Dressing}, C.~D., {et~al.} 2022, \apjl, 927, L5

\bibitem[{{Badenas-Agusti} {et~al.}(2020){Badenas-Agusti}, {G{\"u}nther}, {Daylan}, {Mikal-Evans}, {Vanderburg}, {Huang}, {Matthews}, {Rackham}, {Bieryla}, {Stassun}, {Kane}, {Shporer}, {Fulton}, {Hill}, {Nowak}, {Ribas}, {Pall{\'e}}, {Jenkins}, {Latham}, {Seager}, {Ricker}, {Vanderspek}, {Winn}, {Abril-Pla}, {Collins}, {Serra}, {Niraula}, {Rustamkulov}, {Barclay}, {Crossfield}, {Howell}, {Ciardi}, {Gonzales}, {Schlieder}, {Caldwell}, {Fausnaugh}, {McDermott}, {Paegert}, {Pepper}, {Rose}, \& {Twicken}}]{badenas_agusti_2020}
{Badenas-Agusti}, M., {G{\"u}nther}, M.~N., {Daylan}, T., {et~al.} 2020, \aj, 160, 113

\bibitem[{{Bean} {et~al.}(2021){Bean}, {Raymond}, \& {Owen}}]{bean_2021}
{Bean}, J.~L., {Raymond}, S.~N., \& {Owen}, J.~E. 2021, Journal of Geophysical Research (Planets), 126, e06639

\bibitem[{{Becker} {et~al.}(2019){Becker}, {Vanderburg}, {Rodriguez}, {Omohundro}, {Adams}, {Stassun}, {Yao}, {Hartman}, {Pepper}, {Bakos}, {Barentsen}, {Beatty}, {Bhatti}, {Chontos}, {Collier Cameron}, {Hellier}, {Huber}, {James}, {Kuhn}, {Lund}, {Pollacco}, {Siverd}, {Stevens}, {Cardoso}, \& {West}}]{becker_2019}
{Becker}, J.~C., {Vanderburg}, A., {Rodriguez}, J.~E., {et~al.} 2019, \aj, 157, 19

\bibitem[{{Benz} {et~al.}(2021){Benz}, {Broeg}, {Fortier}, {Rando}, {Beck}, {Beck}, {Queloz}, {Ehrenreich}, {Maxted}, {Isaak}, {Billot}, {Alibert}, {Alonso}, {Ant{\'o}nio}, {Asquier}, {Bandy}, {B{\'a}rczy}, {Barrado}, {Barros}, {Baumjohann}, {Bekkelien}, {Bergomi}, {Biondi}, {Bonfils}, {Borsato}, {Brandeker}, {Busch}, {Cabrera}, {Cessa}, {Charnoz}, {Chazelas}, {Collier Cameron}, {Corral Van Damme}, {Cortes}, {Davies}, {Deleuil}, {Deline}, {Delrez}, {Demangeon}, {Demory}, {Erikson}, {Farinato}, {Fossati}, {Fridlund}, {Futyan}, {Gandolfi}, {Garcia Munoz}, {Gillon}, {Guterman}, {Gutierrez}, {Hasiba}, {Heng}, {Hernandez}, {Hoyer}, {Kiss}, {Kovacs}, {Kuntzer}, {Laskar}, {Lecavelier des Etangs}, {Lendl}, {L{\'o}pez}, {Lora}, {Lovis}, {L{\"u}ftinger}, {Magrin}, {Malvasio}, {Marafatto}, {Michaelis}, {de Miguel}, {Modrego}, {Munari}, {Nascimbeni}, {Olofsson}, {Ottacher}, {Ottensamer}, {Pagano}, {Palacios}, {Pall{\'e}}, {Peter}, {Piazza}, {Piotto}, {Pizarro}, {Pollaco}, {Ragazzoni}, {Ratti}, {Rauer}, {Ribas}, {Rieder},
  {Rohlfs}, {Safa}, {Salatti}, {Santos}, {Scandariato}, {S{\'e}gransan}, {Simon}, {Smith}, {Sordet}, {Sousa}, {Steller}, {Szab{\'o}}, {Szoke}, {Thomas}, {Tschentscher}, {Udry}, {Van Grootel}, {Viotto}, {Walter}, {Walton}, {Wildi}, \& {Wolter}}]{benz_2021}
{Benz}, W., {Broeg}, C., {Fortier}, A., {et~al.} 2021, Experimental Astronomy, 51, 109

\bibitem[{{Berardo} {et~al.}(2019){Berardo}, {Crossfield}, {Werner}, {Petigura}, {Christiansen}, {Ciardi}, {Dressing}, {Fulton}, {Gorjian}, {Greene}, {Hardegree-Ullman}, {Kane}, {Livingston}, {Morales}, \& {Schlieder}}]{berardo_2019}
{Berardo}, D., {Crossfield}, I. J.~M., {Werner}, M., {et~al.} 2019, \aj, 157, 185

\bibitem[{{Blackwell} \& {Shallis}(1977)}]{blackwell_1977}
{Blackwell}, D.~E. \& {Shallis}, M.~J. 1977, \mnras, 180, 177

\bibitem[{{Bonfanti} {et~al.}(2021){Bonfanti}, {Delrez}, {Hooton}, {Wilson}, {Fossati}, {Alibert}, {Hoyer}, {Mustill}, {Osborn}, {Adibekyan}, {Gandolfi}, {Salmon}, {Sousa}, {Tuson}, {Van Grootel}, {Cabrera}, {Nascimbeni}, {Maxted}, {Barros}, {Billot}, {Bonfils}, {Borsato}, {Broeg}, {Davies}, {Deleuil}, {Demangeon}, {Fridlund}, {Lacedelli}, {Lendl}, {Persson}, {Santos}, {Scandariato}, {Szab{\'o}}, {Collier Cameron}, {Udry}, {Benz}, {Beck}, {Ehrenreich}, {Fortier}, {Isaak}, {Queloz}, {Alonso}, {Asquier}, {Bandy}, {B{\'a}rczy}, {Barrado}, {Barrag{\'a}n}, {Baumjohann}, {Beck}, {Bekkelien}, {Bergomi}, {Brandeker}, {Busch}, {Cessa}, {Charnoz}, {Chazelas}, {Corral Van Damme}, {Demory}, {Erikson}, {Farinato}, {Futyan}, {Garcia Mu{\~n}oz}, {Gillon}, {Guedel}, {Guterman}, {Hasiba}, {Heng}, {Hernandez}, {Kiss}, {Kuntzer}, {Laskar}, {Lecavelier des Etangs}, {Lovis}, {Magrin}, {Malvasio}, {Marafatto}, {Michaelis}, {Munari}, {Olofsson}, {Ottacher}, {Ottensamer}, {Pagano}, {Pall{\'e}}, {Peter}, {Piazza}, {Piotto},
  {Pollacco}, {Ragazzoni}, {Rando}, {Ratti}, {Rauer}, {Ribas}, {Rieder}, {Rohlfs}, {Safa}, {Salatti}, {S{\'e}gransan}, {Simon}, {Smith}, {Sordet}, {Steller}, {Thomas}, {Tschentscher}, {Van Eylen}, {Viotto}, {Walter}, {Walton}, {Wildi}, \& {Wolter}}]{bonfanti_2021}
{Bonfanti}, A., {Delrez}, L., {Hooton}, M.~J., {et~al.} 2021, \aap, 646, A157

\bibitem[{{Bonfanti} {et~al.}(2016){Bonfanti}, {Ortolani}, \& {Nascimbeni}}]{bonfanti_2016}
{Bonfanti}, A., {Ortolani}, S., \& {Nascimbeni}, V. 2016, \aap, 585, A5

\bibitem[{{Bonfanti} {et~al.}(2015){Bonfanti}, {Ortolani}, {Piotto}, \& {Nascimbeni}}]{bonfanti_2015}
{Bonfanti}, A., {Ortolani}, S., {Piotto}, G., \& {Nascimbeni}, V. 2015, \aap, 575, A18

\bibitem[{{Borsato} {et~al.}(2024){Borsato}, {Degen}, {Leleu}, {Hooton}, {Egger}, {Bekkelien}, {Brandeker}, {Collier Cameron}, {G{\"u}nther}, {Nascimbeni}, {Persson}, {Bonfanti}, {Wilson}, {Correia}, {Zingales}, {Guillot}, {Triaud}, {Piotto}, {Gandolfi}, {Abe}, {Alibert}, {Alonso}, {B{\'a}rczy}, {Navascues}, {Barros}, {Baumjohann}, {Beck}, {Bendjoya}, {Benz}, {Billot}, {Broeg}, {Busch}, {Csizmadia}, {Cubillos}, {Davies}, {Deleuil}, {Deline}, {Delrez}, {Demangeon}, {Demory}, {Derekas}, {Edwards}, {Ehrenreich}, {Erikson}, {Fortier}, {Fossati}, {Fridlund}, {Gazeas}, {Gillon}, {G{\"u}del}, {Heitzmann}, {Helling}, {Hoyer}, {Isaak}, {Kiss}, {Korth}, {Lam}, {Laskar}, {Lecavelier des Etangs}, {Lendl}, {Magrin}, {Marafatto}, {Maxted}, {Mecina}, {M{\'e}karnia}, {Mordasini}, {Mura}, {Olofsson}, {Ottensamer}, {Pagano}, {Pall{\'e}}, {Peter}, {Pollacco}, {Queloz}, {Ragazzoni}, {Rando}, {Ratti}, {Rauer}, {Ribas}, {Salmon}, {Santos}, {Scandariato}, {S{\'e}gransan}, {Simon}, {Smith}, {Sousa}, {Stalport}, {Suarez}, {Sulis},
  {Szab{\'o}}, {Udry}, {Van Grootel}, {Venturini}, {Villaver}, {Walton}, \& {Wolter}}]{borsato_2024}
{Borsato}, L., {Degen}, D., {Leleu}, A., {et~al.} 2024, \aap, 689, A52

\bibitem[{{Borsato} {et~al.}(2019){Borsato}, {Malavolta}, {Piotto}, {Buchhave}, {Mortier}, {Rice}, {Collier Cameron}, {Coffinet}, {Sozzetti}, {Charbonneau}, {Cosentino}, {Dumusque}, {Figueira}, {Latham}, {Lopez-Morales}, {Mayor}, {Micela}, {Molinari}, {Pepe}, {Phillips}, {Poretti}, {Udry}, \& {Watson}}]{borsato_2019}
{Borsato}, L., {Malavolta}, L., {Piotto}, G., {et~al.} 2019, \mnras, 484, 3233

\bibitem[{{Borsato} {et~al.}(2014){Borsato}, {Marzari}, {Nascimbeni}, {Piotto}, {Granata}, {Bedin}, \& {Malavolta}}]{borsato_2014}
{Borsato}, L., {Marzari}, F., {Nascimbeni}, V., {et~al.} 2014, \aap, 571, A38

\bibitem[{{Borsato} {et~al.}(2022){Borsato}, {Nascimbeni}, {Piotto}, \& {Szab{\'o}}}]{borsato_2022}
{Borsato}, L., {Nascimbeni}, V., {Piotto}, G., \& {Szab{\'o}}, G. 2022, Experimental Astronomy, 53, 635

\bibitem[{{Borsato} {et~al.}(2021){Borsato}, {Piotto}, {Gandolfi}, {Nascimbeni}, {Lacedelli}, {Marzari}, {Billot}, {Maxted}, {Sousa}, {Cameron}, {Bonfanti}, {Wilson}, {Serrano}, {Garai}, {Alibert}, {Alonso}, {Asquier}, {B{\'a}rczy}, {Bandy}, {Barrado}, {Barros}, {Baumjohann}, {Beck}, {Beck}, {Benz}, {Bonfils}, {Brandeker}, {Broeg}, {Cabrera}, {Charnoz}, {Csizmadia}, {Davies}, {Deleuil}, {Delrez}, {Demangeon}, {Demory}, {des Etangs}, {Ehrenreich}, {Erikson}, {Escud{\'e}}, {Fortier}, {Fossati}, {Fridlund}, {Gillon}, {Guedel}, {Hasiba}, {Heng}, {Hoyer}, {Isaak}, {Kiss}, {Kopp}, {Laskar}, {Lendl}, {Lovis}, {Magrin}, {Munari}, {Olofsson}, {Ottensamer}, {Pagano}, {Pall{\'e}}, {Peter}, {Pollacco}, {Queloz}, {Ragazzoni}, {Rando}, {Rauer}, {Ribas}, {S{\'e}gransan}, {Santos}, {Scandariato}, {Simon}, {Smith}, {Steller}, {Szab{\'o}}, {Thomas}, {Udry}, {Van Grootel}, \& {Walton}}]{borsato_2021}
{Borsato}, L., {Piotto}, G., {Gandolfi}, D., {et~al.} 2021, \mnras, 506, 3810

\bibitem[{{Borucki} {et~al.}(2011){Borucki}, {Koch}, {Basri}, {Batalha}, {Brown}, {Bryson}, {Caldwell}, {Christensen-Dalsgaard}, {Cochran}, {DeVore}, {Dunham}, {Gautier}, {Geary}, {Gilliland}, {Gould}, {Howell}, {Jenkins}, {Latham}, {Lissauer}, {Marcy}, {Rowe}, {Sasselov}, {Boss}, {Charbonneau}, {Ciardi}, {Doyle}, {Dupree}, {Ford}, {Fortney}, {Holman}, {Seager}, {Steffen}, {Tarter}, {Welsh}, {Allen}, {Buchhave}, {Christiansen}, {Clarke}, {Das}, {D{\'e}sert}, {Endl}, {Fabrycky}, {Fressin}, {Haas}, {Horch}, {Howard}, {Isaacson}, {Kjeldsen}, {Kolodziejczak}, {Kulesa}, {Li}, {Lucas}, {Machalek}, {McCarthy}, {MacQueen}, {Meibom}, {Miquel}, {Prsa}, {Quinn}, {Quintana}, {Ragozzine}, {Sherry}, {Shporer}, {Tenenbaum}, {Torres}, {Twicken}, {Van Cleve}, {Walkowicz}, {Witteborn}, \& {Still}}]{borucki_2011}
{Borucki}, W.~J., {Koch}, D.~G., {Basri}, G., {et~al.} 2011, \apj, 736, 19

\bibitem[{{Brande} {et~al.}(2024){Brande}, {Crossfield}, {Kreidberg}, {Morley}, {Barman}, {Benneke}, {Christiansen}, {Dragomir}, {Fortney}, {Greene}, {Hardegree-Ullman}, {Howard}, {Knutson}, {Lothringer}, \& {Mikal-Evans}}]{brande_2024}
{Brande}, J., {Crossfield}, I. J.~M., {Kreidberg}, L., {et~al.} 2024, \apjl, 961, L23

\bibitem[{{Bryant} {et~al.}(2021){Bryant}, {Bayliss}, {Santerne}, {Wheatley}, {Nascimbeni}, {Ducrot}, {Burdanov}, {Acton}, {Alves}, {Anderson}, {Armstrong}, {Awiphan}, {Cooke}, {Burleigh}, {Casewell}, {Delrez}, {Demory}, {Eigm{\"u}ller}, {Fukui}, {Gan}, {Gill}, {Gillon}, {Goad}, {Tan}, {G{\"u}nther}, {Hardee}, {Henderson}, {Jehin}, {Jenkins}, {Kosiarek}, {Lendl}, {Moyano}, {Murray}, {Narita}, {Niraula}, {Odden}, {Palle}, {Parviainen}, {Pedersen}, {Pozuelos}, {Rackham}, {Sebastian}, {Stockdale}, {Tilbrook}, {Thompson}, {Triaud}, {Udry}, {Vines}, {West}, \& {de Wit}}]{bryant_2021}
{Bryant}, E.~M., {Bayliss}, D., {Santerne}, A., {et~al.} 2021, \mnras, 504, L45

\bibitem[{{Butler} {et~al.}(1997){Butler}, {Marcy}, {Williams}, {Hauser}, \& {Shirts}}]{butler_1997}
{Butler}, R.~P., {Marcy}, G.~W., {Williams}, E., {Hauser}, H., \& {Shirts}, P. 1997, 474, L115

\bibitem[{Cabrera {et~al.}(2013)Cabrera, Csizmadia, Lehmann, Dvorak, Gandolfi, Rauer, Erikson, Dreyer, Eigmüller, \& Hatzes}]{cabrera_2014}
Cabrera, J., Csizmadia, S., Lehmann, H., {et~al.} 2013, The Astrophysical Journal, 781, 18

\bibitem[{{Campante} {et~al.}(2015){Campante}, {Barclay}, {Swift}, {Huber}, {Adibekyan}, {Cochran}, {Burke}, {Isaacson}, {Quintana}, {Davies}, {Silva Aguirre}, {Ragozzine}, {Riddle}, {Baranec}, {Basu}, {Chaplin}, {Christensen-Dalsgaard}, {Metcalfe}, {Bedding}, {Handberg}, {Stello}, {Brewer}, {Hekker}, {Karoff}, {Kolbl}, {Law}, {Lundkvist}, {Miglio}, {Rowe}, {Santos}, {Van Laerhoven}, {Arentoft}, {Elsworth}, {Fischer}, {Kawaler}, {Kjeldsen}, {Lund}, {Marcy}, {Sousa}, {Sozzetti}, \& {White}}]{campante_2015}
{Campante}, T.~L., {Barclay}, T., {Swift}, J.~J., {et~al.} 2015, \apj, 799, 170

\bibitem[{{Castelli} \& {Kurucz}(2003)}]{castelli_2003}
{Castelli}, F. \& {Kurucz}, R.~L. 2003, in IAU Symposium, Vol. 210, Modelling of Stellar Atmospheres, ed. N.~{Piskunov}, W.~W. {Weiss}, \& D.~F. {Gray}, A20

\bibitem[{{Cincotta} {et~al.}(2003){Cincotta}, {Giordano}, \& {Sim{\'o}}}]{cincotta_2003}
{Cincotta}, P.~M., {Giordano}, C.~M., \& {Sim{\'o}}, C. 2003, Physica D Nonlinear Phenomena, 182, 151

\bibitem[{{Cincotta} \& {Sim{\'o}}(2000)}]{cincotta_simo_2000}
{Cincotta}, P.~M. \& {Sim{\'o}}, C. 2000, \aaps, 147, 205

\bibitem[{{Cosentino} {et~al.}(2014){Cosentino}, {Lovis}, {Pepe}, {Collier Cameron}, {Latham}, {Molinari}, {Udry}, {Bezawada}, {Buchschacher}, {Figueira}, {Fleury}, {Ghedina}, {Glenday}, {Gonzalez}, {Guerra}, {Henry}, {Hughes}, {Maire}, {Motalebi}, \& {Phillips}}]{cosentino_2014}
{Cosentino}, R., {Lovis}, C., {Pepe}, F., {et~al.} 2014, in Society of Photo-Optical Instrumentation Engineers (SPIE) Conference Series, Vol. 9147, Ground-based and Airborne Instrumentation for Astronomy V, ed. S.~K. {Ramsay}, I.~S. {McLean}, \& H.~{Takami}, 91478C

\bibitem[{{Crane} {et~al.}(2006){Crane}, {Shectman}, \& {Butler}}]{crane_2006}
{Crane}, J.~D., {Shectman}, S.~A., \& {Butler}, R.~P. 2006, in Society of Photo-Optical Instrumentation Engineers (SPIE) Conference Series, Vol. 6269, Ground-based and Airborne Instrumentation for Astronomy, ed. I.~S. {McLean} \& M.~{Iye}, 626931

\bibitem[{{Crane} {et~al.}(2010){Crane}, {Shectman}, {Butler}, {Thompson}, {Birk}, {Jones}, \& {Burley}}]{crane_2010}
{Crane}, J.~D., {Shectman}, S.~A., {Butler}, R.~P., {et~al.} 2010, in Society of Photo-Optical Instrumentation Engineers (SPIE) Conference Series, Vol. 7735, Ground-based and Airborne Instrumentation for Astronomy III, ed. I.~S. {McLean}, S.~K. {Ramsay}, \& H.~{Takami}, 773553

\bibitem[{{Crane} {et~al.}(2008){Crane}, {Shectman}, {Butler}, {Thompson}, \& {Burley}}]{crane_2008}
{Crane}, J.~D., {Shectman}, S.~A., {Butler}, R.~P., {Thompson}, I.~B., \& {Burley}, G.~S. 2008, in Society of Photo-Optical Instrumentation Engineers (SPIE) Conference Series, Vol. 7014, Ground-based and Airborne Instrumentation for Astronomy II, ed. I.~S. {McLean} \& M.~M. {Casali}, 701479

\bibitem[{{Delisle}(2017)}]{delisle_2017}
{Delisle}, J.~B. 2017, \aap, 605, A96

\bibitem[{{Delrez} {et~al.}(2023){Delrez}, {Leleu}, {Brandeker}, {Gillon}, {Hooton}, {Collier Cameron}, {Deline}, {Fortier}, {Queloz}, {Bonfanti}, {Van Grootel}, {Wilson}, {Egger}, {Alibert}, {Alonso}, {Anglada}, {Asquier}, {B{\'a}rczy}, {Barrado y Navascues}, {Barros}, {Baumjohann}, {Beck}, {Beck}, {Benz}, {Billot}, {Bonf{\i}ls}, {Borsato}, {Broeg}, {Buder}, {Cabrera}, {Cessa}, {Charnoz}, {Csizmadia}, {Cubillos}, {Davies}, {Deleuil}, {Demangeon}, {Demory}, {Ehrenreich}, {Erikson}, {Fossati}, {Fridlund}, {Gandolfi}, {G{\"u}del}, {Hasiba}, {Hoyer}, {Isaak}, {Jenkins}, {Kiss}, {Laskar}, {Latham}, {Lecavelier des Etangs}, {Lendl}, {Lovis}, {Luque}, {Magrin}, {Maxted}, {Mordasini}, {Nascimbeni}, {Olofsson}, {Ottensamer}, {Pagano}, {Pall{\'e}}, {Peter}, {Piotto}, {Pollacco}, {Ragazzoni}, {Rando}, {Rauer}, {Ribas}, {Ricker}, {Santos}, {Scandariato}, {Seager}, {S{\'e}gransan}, {Simon}, {Smith}, {Sousa}, {Steller}, {Szab{\'o}}, {Thomas}, {Udry}, {Vanderspek}, {Venturini}, {Viotto}, {Walton}, \& {Winn}}]{delrez_2023}
{Delrez}, L., {Leleu}, A., {Brandeker}, A., {et~al.} 2023, \aap, 678, A200

\bibitem[{{Dorn} \& {Lichtenberg}(2021)}]{dorn_2021}
{Dorn}, C. \& {Lichtenberg}, T. 2021, \apjl, 922, L4

\bibitem[{{Eastman} {et~al.}(2010){Eastman}, {Siverd}, \& {Gaudi}}]{eastman_2010}
{Eastman}, J., {Siverd}, R., \& {Gaudi}, B.~S. 2010, \pasp, 122, 935

\bibitem[{{Edwards} {et~al.}(2023){Edwards}, {Changeat}, {Tsiaras}, {Allan}, {Behr}, {Hagey}, {Himes}, {Ma}, {Stassun}, {Thomas}, {Thompson}, {Boley}, {Booth}, {Bouwman}, {France}, {Lowson}, {Meech}, {Phillips}, {Vidotto}, {Yip}, {Bieger}, {Gressier}, {Janin}, {Jiang}, {Leonardi}, {Sarkar}, {Skaf}, {Taylor}, {Yang}, \& {Ward-Thompson}}]{Edwards_2023a}
{Edwards}, B., {Changeat}, Q., {Tsiaras}, A., {et~al.} 2023, \aj, 166, 158

\bibitem[{Edwards {et~al.}(2023)Edwards, Changeat, Tsiaras, Yip, Al-Refaie, Anisman, Bieger, Gressier, Shibata, Skaf, Bouwman, Cho, Ikoma, Venot, Waldmann, Lagage, \& Tinetti}]{Edwards_2023b}
Edwards, B., Changeat, Q., Tsiaras, A., {et~al.} 2023, The Astrophysical Journal Supplement Series, 269, 31

\bibitem[{{Egger} {et~al.}(2024){Egger}, {Osborn}, {Kubyshkina}, {Mordasini}, {Alibert}, {G{\"u}nther}, {Lendl}, {Brandeker}, {Heitzmann}, {Leleu}, {Damasso}, {Bonfanti}, {Wilson}, {Sousa}, {Haldemann}, {Delrez}, {Hooton}, {Zingales}, {Luque}, {Alonso}, {Asquier}, {B{\'a}rczy}, {Navascues}, {Barros}, {Baumjohann}, {Benz}, {Billot}, {Borsato}, {Broeg}, {Buder}, {Castro-Gonz{\'a}lez}, {Cameron}, {Correia}, {Cortes}, {Csizmadia}, {Cubillos}, {Davies}, {Deleuil}, {Deline}, {Demangeon}, {Demory}, {Derekas}, {Edwards}, {Ehrenreich}, {Erikson}, {Fortier}, {Fossati}, {Fridlund}, {Gandolfi}, {Gazeas}, {Gillon}, {G{\"u}del}, {Helling}, {Isaak}, {Kiss}, {Korth}, {Lam}, {Laskar}, {Lavie}, {des Etangs}, {Lovis}, {Luntzer}, {Magrin}, {Maxted}, {Mer{\'\i}n}, {Munari}, {Nascimbeni}, {Olofsson}, {Ottensamer}, {Pagano}, {Pall{\'e}}, {Peter}, {Piazza}, {Piotto}, {Pollacco}, {Queloz}, {Ragazzoni}, {Rando}, {Rauer}, {Ribas}, {Rodrigues}, {Santos}, {Scandariato}, {S{\'e}gransan}, {Simon}, {Smith}, {Stalport}, {Sulis}, {Szab{\'o}},
  {Udry}, {Van Grootel}, {Venturini}, {Villaver}, \& {Walton}}]{egger_2024}
{Egger}, J.~A., {Osborn}, H.~P., {Kubyshkina}, D., {et~al.} 2024, \aap, 688, A223

\bibitem[{{Fang} \& {Margot}(2012)}]{fang_margot_2012}
{Fang}, J. \& {Margot}, J.-L. 2012, \apj, 761, 92

\bibitem[{{Foreman-Mackey} {et~al.}(2013){Foreman-Mackey}, {Hogg}, {Lang}, \& {Goodman}}]{foremanmackey_2013}
{Foreman-Mackey}, D., {Hogg}, D.~W., {Lang}, D., \& {Goodman}, J. 2013, \pasp, 125, 306

\bibitem[{{Fortier} {et~al.}(2024){Fortier}, {Simon}, {Broeg}, {Olofsson}, {Deline}, {Wilson}, {Maxted}, {Brandeker}, {Collier Cameron}, {Beck}, {Bekkelien}, {Billot}, {Bonfanti}, {Bruno}, {Cabrera}, {Delrez}, {Demory}, {Futyan}, {Flor{\'e}n}, {G{\"u}nther}, {Heitzmann}, {Hoyer}, {Isaak}, {Sousa}, {Stalport}, {Turin}, {Verhoeve}, {Akinsanmi}, {Alibert}, {Alonso}, {B{\'a}nhidi}, {B{\'a}rczy}, {Barrado}, {Barros}, {Baumjohann}, {Baycroft}, {Beck}, {Benz}, {B{\'\i}r{\'o}}, {B{\'o}di}, {Bonfils}, {Borsato}, {Charnoz}, {Cseh}, {Csizmadia}, {Cs{\'a}nyi}, {Cubillos}, {Davies}, {Davis}, {Deleuil}, {Demangeon}, {Derekas}, {Dransfield}, {Ducrot}, {Ehrenreich}, {Erikson}, {Fari{\~n}a}, {Fossati}, {Fridlund}, {Gandolfi}, {Garai}, {Garcia}, {Gillon}, {G{\'o}mez Maqueo Chew}, {G{\'o}mez-Mu{\~n}oz}, {Granata}, {G{\"u}del}, {Guterman}, {Heged{\"u}s}, {Helling}, {Jehin}, {Kalup}, {Kilkenny}, {Kiss}, {Kriskovics}, {Lam}, {Laskar}, {Lecavelier des Etangs}, {Lendl}, {Lopez Pina}, {Luntzer}, {Magrin}, {Miller}, {Modrego
  Contreras}, {Mordasini}, {Munari}, {Murray}, {Nascimbeni}, {Ottacher}, {Ottensamer}, {Pagano}, {P{\'a}l}, {Pall{\'e}}, {Pasetti}, {Pedersen}, {Peter}, {Petrucci}, {Piotto}, {Pizarro-Rubio}, {Pollacco}, {Pribulla}, {Queloz}, {Ragazzoni}, {Rando}, {Rauer}, {Ribas}, {Sabin}, {Santos}, {Scandariato}, {Schanche}, {Schroffenegger}, {Scutt}, {Sebastian}, {S{\'e}gransan}, {Seli}, {Smith}, {Southworth}, {Standing}, {Szab{\'o}}, {Szak{\'a}ts}, {Thomas}, {Timmermans}, {Triaud}, {Udry}, {Van Grootel}, {Venturini}, {Villaver}, {Vink{\'o}}, {Walton}, {Wells}, \& {Wolter}}]{Fortier_2024}
{Fortier}, A., {Simon}, A.~E., {Broeg}, C., {et~al.} 2024, \aap, 687, A302

\bibitem[{{Gaia Collaboration} {et~al.}(2023){Gaia Collaboration}, {Vallenari}, {Brown}, {Prusti}, {de Bruijne}, {Arenou}, {Babusiaux}, {Biermann}, {Creevey}, {Ducourant}, {Evans}, {Eyer}, {Guerra}, {Hutton}, {Jordi}, {Klioner}, {Lammers}, {Lindegren}, {Luri}, {Mignard}, {Panem}, {Pourbaix}, {Randich}, {Sartoretti}, {Soubiran}, {Tanga}, {Walton}, {Bailer-Jones}, {Bastian}, {Drimmel}, {Jansen}, {Katz}, {Lattanzi}, {van Leeuwen}, {Bakker}, {Cacciari}, {Casta{\~n}eda}, {De Angeli}, {Fabricius}, {Fouesneau}, {Fr{\'e}mat}, {Galluccio}, {Guerrier}, {Heiter}, {Masana}, {Messineo}, {Mowlavi}, {Nicolas}, {Nienartowicz}, {Pailler}, {Panuzzo}, {Riclet}, {Roux}, {Seabroke}, {Sordo}, {Th{\'e}venin}, {Gracia-Abril}, {Portell}, {Teyssier}, {Altmann}, {Andrae}, {Audard}, {Bellas-Velidis}, {Benson}, {Berthier}, {Blomme}, {Burgess}, {Busonero}, {Busso}, {C{\'a}novas}, {Carry}, {Cellino}, {Cheek}, {Clementini}, {Damerdji}, {Davidson}, {de Teodoro}, {Nu{\~n}ez Campos}, {Delchambre}, {Dell'Oro}, {Esquej},
  {Fern{\'a}ndez-Hern{\'a}ndez}, {Fraile}, {Garabato}, {Garc{\'\i}a-Lario}, {Gosset}, {Haigron}, {Halbwachs}, {Hambly}, {Harrison}, {Hern{\'a}ndez}, {Hestroffer}, {Hodgkin}, {Holl}, {Jan{\ss}en}, {Jevardat de Fombelle}, {Jordan}, {Krone-Martins}, {Lanzafame}, {L{\"o}ffler}, {Marchal}, {Marrese}, {Moitinho}, {Muinonen}, {Osborne}, {Pancino}, {Pauwels}, {Recio-Blanco}, {Reyl{\'e}}, {Riello}, {Rimoldini}, {Roegiers}, {Rybizki}, {Sarro}, {Siopis}, {Smith}, {Sozzetti}, {Utrilla}, {van Leeuwen}, {Abbas}, {{\'A}brah{\'a}m}, {Abreu Aramburu}, {Aerts}, {Aguado}, {Ajaj}, {Aldea-Montero}, {Altavilla}, {{\'A}lvarez}, {Alves}, {Anders}, {Anderson}, {Anglada Varela}, {Antoja}, {Baines}, {Baker}, {Balaguer-N{\'u}{\~n}ez}, {Balbinot}, {Balog}, {Barache}, {Barbato}, {Barros}, {Barstow}, {Bartolom{\'e}}, {Bassilana}, {Bauchet}, {Becciani}, {Bellazzini}, {Berihuete}, {Bernet}, {Bertone}, {Bianchi}, {Binnenfeld}, {Blanco-Cuaresma}, {Blazere}, {Boch}, {Bombrun}, {Bossini}, {Bouquillon}, {Bragaglia}, {Bramante}, {Breedt},
  {Bressan}, {Brouillet}, {Brugaletta}, {Bucciarelli}, {Burlacu}, {Butkevich}, {Buzzi}, {Caffau}, {Cancelliere}, {Cantat-Gaudin}, {Carballo}, {Carlucci}, {Carnerero}, {Carrasco}, {Casamiquela}, {Castellani}, {Castro-Ginard}, {Chaoul}, {Charlot}, {Chemin}, {Chiaramida}, {Chiavassa}, {Chornay}, {Comoretto}, {Contursi}, {Cooper}, {Cornez}, {Cowell}, {Crifo}, {Cropper}, {Crosta}, {Crowley}, {Dafonte}, {Dapergolas}, {David}, {David}, {de Laverny}, {De Luise}, \& {De March}}]{gaiacollaboration_2022}
{Gaia Collaboration}, {Vallenari}, A., {Brown}, A.~G.~A., {et~al.} 2023, \aap, 674, A1

\bibitem[{Gelman \& Rubin(1992)}]{gelmanrubin_1992}
Gelman, A. \& Rubin, D.~B. 1992, Statistical Science, 7, 457

\bibitem[{Geweke(1991)}]{geweke_1991}
Geweke, J.~F. 1991, {Evaluating the accuracy of sampling-based approaches to the calculation of posterior moments}, Staff Report 148, Federal Reserve Bank of Minneapolis

\bibitem[{{Gillon} {et~al.}(2017){Gillon}, {Demory}, {Van Grootel}, {Motalebi}, {Lovis}, {Collier Cameron}, {Charbonneau}, {Latham}, {Molinari}, {Pepe}, {S{\'e}gransan}, {Sasselov}, {Udry}, {Mayor}, {Micela}, {Piotto}, \& {Sozzetti}}]{gillon_2017b}
{Gillon}, M., {Demory}, B.-O., {Van Grootel}, V., {et~al.} 2017, Nature Astronomy, 1, 0056

\bibitem[{Goodman \& Weare(2010)}]{goodman_weare_2010}
Goodman, J. \& Weare, J. 2010, Communications in Applied Mathematics and Computational Science, 5, 65

\bibitem[{{Haldemann} {et~al.}(2024){Haldemann}, {Dorn}, {Venturini}, {Alibert}, \& {Benz}}]{haldemann_2024}
{Haldemann}, J., {Dorn}, C., {Venturini}, J., {Alibert}, Y., \& {Benz}, W. 2024, \aap, 681, A96

\bibitem[{{Holman} \& {Murray}(2005)}]{holman_2005}
{Holman}, M.~J. \& {Murray}, N.~W. 2005, Science, 307, 1288

\bibitem[{Howe {et~al.}(2025)Howe, Becker, Stark, \& Adams}]{howe_2025}
Howe, A.~R., Becker, J.~C., Stark, C.~C., \& Adams, F.~C. 2025, The Astronomical Journal, 169, 149

\bibitem[{{Howell} {et~al.}(2014){Howell}, {Sobeck}, {Haas}, {Still}, {Barclay}, {Mullally}, {Troeltzsch}, {Aigrain}, {Bryson}, {Caldwell}, {Chaplin}, {Cochran}, {Huber}, {Marcy}, {Miglio}, {Najita}, {Smith}, {Twicken}, \& {Fortney}}]{howel_2014}
{Howell}, S.~B., {Sobeck}, C., {Haas}, M., {et~al.} 2014, \pasp, 126, 398

\bibitem[{{Hoyer} {et~al.}(2020){Hoyer}, {Guterman}, {Demangeon}, {Sousa}, {Deleuil}, {Meunier}, \& {Benz}}]{hoyer_2020}
{Hoyer}, S., {Guterman}, P., {Demangeon}, O., {et~al.} 2020, \aap, 635, A24

\bibitem[{{Husser} {et~al.}(2013){Husser}, {Wende-von Berg}, {Dreizler}, {Homeier}, {Reiners}, {Barman}, \& {Hauschildt}}]{Husser_2013}
{Husser}, T.~O., {Wende-von Berg}, S., {Dreizler}, S., {et~al.} 2013, \aap, 553, A6

\bibitem[{{Izidoro} {et~al.}(2017){Izidoro}, {Ogihara}, {Raymond}, {Morbidelli}, {Pierens}, {Bitsch}, {Cossou}, \& {Hersant}}]{izidoro_2017}
{Izidoro}, A., {Ogihara}, M., {Raymond}, S.~N., {et~al.} 2017, \mnras, 470, 1750

\bibitem[{{Jenkins} {et~al.}(2016){Jenkins}, {Twicken}, {McCauliff}, {Campbell}, {Sanderfer}, {Lung}, {Mansouri-Samani}, {Girouard}, {Tenenbaum}, {Klaus}, {Smith}, {Caldwell}, {Chacon}, {Henze}, {Heiges}, {Latham}, {Morgan}, {Swade}, {Rinehart}, \& {Vanderspek}}]{jenkins_2016}
{Jenkins}, J.~M., {Twicken}, J.~D., {McCauliff}, S., {et~al.} 2016, in Society of Photo-Optical Instrumentation Engineers (SPIE) Conference Series, Vol. 9913, Software and Cyberinfrastructure for Astronomy IV, ed. G.~{Chiozzi} \& J.~C. {Guzman}, 99133E

\bibitem[{{Kass} \& {Raftery}(1995)}]{kass_95}
{Kass}, R.~E. \& {Raftery}, A.~E. 1995, Journal of the American Statistical Association, 90, 773

\bibitem[{Kipping(2013)}]{kipping_2013}
Kipping, D.~M. 2013, Monthly Notices of the Royal Astronomical Society, 435, 2152

\bibitem[{{Kley} \& {Nelson}(2012)}]{kley_nelson_2012}
{Kley}, W. \& {Nelson}, R.~P. 2012, \araa, 50, 211

\bibitem[{Kopparapu {et~al.}(2018)Kopparapu, Hébrard, Belikov, Batalha, Mulders, Stark, Teal, Domagal-Goldman, \& Mandell}]{kopparapu_2018}
Kopparapu, R.~K., Hébrard, E., Belikov, R., {et~al.} 2018, The Astrophysical Journal, 856, 122

\bibitem[{{Kreidberg} {et~al.}(2015){Kreidberg}, {Line}, {Bean}, {Stevenson}, {D{\'e}sert}, {Madhusudhan}, {Fortney}, {Barstow}, {Henry}, {Williamson}, \& {Showman}}]{kreidberg_2015}
{Kreidberg}, L., {Line}, M.~R., {Bean}, J.~L., {et~al.} 2015, \apj, 814, 66

\bibitem[{{Kurucz}(1993)}]{Kurucz1993}
{Kurucz}, R.~L. 1993, {SYNTHE spectrum synthesis programs and line data} (Smithsonian Astrophysical Observatory)

\bibitem[{{Laskar}(1997)}]{laskar_1997}
{Laskar}, J. 1997, \aap, 317, L75

\bibitem[{{Laskar}(2000)}]{laskar_2000}
{Laskar}, J. 2000, \prl, 84, 3240

\bibitem[{{Laskar} \& {Petit}(2017)}]{laskar_petit_2017}
{Laskar}, J. \& {Petit}, A.~C. 2017, \aap, 605, A72

\bibitem[{{Latham} {et~al.}(2011){Latham}, {Rowe}, {Quinn}, {Batalha}, {Borucki}, {Brown}, {Bryson}, {Buchhave}, {Caldwell}, {Carter}, {Christiansen}, {Ciardi}, {Cochran}, {Dunham}, {Fabrycky}, {Ford}, {Gautier}, {Gilliland}, {Holman}, {Howell}, {Ibrahim}, {Isaacson}, {Jenkins}, {Koch}, {Lissauer}, {Marcy}, {Quintana}, {Ragozzine}, {Sasselov}, {Shporer}, {Steffen}, {Welsh}, \& {Wohler}}]{latham_2011}
{Latham}, D.~W., {Rowe}, J.~F., {Quinn}, S.~N., {et~al.} 2011, \apjl, 732, L24

\bibitem[{{Leleu} {et~al.}(2021){Leleu}, {Alibert}, {Hara}, {Hooton}, {Wilson}, {Robutel}, {Delisle}, {Laskar}, {Hoyer}, {Lovis}, {Bryant}, {Ducrot}, {Cabrera}, {Delrez}, {Acton}, {Adibekyan}, {Allart}, {Allende Prieto}, {Alonso}, {Alves}, {Anderson}, {Angerhausen}, {Anglada Escud{\'e}}, {Asquier}, {Barrado}, {Barros}, {Baumjohann}, {Bayliss}, {Beck}, {Beck}, {Bekkelien}, {Benz}, {Billot}, {Bonfanti}, {Bonfils}, {Bouchy}, {Bourrier}, {Bou{\'e}}, {Brandeker}, {Broeg}, {Buder}, {Burdanov}, {Burleigh}, {B{\'a}rczy}, {Cameron}, {Chamberlain}, {Charnoz}, {Cooke}, {Corral Van Damme}, {Correia}, {Cristiani}, {Damasso}, {Davies}, {Deleuil}, {Demangeon}, {Demory}, {Di Marcantonio}, {Di Persio}, {Dumusque}, {Ehrenreich}, {Erikson}, {Figueira}, {Fortier}, {Fossati}, {Fridlund}, {Futyan}, {Gandolfi}, {Garc{\'\i}a Mu{\~n}oz}, {Garcia}, {Gill}, {Gillen}, {Gillon}, {Goad}, {Gonz{\'a}lez Hern{\'a}ndez}, {Guedel}, {G{\"u}nther}, {Haldemann}, {Henderson}, {Heng}, {Hogan}, {Isaak}, {Jehin}, {Jenkins}, {Jord{\'a}n}, {Kiss},
  {Kristiansen}, {Lam}, {Lavie}, {Lecavelier des Etangs}, {Lendl}, {Lillo-Box}, {Lo Curto}, {Magrin}, {Martins}, {Maxted}, {McCormac}, {Mehner}, {Micela}, {Molaro}, {Moyano}, {Murray}, {Nascimbeni}, {Nunes}, {Olofsson}, {Osborn}, {Oshagh}, {Ottensamer}, {Pagano}, {Pall{\'e}}, {Pedersen}, {Pepe}, {Persson}, {Peter}, {Piotto}, {Polenta}, {Pollacco}, {Poretti}, {Pozuelos}, {Queloz}, {Ragazzoni}, {Rando}, {Ratti}, {Rauer}, {Raynard}, {Rebolo}, {Reimers}, {Ribas}, {Santos}, {Scandariato}, {Schneider}, {Sebastian}, {Sestovic}, {Simon}, {Smith}, {Sousa}, {Sozzetti}, {Steller}, {Su{\'a}rez Mascare{\~n}o}, {Szab{\'o}}, {S{\'e}gransan}, {Thomas}, {Thompson}, {Tilbrook}, {Triaud}, {Turner}, {Udry}, {Van Grootel}, {Venus}, {Verrecchia}, {Vines}, {Walton}, {West}, {Wheatley}, {Wolter}, \& {Zapatero Osorio}}]{leleu_2021}
{Leleu}, A., {Alibert}, Y., {Hara}, N.~C., {et~al.} 2021, \aap, 649, A26

\bibitem[{{Leleu} {et~al.}(2024){Leleu}, {Delisle}, {Delrez}, {Bryant}, {Brandeker}, {Osborn}, {Hara}, {Wilson}, {Billot}, {Lendl}, {Ehrenreich}, {Chakraborty}, {G{\"u}nther}, {Hooton}, {Alibert}, {Alonso}, {Alves}, {Anderson}, {Apergis}, {Armstrong}, {B{\'a}rczy}, {Barrado Navascues}, {Barros}, {Battley}, {Baumjohann}, {Bayliss}, {Beck}, {Benz}, {Borsato}, {Broeg}, {Burleigh}, {Casewell}, {Collier Cameron}, {Correia}, {Csizmadia}, {Cubillos}, {Davies}, {Deleuil}, {Deline}, {Demangeon}, {Demory}, {Derekas}, {Edwards}, {Erikson}, {Fortier}, {Fossati}, {Fridlund}, {Gandolfi}, {Gazeas}, {Gillen}, {Gillon}, {Goad}, {G{\"u}del}, {Hawthorn}, {Heitzmann}, {Helling}, {Isaak}, {Jenkins}, {Jenkins}, {Kendall}, {Kiss}, {Korth}, {Lam}, {Laskar}, {Latham}, {Lecavelier des Etangs}, {Magrin}, {Maxted}, {McCormac}, {Mordasini}, {Moyano}, {Nascimbeni}, {Olofsson}, {Osborn}, {Ottensamer}, {Pagano}, {Pall{\'e}}, {Peter}, {Piotto}, {Pollacco}, {Queloz}, {Ragazzoni}, {Rando}, {Rauer}, {Ribas}, {Ricker}, {Saha}, {Santos},
  {Scandariato}, {Seager}, {S{\'e}gransan}, {Simon}, {Smith}, {Sousa}, {Stalport}, {Sulis}, {Szab{\'o}}, {Udry}, {Ulmer-Moll}, {Van Grootel}, {Vanderspek}, {Venturini}, {Villaver}, {Vin{\'e}s}, {Walton}, {West}, {Wheatley}, {Winn}, \& {Zivave}}]{leleu_2024}
{Leleu}, A., {Delisle}, J.~B., {Delrez}, L., {et~al.} 2024, \aap, 688, A211

\bibitem[{{Lindegren} {et~al.}(2021){Lindegren}, {Bastian}, {Biermann}, {Bombrun}, {de Torres}, {Gerlach}, {Geyer}, {Hern{\'a}ndez}, {Hilger}, {Hobbs}, {Klioner}, {Lammers}, {McMillan}, {Ramos-Lerate}, {Steidelm{\"u}ller}, {Stephenson}, \& {van Leeuwen}}]{lindegren_2021}
{Lindegren}, L., {Bastian}, U., {Biermann}, M., {et~al.} 2021, \aap, 649, A4

\bibitem[{Lithwick {et~al.}(2012)Lithwick, Xie, \& Wu}]{lithwick_2012}
Lithwick, Y., Xie, J., \& Wu, Y. 2012, The Astrophysical Journal, 761, 122

\bibitem[{{Lopez} \& {Fortney}(2014)}]{lopez_fortney_2014}
{Lopez}, E.~D. \& {Fortney}, J.~J. 2014, \apj, 792, 1

\bibitem[{{Luger} {et~al.}(2016){Luger}, {Agol}, {Kruse}, {Barnes}, {Becker}, {Foreman-Mackey}, \& {Deming}}]{luger_2016}
{Luger}, R., {Agol}, E., {Kruse}, E., {et~al.} 2016, \aj, 152, 100

\bibitem[{{Luger} {et~al.}(2018){Luger}, {Kruse}, {Foreman-Mackey}, {Agol}, \& {Saunders}}]{luger_2018}
{Luger}, R., {Kruse}, E., {Foreman-Mackey}, D., {Agol}, E., \& {Saunders}, N. 2018, \aj, 156, 99

\bibitem[{Lund {et~al.}(2019)Lund, Knudstrup, Aguirre, Basu, Chontos, Von~Essen, Chaplin, Bieryla, Casagrande, Vanderburg, Huber, Kane, Albrecht, Latham, Davies, Becker, \& Rodriguez}]{Lund_2019}
Lund, M.~N., Knudstrup, E., Aguirre, V.~S., {et~al.} 2019, The Astronomical Journal, 158, 248

\bibitem[{{Luque} {et~al.}(2023){Luque}, {Osborn}, {Leleu}, {Pall{\'e}}, {Bonfanti}, {Barrag{\'a}n}, {Wilson}, {Broeg}, {Cameron}, {Lendl}, {Maxted}, {Alibert}, {Gandolfi}, {Delisle}, {Hooton}, {Egger}, {Nowak}, {Lafarga}, {Rapetti}, {Twicken}, {Morales}, {Carleo}, {Orell-Miquel}, {Adibekyan}, {Alonso}, {Alqasim}, {Amado}, {Anderson}, {Anglada-Escud{\'e}}, {Bandy}, {B{\'a}rczy}, {Barrado Navascues}, {Barros}, {Baumjohann}, {Bayliss}, {Bean}, {Beck}, {Beck}, {Benz}, {Billot}, {Bonfils}, {Borsato}, {Boyle}, {Brandeker}, {Bryant}, {Cabrera}, {Carrazco-Gaxiola}, {Charbonneau}, {Charnoz}, {Ciardi}, {Cochran}, {Collins}, {Crossfield}, {Csizmadia}, {Cubillos}, {Dai}, {Davies}, {Deeg}, {Deleuil}, {Deline}, {Delrez}, {Demangeon}, {Demory}, {Ehrenreich}, {Erikson}, {Esparza-Borges}, {Falk}, {Fortier}, {Fossati}, {Fridlund}, {Fukui}, {Garcia-Mejia}, {Gill}, {Gillon}, {Goffo}, {G{\'o}mez Maqueo Chew}, {G{\"u}del}, {Guenther}, {G{\"u}nther}, {Hatzes}, {Helling}, {Hesse}, {Howell}, {Hoyer}, {Ikuta}, {Isaak}, {Jenkins},
  {Kagetani}, {Kiss}, {Kodama}, {Korth}, {Lam}, {Laskar}, {Latham}, {Lecavelier des Etangs}, {Leon}, {Livingston}, {Magrin}, {Matson}, {Matthews}, {Mordasini}, {Mori}, {Moyano}, {Munari}, {Murgas}, {Narita}, {Nascimbeni}, {Olofsson}, {Osborne}, {Ottensamer}, {Pagano}, {Parviainen}, {Peter}, {Piotto}, {Pollacco}, {Queloz}, {Quinn}, {Quirrenbach}, {Ragazzoni}, {Rando}, {Ratti}, {Rauer}, {Redfield}, {Ribas}, {Ricker}, {Rudat}, {Sabin}, {Salmon}, {Santos}, {Scandariato}, {Schanche}, {Schlieder}, {Seager}, {S{\'e}gransan}, {Shporer}, {Simon}, {Smith}, {Sousa}, {Stalport}, {Szab{\'o}}, {Thomas}, {Tuson}, {Udry}, {Vanderburg}, {Van Eylen}, {Van Grootel}, {Venturini}, {Walter}, {Walton}, {Watanabe}, {Winn}, \& {Zingales}}]{luque_2023}
{Luque}, R., {Osborn}, H.~P., {Leleu}, A., {et~al.} 2023, \nat, 623, 932

\bibitem[{{Luque} \& {Pall{\'e}}(2022)}]{luque_2022}
{Luque}, R. \& {Pall{\'e}}, E. 2022, Science, 377, 1211

\bibitem[{Léger {et~al.}(2004)Léger, Selsis, Sotin, Guillot, Despois, Mawet, Ollivier, Labèque, Valette, Brachet, Chazelas, \& Lammer}]{leger_2004}
Léger, A., Selsis, F., Sotin, C., {et~al.} 2004, Icarus, 169, 499

\bibitem[{{MacDonald} \& {Dawson}(2018)}]{Macdonald_Dawson_2018}
{MacDonald}, M.~G. \& {Dawson}, R.~I. 2018, \aj, 156, 228

\bibitem[{{Madhusudhan}(2019)}]{Madhusudhan_2019}
{Madhusudhan}, N. 2019, \araa, 57, 617

\bibitem[{{Malavolta} {et~al.}(2018){Malavolta}, {Mayo}, {Louden}, {Rajpaul}, {Bonomo}, {Buchhave}, {Kreidberg}, {Kristiansen}, {Lopez-Morales}, {Mortier}, {Vanderburg}, {Coffinet}, {Ehrenreich}, {Lovis}, {Bouchy}, {Charbonneau}, {Ciardi}, {Collier Cameron}, {Cosentino}, {Crossfield}, {Damasso}, {Dressing}, {Dumusque}, {Everett}, {Figueira}, {Fiorenzano}, {Gonzales}, {Haywood}, {Harutyunyan}, {Hirsch}, {Howell}, {Johnson}, {Latham}, {Lopez}, {Mayor}, {Micela}, {Molinari}, {Nascimbeni}, {Pepe}, {Phillips}, {Piotto}, {Rice}, {Sasselov}, {S{\'e}gransan}, {Sozzetti}, {Udry}, \& {Watson}}]{malavolta_2018}
{Malavolta}, L., {Mayo}, A.~W., {Louden}, T., {et~al.} 2018, \aj, 155, 107

\bibitem[{{Malavolta} {et~al.}(2016){Malavolta}, {Nascimbeni}, {Piotto}, {Quinn}, {Borsato}, {Granata}, {Bonomo}, {Marzari}, {Bedin}, {Rainer}, {Desidera}, {Lanza}, {Poretti}, {Sozzetti}, {White}, {Latham}, {Cunial}, {Libralato}, {Nardiello}, {Boccato}, {Claudi}, {Cosentino}, {Covino}, {Gratton}, {Maggio}, {Micela}, {Molinari}, {Pagano}, {Smareglia}, {Affer}, {Andreuzzi}, {Aparicio}, {Benatti}, {Bignamini}, {Borsa}, {Damasso}, {Di Fabrizio}, {Harutyunyan}, {Esposito}, {Fiorenzano}, {Gandolfi}, {Giacobbe}, {Gonz{\'a}lez Hern{\'a}ndez}, {Maldonado}, {Masiero}, {Molinaro}, {Pedani}, \& {Scandariato}}]{malavolta_2016}
{Malavolta}, L., {Nascimbeni}, V., {Piotto}, G., {et~al.} 2016, \aap, 588, A118

\bibitem[{{Malhotra}(1993)}]{malholtra_1993}
{Malhotra}, R. 1993, \nat, 365, 819

\bibitem[{{Marigo} {et~al.}(2017){Marigo}, {Girardi}, {Bressan}, {Rosenfield}, {Aringer}, {Chen}, {Dussin}, {Nanni}, {Pastorelli}, {Rodrigues}, {Trabucchi}, {Bladh}, {Dalcanton}, {Groenewegen}, {Montalb{\'a}n}, \& {Wood}}]{marigo_2017}
{Marigo}, P., {Girardi}, L., {Bressan}, A., {et~al.} 2017, \apj, 835, 77

\bibitem[{{Marinelli} \& {Green}(2024)}]{marinelli_2024}
{Marinelli}, M. \& {Green}, J. 2024, in WFC3 Instrument Handbook for Cycle 33 v. 17, Vol.~17, 17

\bibitem[{{Mayor} {et~al.}(2003){Mayor}, {Pepe}, {Queloz}, {Bouchy}, {Rupprecht}, {Lo Curto}, {Avila}, {Benz}, {Bertaux}, {Bonfils}, {Dall}, {Dekker}, {Delabre}, {Eckert}, {Fleury}, {Gilliotte}, {Gojak}, {Guzman}, {Kohler}, {Lizon}, {Longinotti}, {Lovis}, {Megevand}, {Pasquini}, {Reyes}, {Sivan}, {Sosnowska}, {Soto}, {Udry}, {van Kesteren}, {Weber}, \& {Weilenmann}}]{Mayor_2003}
{Mayor}, M., {Pepe}, F., {Queloz}, D., {et~al.} 2003, The Messenger, 114, 20

\bibitem[{{Nardiello} {et~al.}(2025){Nardiello}, {Akana Murphy}, {Spinelli}, {Baratella}, {Desidera}, {Nascimbeni}, {Malavolta}, {Biazzo}, {Maggio}, {Locci}, {Benatti}, {Batalha}, {D'Orazi}, {Borsato}, {Piotto}, {Oelkers}, {Mallonn}, {Sozzetti}, {Bedin}, {Mantovan}, {Zingales}, {Affer}, {Bignamini}, {Bonomo}, {Cabona}, {Collins}, {Damasso}, {Filomeno}, {Ghedina}, {Harutyunyan}, {Lanza}, {Mancini}, {Rainer}, {Scandariato}, {Schwarz}, {Sefako}, \& {Srdoc}}]{nardiello_2025}
{Nardiello}, D., {Akana Murphy}, J.~M., {Spinelli}, R., {et~al.} 2025, \aap, 693, A32

\bibitem[{{Nardiello} {et~al.}(2021){Nardiello}, {Deleuil}, {Mantovan}, {Malavolta}, {Lacedelli}, {Libralato}, {Bedin}, {Borsato}, {Granata}, \& {Piotto}}]{nardiello_2021}
{Nardiello}, D., {Deleuil}, M., {Mantovan}, G., {et~al.} 2021, \mnras, 505, 3767

\bibitem[{{Nardiello} {et~al.}(2022){Nardiello}, {Malavolta}, {Desidera}, {Baratella}, {D'Orazi}, {Messina}, {Biazzo}, {Benatti}, {Damasso}, {Rajpaul}, {Bonomo}, {Capuzzo Dolcetta}, {Mallonn}, {Cale}, {Plavchan}, {El Mufti}, {Bignamini}, {Borsa}, {Carleo}, {Claudi}, {Covino}, {Lanza}, {Maldonado}, {Mancini}, {Micela}, {Molinari}, {Pinamonti}, {Piotto}, {Poretti}, {Scandariato}, {Sozzetti}, {Andreuzzi}, {Boschin}, {Cosentino}, {Fiorenzano}, {Harutyunyan}, {Knapic}, {Pedani}, {Affer}, {Maggio}, \& {Rainer}}]{nardiello_2022}
{Nardiello}, D., {Malavolta}, L., {Desidera}, S., {et~al.} 2022, \aap, 664, A163

\bibitem[{{Nascimbeni} {et~al.}(2024){Nascimbeni}, {Borsato}, {Leonardi}, {Sousa}, {Wilson}, {Fortier}, {Heitzmann}, {Mantovan}, {Luque}, {Zingales}, {Piotto}, {Alibert}, {Alonso}, {B{\'a}rczy}, {Barrado Navascues}, {Barros}, {Baumjohann}, {Beck}, {Benz}, {Billot}, {Biondi}, {Brandeker}, {Broeg}, {Busch}, {Collier Cameron}, {Correia}, {Csizmadia}, {Cubillos}, {Davies}, {Deleuil}, {Deline}, {Delrez}, {Demangeon}, {Demory}, {Derekas}, {Edwards}, {Ehrenreich}, {Erikson}, {Fossati}, {Fridlund}, {Gandolfi}, {Gazeas}, {Gillon}, {G{\"u}del}, {G{\"u}nther}, {Helling}, {Isaak}, {Kerschbaum}, {Kiss}, {Korth}, {Lam}, {Laskar}, {Lecavelier des Etangs}, {Leleu}, {Lendl}, {Magrin}, {Maxted}, {Mer{\'\i}n}, {Mordasini}, {Olofsson}, {Ottensamer}, {Pagano}, {Pall{\'e}}, {Peter}, {Pollacco}, {Queloz}, {Ragazzoni}, {Rando}, {Rauer}, {Ribas}, {Santos}, {Scandariato}, {S{\'e}gransan}, {Simon}, {Smith}, {Southworth}, {Stalport}, {Sulis}, {Szab{\'o}}, {Udry}, {Ulmer}, {Van Grootel}, {Venturini}, {Villaver}, \&
  {Walton}}]{nascimbeni_2024}
{Nascimbeni}, V., {Borsato}, L., {Leonardi}, P., {et~al.} 2024, \aap, 690, A349

\bibitem[{{Nascimbeni} {et~al.}(2023){Nascimbeni}, {Borsato}, {Zingales}, {Piotto}, {Pagano}, {Beck}, {Broeg}, {Ehrenreich}, {Hoyer}, {Majidi}, {Granata}, {Sousa}, {Wilson}, {Van Grootel}, {Bonfanti}, {Salmon}, {Mustill}, {Delrez}, {Alibert}, {Alonso}, {Anglada}, {B{\'a}rczy}, {Barrado}, {Barros}, {Baumjohann}, {Beck}, {Benz}, {Bergomi}, {Billot}, {Bonfils}, {Brandeker}, {Cabrera}, {Charnoz}, {Collier Cameron}, {Csizmadia}, {Cubillos}, {Davies}, {Deleuil}, {Deline}, {Demangeon}, {Demory}, {Erikson}, {Fortier}, {Fossati}, {Fridlund}, {Gandolfi}, {Gillon}, {G{\"u}del}, {Isaak}, {Kiss}, {Laskar}, {Lecavelier des Etangs}, {Lendl}, {Lovis}, {Luque}, {Magrin}, {Maxted}, {Mordasini}, {Olofsson}, {Ottensamer}, {Pall{\'e}}, {Peter}, {Piazza}, {Pollacco}, {Queloz}, {Ragazzoni}, {Rando}, {Ratti}, {Rauer}, {Ribas}, {Santos}, {Scandariato}, {S{\'e}gransan}, {Simon}, {Smith}, {Steinberger}, {Steller}, {Szab{\'o}}, {Thomas}, {Udry}, {Venturini}, {Walton}, \& {Wolter}}]{nascimbeni_2023}
{Nascimbeni}, V., {Borsato}, L., {Zingales}, T., {et~al.} 2023, \aap, 673, A42

\bibitem[{{Nelson} {et~al.}(2014){Nelson}, {Ford}, \& {Payne}}]{nelson_2014}
{Nelson}, B., {Ford}, E.~B., \& {Payne}, M.~J. 2014, \apjs, 210, 11

\bibitem[{{Orell-Miquel} {et~al.}(2023){Orell-Miquel}, {Nowak}, {Murgas}, {Palle}, {Morello}, {Luque}, {Badenas-Agusti}, {Ribas}, {Lafarga}, {Espinoza}, {Morales}, {Zechmeister}, {Alqasim}, {Cochran}, {Gandolfi}, {Goffo}, {Kab{\'a}th}, {Korth}, {Lam}, {Livingston}, {Muresan}, {Persson}, \& {Van Eylen}}]{orell_miquel_2023}
{Orell-Miquel}, J., {Nowak}, G., {Murgas}, F., {et~al.} 2023, \aap, 669, A40

\bibitem[{{Otegi} {et~al.}(2022){Otegi}, {Helled}, \& {Bouchy}}]{otegi_2022}
{Otegi}, J.~F., {Helled}, R., \& {Bouchy}, F. 2022, \aap, 658, A107

\bibitem[{{Owen} \& {Wu}(2017)}]{Owen_Wu_2017}
{Owen}, J.~E. \& {Wu}, Y. 2017, \apj, 847, 29

\bibitem[{Parviainen \& Aigrain(2015)}]{Parviainen_2015}
Parviainen, H. \& Aigrain, S. 2015, MNRAS, 453, 3821

\bibitem[{Parviainen {et~al.}(2023)Parviainen, Luque, \& Palle}]{parviainen_2024}
Parviainen, H., Luque, R., \& Palle, E. 2023, Monthly Notices of the Royal Astronomical Society, 527, 5693

\bibitem[{{Parviainen} {et~al.}(2016){Parviainen}, {Pall{\'e}}, {Nortmann}, {Nowak}, {Iro}, {Murgas}, \& {Aigrain}}]{Parviainen_2016}
{Parviainen}, H., {Pall{\'e}}, E., {Nortmann}, L., {et~al.} 2016, \aap, 585, A114

\bibitem[{{Petit} {et~al.}(2018){Petit}, {Laskar}, \& {Bou{\'e}}}]{petit_2018}
{Petit}, A.~C., {Laskar}, J., \& {Bou{\'e}}, G. 2018, \aap, 617, A93

\bibitem[{Piro \& Vissapragada(2020)}]{piro_vissapragada_2020}
Piro, A.~L. \& Vissapragada, S. 2020, The Astronomical Journal, 159, 131

\bibitem[{{Rein} \& {Liu}(2012)}]{rein_liu_2012}
{Rein}, H. \& {Liu}, S.~F. 2012, \aap, 537, A128

\bibitem[{{Rein} \& {Tamayo}(2016)}]{rein_tamayo_2016}
{Rein}, H. \& {Tamayo}, D. 2016, \mnras, 459, 2275

\bibitem[{{Ricker} {et~al.}(2015){Ricker}, {Winn}, {Vanderspek}, {Latham}, {Bakos}, {Bean}, {Berta-Thompson}, {Brown}, {Buchhave}, {Butler}, {Butler}, {Chaplin}, {Charbonneau}, {Christensen-Dalsgaard}, {Clampin}, {Deming}, {Doty}, {De Lee}, {Dressing}, {Dunham}, {Endl}, {Fressin}, {Ge}, {Henning}, {Holman}, {Howard}, {Ida}, {Jenkins}, {Jernigan}, {Johnson}, {Kaltenegger}, {Kawai}, {Kjeldsen}, {Laughlin}, {Levine}, {Lin}, {Lissauer}, {MacQueen}, {Marcy}, {McCullough}, {Morton}, {Narita}, {Paegert}, {Palle}, {Pepe}, {Pepper}, {Quirrenbach}, {Rinehart}, {Sasselov}, {Sato}, {Seager}, {Sozzetti}, {Stassun}, {Sullivan}, {Szentgyorgyi}, {Torres}, {Udry}, \& {Villasenor}}]{ricker_2015}
{Ricker}, G.~R., {Winn}, J.~N., {Vanderspek}, R., {et~al.} 2015, Journal of Astronomical Telescopes, Instruments, and Systems, 1, 014003

\bibitem[{{Rogers} {et~al.}(2023){Rogers}, {Schlichting}, \& {Owen}}]{rogers_2023}
{Rogers}, J.~G., {Schlichting}, H.~E., \& {Owen}, J.~E. 2023, \apjl, 947, L19

\bibitem[{{Salmon} {et~al.}(2021){Salmon}, {Van Grootel}, {Buldgen}, {Dupret}, \& {Eggenberger}}]{salmon_2021}
{Salmon}, S.~J.~A.~J., {Van Grootel}, V., {Buldgen}, G., {Dupret}, M.~A., \& {Eggenberger}, P. 2021, \aap, 646, A7

\bibitem[{{Santerne} {et~al.}(2019){Santerne}, {Malavolta}, {Kosiarek}, {Dai}, {Dressing}, {Dumusque}, {Hara}, {Lopez}, {Mortier}, {Vanderburg}, {Adibekyan}, {Armstrong}, {Barrado}, {Barros}, {Bayliss}, {Berardo}, {Boisse}, {Bonomo}, {Bouchy}, {Brown}, {Buchhave}, {Butler}, {Collier Cameron}, {Cosentino}, {Crane}, {Crossfield}, {Damasso}, {Deleuil}, {Delgado Mena}, {Demangeon}, {D{\'\i}az}, {Donati}, {Figueira}, {Fulton}, {Ghedina}, {Harutyunyan}, {H{\'e}brard}, {Hirsch}, {Hojjatpanah}, {Howard}, {Isaacson}, {Latham}, {Lillo-Box}, {L{\'o}pez-Morales}, {Lovis}, {Martinez Fiorenzano}, {Molinari}, {Mousis}, {Moutou}, {Nava}, {Nielsen}, {Osborn}, {Petigura}, {Phillips}, {Pollacco}, {Poretti}, {Rice}, {Santos}, {S{\'e}gransan}, {Shectman}, {Sinukoff}, {Sousa}, {Sozzetti}, {Teske}, {Udry}, {Vigan}, {Wang}, {Watson}, {Weiss}, {Wheatley}, \& {Winn}}]{santerne_2019}
{Santerne}, A., {Malavolta}, L., {Kosiarek}, M.~R., {et~al.} 2019, arXiv e-prints, arXiv:1911.07355

\bibitem[{{Santos} {et~al.}(2013){Santos}, {Sousa}, {Mortier}, {Neves}, {Adibekyan}, {Tsantaki}, {Delgado Mena}, {Bonfils}, {Israelian}, {Mayor}, \& {Udry}}]{Santos2013}
{Santos}, N.~C., {Sousa}, S.~G., {Mortier}, A., {et~al.} 2013, \aap, 556, A150

\bibitem[{{Schanche} {et~al.}(2020){Schanche}, {H{\'e}brard}, {Collier Cameron}, {Dalal}, {Smalley}, {Wilson}, {Boisse}, {Bouchy}, {Brown}, {Demangeon}, {Haswell}, {Hellier}, {Kolb}, {Lopez}, {Maxted}, {Pollacco}, {West}, \& {Wheatley}}]{schanche_2020}
{Schanche}, N., {H{\'e}brard}, G., {Collier Cameron}, A., {et~al.} 2020, \mnras, 499, 428

\bibitem[{{Schwarz}(1978)}]{schwarz_1978}
{Schwarz}, G. 1978, Annals of Statistics, 6, 461

\bibitem[{{Scuflaire} {et~al.}(2008){Scuflaire}, {Th{\'e}ado}, {Montalb{\'a}n}, {Miglio}, {Bourge}, {Godart}, {Thoul}, \& {Noels}}]{scuflaire_2008}
{Scuflaire}, R., {Th{\'e}ado}, S., {Montalb{\'a}n}, J., {et~al.} 2008, \apss, 316, 83

\bibitem[{{Shen} \& {Gonz{\'a}lez}(2021)}]{Shen_Gonzalez_2021}
{Shen}, N. \& {Gonz{\'a}lez}, B. 2021, arXiv e-prints, arXiv:2104.14725

\bibitem[{{Siegel} \& {Fabrycky}(2021)}]{Siegel_Fabrycky_2021}
{Siegel}, J.~C. \& {Fabrycky}, D. 2021, \aj, 161, 290

\bibitem[{Skrutskie {et~al.}(2006)Skrutskie, Cutri, Stiening, Weinberg, Schneider, Carpenter, Beichman, Capps, Chester, Elias, Huchra, Liebert, Lonsdale, Monet, Price, Seitzer, Jarrett, Kirkpatrick, Gizis, Howard, Evans, Fowler, Fullmer, Hurt, Light, Kopan, Marsh, McCallon, Tam, Dyk, \& Wheelock}]{skrutskie_2006}
Skrutskie, M.~F., Cutri, R.~M., Stiening, R., {et~al.} 2006, The Astronomical Journal, 131, 1163

\bibitem[{{Sneden}(1973)}]{Sneden1973}
{Sneden}, C.~A. 1973, PhD thesis, University of Texas, Austin

\bibitem[{{Sousa}(2014)}]{Sousa2014}
{Sousa}, S.~G. 2014, in Determination of Atmospheric Parameters of B (Springer), 297--310

\bibitem[{{Sousa} {et~al.}(2021){Sousa}, {Adibekyan}, {Delgado-Mena}, {Santos}, {Rojas-Ayala}, {Soares}, {Legoinha}, {Ulmer-Moll}, {Camacho}, {Barros}, {Demangeon}, {Hoyer}, {Israelian}, {Mortier}, {Tsantaki}, \& {Monteiro}}]{Sousa2021}
{Sousa}, S.~G., {Adibekyan}, V., {Delgado-Mena}, E., {et~al.} 2021, \aap, 656, A53

\bibitem[{{Sousa} {et~al.}(2015){Sousa}, {Santos}, {Adibekyan}, {Delgado-Mena}, \& {Israelian}}]{Sousa2015}
{Sousa}, S.~G., {Santos}, N.~C., {Adibekyan}, V., {Delgado-Mena}, E., \& {Israelian}, G. 2015, \aap, 577, A67

\bibitem[{{Sousa} {et~al.}(2007){Sousa}, {Santos}, {Israelian}, {Mayor}, \& {Monteiro}}]{Sousa2007}
{Sousa}, S.~G., {Santos}, N.~C., {Israelian}, G., {Mayor}, M., \& {Monteiro}, M.~J.~P.~F.~G. 2007, \aap, 469, 783

\bibitem[{{Sousa} {et~al.}(2008){Sousa}, {Santos}, {Mayor}, {Udry}, {Casagrande}, {Israelian}, {Pepe}, {Queloz}, \& {Monteiro}}]{Sousa2008}
{Sousa}, S.~G., {Santos}, N.~C., {Mayor}, M., {et~al.} 2008, \aap, 487, 373

\bibitem[{Steffen {et~al.}(2012)Steffen, Ford, Rowe, Fabrycky, Holman, Welsh, Batalha, Borucki, Bryson, Caldwell, Ciardi, Jenkins, Kjeldsen, Koch, Prša, Sanderfer, Seader, \& Twicken}]{steffen_2012}
Steffen, J.~H., Ford, E.~B., Rowe, J.~F., {et~al.} 2012, The Astrophysical Journal, 756, 186

\bibitem[{Storn \& Price(1997)}]{Storn_97}
Storn, R. \& Price, K.~V. 1997, J. Glob. Optim., 11, 341

\bibitem[{{Sulis} {et~al.}(2024){Sulis}, {Borsato}, {Grouffal}, {Osborn}, {Santerne}, {Brandeker}, {G{\"u}nther}, {Heitzmann}, {Lendl}, {Fridlund}, {Gandolfi}, {Alibert}, {Alonso}, {B{\'a}rczy}, {Barrado Navascues}, {Barros}, {Baumjohann}, {Beck}, {Benz}, {Bergomi}, {Billot}, {Bonfanti}, {Broeg}, {Collier Cameron}, {Corral van Damme}, {Correia}, {Csizmadia}, {Cubillos}, {Davies}, {Deleuil}, {Deline}, {Delrez}, {Demangeon}, {Demory}, {Derekas}, {Edwards}, {Ehrenreich}, {Erikson}, {Fortier}, {Fossati}, {Gazeas}, {Gillon}, {G{\"u}del}, {Helling}, {Hoyer}, {Isaak}, {Kiss}, {Korth}, {Lam}, {Laskar}, {Lecavelier des Etangs}, {Magrin}, {Maxted}, {Mordasini}, {Nascimbeni}, {Olofsson}, {Ottensamer}, {Pagano}, {Pall{\'e}}, {Peter}, {Piazza}, {Piotto}, {Pollacco}, {Queloz}, {Ragazzoni}, {Rando}, {Rauer}, {Ribas}, {Santos}, {Scandariato}, {S{\'e}gransan}, {Simon}, {Smith}, {Sousa}, {Stalport}, {Steinberger}, {Szab{\'o}}, {Tuson}, {Udry}, {Ulmer-Moll}, {Van Grootel}, {Venturini}, {Villaver}, {Walton}, {Wilson}, {Wolter},
  \& {Zingales}}]{sulis_2024}
{Sulis}, S., {Borsato}, L., {Grouffal}, S., {et~al.} 2024, \aap, 686, L18

\bibitem[{ter Braak \& Vrugt(2008)}]{terbraak_2008}
ter Braak, C. J.~F. \& Vrugt, J.~A. 2008, Statistics and Computing, 18, 435

\bibitem[{{Thiabaud} {et~al.}(2015){Thiabaud}, {Marboeuf}, {Alibert}, {Leya}, \& {Mezger}}]{thiabaud_2015}
{Thiabaud}, A., {Marboeuf}, U., {Alibert}, Y., {Leya}, I., \& {Mezger}, K. 2015, \aap, 574, A138

\bibitem[{{Tsiaras} {et~al.}(2016{\natexlab{a}}){Tsiaras}, {Rocchetto}, {Waldmann}, {Venot}, {Varley}, {Morello}, {Damiano}, {Tinetti}, {Barton}, {Yurchenko}, \& {Tennyson}}]{Tsiaras_2016a}
{Tsiaras}, A., {Rocchetto}, M., {Waldmann}, I.~P., {et~al.} 2016{\natexlab{a}}, \apj, 820, 99

\bibitem[{{Tsiaras} {et~al.}(2016{\natexlab{b}}){Tsiaras}, {Waldmann}, {Rocchetto}, {Varley}, {Morello}, {Damiano}, \& {Tinetti}}]{Tsiaras_2016b}
{Tsiaras}, A., {Waldmann}, I.~P., {Rocchetto}, M., {et~al.} 2016{\natexlab{b}}, \apj, 832, 202

\bibitem[{{Tsiaras} {et~al.}(2018){Tsiaras}, {Waldmann}, {Zingales}, {Rocchetto}, {Morello}, {Damiano}, {Karpouzas}, {Tinetti}, {McKemmish}, {Tennyson}, \& {Yurchenko}}]{Tsiaras_2018}
{Tsiaras}, A., {Waldmann}, I.~P., {Zingales}, T., {et~al.} 2018, \aj, 155, 156

\bibitem[{{Van Eylen} {et~al.}(2019){Van Eylen}, {Albrecht}, {Huang}, {MacDonald}, {Dawson}, {Cai}, {Foreman-Mackey}, {Lundkvist}, {Silva Aguirre}, {Snellen}, \& {Winn}}]{vaneylen_2019}
{Van Eylen}, V., {Albrecht}, S., {Huang}, X., {et~al.} 2019, \aj, 157, 61

\bibitem[{{Vanderburg} {et~al.}(2016){Vanderburg}, {Becker}, {Kristiansen}, {Bieryla}, {Duev}, {Jensen-Clem}, {Morton}, {Latham}, {Adams}, {Baranec}, {Berlind}, {Calkins}, {Esquerdo}, {Kulkarni}, {Law}, {Riddle}, {Salama}, \& {Schmitt}}]{vanderburg_2016}
{Vanderburg}, A., {Becker}, J.~C., {Kristiansen}, M.~H., {et~al.} 2016, \apjl, 827, L10

\bibitem[{Weiss {et~al.}(2018)Weiss, Isaacson, Marcy, Howard, Petigura, Fulton, Winn, Hirsch, Sinukoff, Rowe, \& Survey}]{weiss_2018}
Weiss, L.~M., Isaacson, H.~T., Marcy, G.~W., {et~al.} 2018, The Astronomical Journal, 156, 254

\bibitem[{{Weiss} {et~al.}(2023){Weiss}, {Millholland}, {Petigura}, {Adams}, {Batygin}, {Block}, \& {Mordasini}}]{weiss_2023}
{Weiss}, L.~M., {Millholland}, S.~C., {Petigura}, E.~A., {et~al.} 2023, in Astronomical Society of the Pacific Conference Series, Vol. 534, Protostars and Planets VII, ed. S.~{Inutsuka}, Y.~{Aikawa}, T.~{Muto}, K.~{Tomida}, \& M.~{Tamura}, 863

\bibitem[{{Werner} {et~al.}(2004){Werner}, {Roellig}, {Low}, {Rieke}, {Rieke}, {Hoffmann}, {Young}, {Houck}, {Brandl}, {Fazio}, {Hora}, {Gehrz}, {Helou}, {Soifer}, {Stauffer}, {Keene}, {Eisenhardt}, {Gallagher}, {Gautier}, {Irace}, {Lawrence}, {Simmons}, {Van Cleve}, {Jura}, {Wright}, \& {Cruikshank}}]{werner_2004}
{Werner}, M.~W., {Roellig}, T.~L., {Low}, F.~J., {et~al.} 2004, \apjs, 154, 1

\bibitem[{{Winn}(2010)}]{winn_2010}
{Winn}, J.~N. 2010, in Exoplanets, ed. S.~{Seager}, 55--77

\bibitem[{Wong \& Lee(2024)}]{wong_2024}
Wong, K.~H. \& Lee, M.~H. 2024, The Astronomical Journal, 167, 112

\bibitem[{{Wright} {et~al.}(2010){Wright}, {Eisenhardt}, {Mainzer}, {Ressler}, {Cutri}, {Jarrett}, {Kirkpatrick}, {Padgett}, {McMillan}, {Skrutskie}, {Stanford}, {Cohen}, {Walker}, {Mather}, {Leisawitz}, {Gautier}, {McLean}, {Benford}, {Lonsdale}, {Blain}, {Mendez}, {Irace}, {Duval}, {Liu}, {Royer}, {Heinrichsen}, {Howard}, {Shannon}, {Kendall}, {Walsh}, {Larsen}, {Cardon}, {Schick}, {Schwalm}, {Abid}, {Fabinsky}, {Naes}, \& {Tsai}}]{wright_2010}
{Wright}, E.~L., {Eisenhardt}, P. R.~M., {Mainzer}, A.~K., {et~al.} 2010, \aj, 140, 1868

\bibitem[{{Zeng} {et~al.}(2019){Zeng}, {Jacobsen}, {Sasselov}, {Petaev}, {Vanderburg}, {Lopez-Morales}, {Perez-Mercader}, {Mattsson}, {Li}, {Heising}, {Bonomo}, {Damasso}, {Berger}, {Cao}, {Levi}, \& {Wordsworth}}]{zeng_2019}
{Zeng}, L., {Jacobsen}, S.~B., {Sasselov}, D.~D., {et~al.} 2019, Proceedings of the National Academy of Science, 116, 9723

\end{thebibliography}

\begin{appendix}
\label{appendix}
\clearpage

\clearpage
\onecolumn

\section{Additional tables and plots}

\subsection{CHEOPS observations log}

\begin{table*}[!h]
    \centering
    \caption{\label{table:log_CHEOPS}Log of CHEOPS observations.}
    \small
    \begin{tabular}{cclcrccc}
        \hline\hline
        \noalign{\smallskip}
    VISIT ID & Planet & File Key  & Start date & Duration & Number of Frames   & Efficiency\\  
     & &          & (UTC)      & (h)      &         & (\%)       &          \\  
        \noalign{\smallskip}
        \hline
        \noalign{\smallskip}
    1 & -b & CH\_PR100025\_TG005501\_V0300 & 2020-12-23T21:03 & 10.7 & 588 & 55.7  \\
    2 & -b & CH\_PR100025\_TG005201\_V0300 & 2021-01-08T20:08 & 6.49 & 366 & 58.21 \\
    3 & -b & CH\_PR100025\_TG005701\_V0300 & 2021-02-08T17:20 & 10.9 & 672 & 63.83 \\
    4 & -b & CH\_PR100025\_TG005702\_V0300 & 2021-03-11T22:16 & 12.04 & 671 & 55.18 \\
    5 & -b & CH\_PR100025\_TG006501\_V0300 & 2022-01-01T20:18 &  10.77 & 635 & 60.10 \\
    6 & -b & CH\_PR100025\_TG006502\_V0300 & 2022-01-17T07:26 & 10.68 &  616 & 59.68\\
    7 & -b & CH\_PR100025\_TG006503\_V0300 & 2022-02-17T10:51 &  11.42 & 648 & 54.71 \\
    8 & -c & CH\_PR100025\_TG006901\_V0300 & 2022-02-26T00:11 & 10.84 & 662  & 62.90\\
    9 & -b & CH\_PR100025\_TG006801\_V0300 & 2022-03-20T15:25 & 11.40 & 536 & 46.11\\
    10 & -c & CH\_PR100025\_TG006902\_V0300 & 2022-12-08T10:11 & 11.42 & 562 & 49.17\\
    11 & -b & CH\_PR100025\_TG006802\_V0300 & 2023-01-10T12:28 & 10.68 & 653 & 62.75 \\ 
    12 & -b & CH\_PR100025\_TG006803\_V0300  & 2023-01-25T23:11 & 10.87 & 650 & 60.49 \\
    13 & -c & CH\_PR100025\_TG006903\_V0300 & 2023-02-09T20:51 & 10.92 & 681 & 60.64 \\
    14 & -b & CH\_PR100025\_TG006904\_V0300 & 2023-03-13T14:31 & 15.69 & 823 & 52.09 \\ 
    15 & -c & CH\_PR140080\_TG008101\_V0300 & 2025-03-11T15:24 & 13.03 & 724 & 44.85 \\
        \noalign{\smallskip}
    \hline 
    \end{tabular}
\end{table*}

\clearpage

\subsection{Global transit light curve analysis}

\begin{table*}[!h]
    \centering\small\centering\renewcommand{\arraystretch}{1.3} 
    \caption{Posteriors and derived orbital parameters for HIP\,41378\,b \& c from the global photometric analysis with PyORBIT, presented in Section \ref{section:pyorbit_analysis} }
    \begin{tabular}{l c c r}
    \hline\hline
    Parameter & Unit &  Prior &  Posterior value \\
    \hline
    \emph{HIP\,41378\,b} \rule{0pt}{12pt} & & & \\
    &&& \\
    \textsc{Fitted Parameters} & & &\\
    Orbital Period ($P$) & [days] &  (fixed) & $15.571893_{-0.000053}^{+0.000068}$ \\
    Impact Parameter ($b$) &-- & \unif{0}{1+ (R_{\star}/R_{p})/ 2}  & ${0.445}_{-0.023}^{+0.020}$ \\
    Planet-star radius ratio ($R_{p}/R_{\star}$) & --& \unif{0}{0.5} & ${0.01705}_{-0.00015}^{+0.00015}$\\
    \\
    \textsc{Derived Parameters} &  & & \\ 
    Semimajor axis ($a$) & [au] & --  & ${0.1303}_{-0.0013}^{+0.0013}$ \\
    Scaled semimajor axis ($a/R_{s}$) & -- & -- &  ${21.54}_{-0.16}^{+0.16}$ \\
    Inclination ($i$) & [deg] & --  & ${88.816}_{-0.061}^{+0.065}$ \\ 
    Planet Radius ($R_{p}$) & [$R_{\oplus}$]  & --  & ${2.419}_{-0.027}^{+0.027}$ \\ 
    Transit duration ($T_{14}$) & [days] & --& ${0.2099}_{-0.0012}^{+0.0014}$ \\
    Equilibrium temperature ($T_{eq}$) & [K] & -- &   ${970.67}_{-10.54}^{+10.54}$  \\ 
    Stellar Insolation Flux ($S$)&  [$S_{\oplus}$] & -- & ${148.50}_{-7.12}^{+7.12}$ \\
    \hline
    \emph{HIP\,41378\,c} \rule{0pt}{12pt} & & & \\
    &&& \\
    \textsc{Fitted Parameters} & & &\\
    Orbital Period ($P$) & [days] &  (fixed) & $31.708380_{-0.00041}^{+0.00039}$ \\
    Impact Parameter ($b$) &-- & \unif{0}{1+ (R_{\star}/R_{p})/ 2} & ${0.9290}_{-0.0044}^{+0.0041}$ \\
    Planet-star radius ratio ($R_{p}/R_{\star}$) & -- & \unif{0}{0.5} & ${0.01766}_{-0.00042}^{+0.00043}$\\
    \\
    \textsc{Derived Parameters} &  & & \\ 
    Semimajor axis ($a$) & [au] & --  & ${0.2093}_{-0.0022}^{+0.0022}$ \\
    Scaled semimajor axis ($a/R_{s}$) & -- & -- & ${34.60}_{-0.26}^{+0.26}$ \\
    Inclination ($i$) & [deg] &--  & ${88.462}_{-0.015}^{+0.015}$ \\ 
    Planet Radius ($R_{p}$) & [$R_{\oplus}$]  & --  & ${2.505}_{-0.056}^{+0.057}$ \\ 
    Transit duration ($T_{14}$) & [days] & -- & ${0.1212}_{-0.0025}^{0.0027}$ \\
    Equilibrium temperature ($T_{eq}$) & [K] & -- &   ${765.87}_{-8.33}^{+8.33}$  \\
    Stellar Insolation Flux ($S$)&  [$S_{\oplus}$] & -- & ${57.55}_{-2.79}^{+2.79}$ \\
    \hline
    \textsc{Host star HIP\,41378} \\
    Stellar density  & [$\rho_{\odot}$] & \gauss{0.5539}{0.0126}& $0.554_{-0.013}^{+0.013}$ \\
    \hline
    \emph{Limb Darkening Coefficients} \rule{0pt}{12pt} & & & \\
    Quadratic Limb Darkening term $c_{1}$ (\textit{K2}) &-- & \gauss{0.4834}{0.05} & $0.447_{-0.043}^{+0.043}$\\ 
    Quadratic Limb Darkening term $c_{2}$ (\textit{K2}) & --& \gauss{0.1535}{0.05}  &  $0.157_{-0.046}^{+0.046}$\\ 
    Quadratic Limb Darkening term $c_{1}$ (TESS) & --&  \gauss{0.3822}{0.05} & $0.401_{-0.047}^{+0.047}$ \\ 
    Quadratic Limb Darkening term $c_{2}$ (TESS) & --&  \gauss{0.1429}{0.05}  & $0.171_{-0.048}^{+0.048}$\\
    Quadratic Limb Darkening term $c_{1}$ (HST) & --& \gauss{0.228}{0.05}  & $0.183_{-0.042}^{+0.041}$ \\ 
    Quadratic Limb Darkening term $c_{2}$ (HST) & --&  \gauss{0.1398}{0.05}  & $0.127_{-0.045}^{+0.046}$\\ 
    Quadratic Limb Darkening term $c_{1}$ (\textit{Spitzer}) & --& \gauss{0.0969}{0.05} & $0.108_{-0.047}^{+0.049}$ \\ 
    Quadratic Limb Darkening term $c_{2}$ (\textit{Spitzer}) & --&  \gauss{0.0508}{0.05} & $0.058_{-0.047}^{+0.049}$\\
    Quadratic Limb Darkening term $c_{1}$ (CHEOPS) & --&  \gauss{0.4947}{0.05} & $0.463_{-0.041}^{+0.040}$ \\ 
    Quadratic Limb Darkening term $c_{2}$ (CHEOPS) & --&   \gauss{0.1519}{0.05} & $0.150_{-0.045}^{+0.045}$\\ 
    \hline
    \end{tabular}
    \label{table:fit_orbital_parameters}
    \tablefoot{The listed best-fit values and uncertainties are the medians and 15.865th-84.135th percentiles of the posterior distributions, respectively.}
\end{table*}

\begin{figure*}[!h]
    \centering
    \includegraphics[width=\linewidth,keepaspectratio] {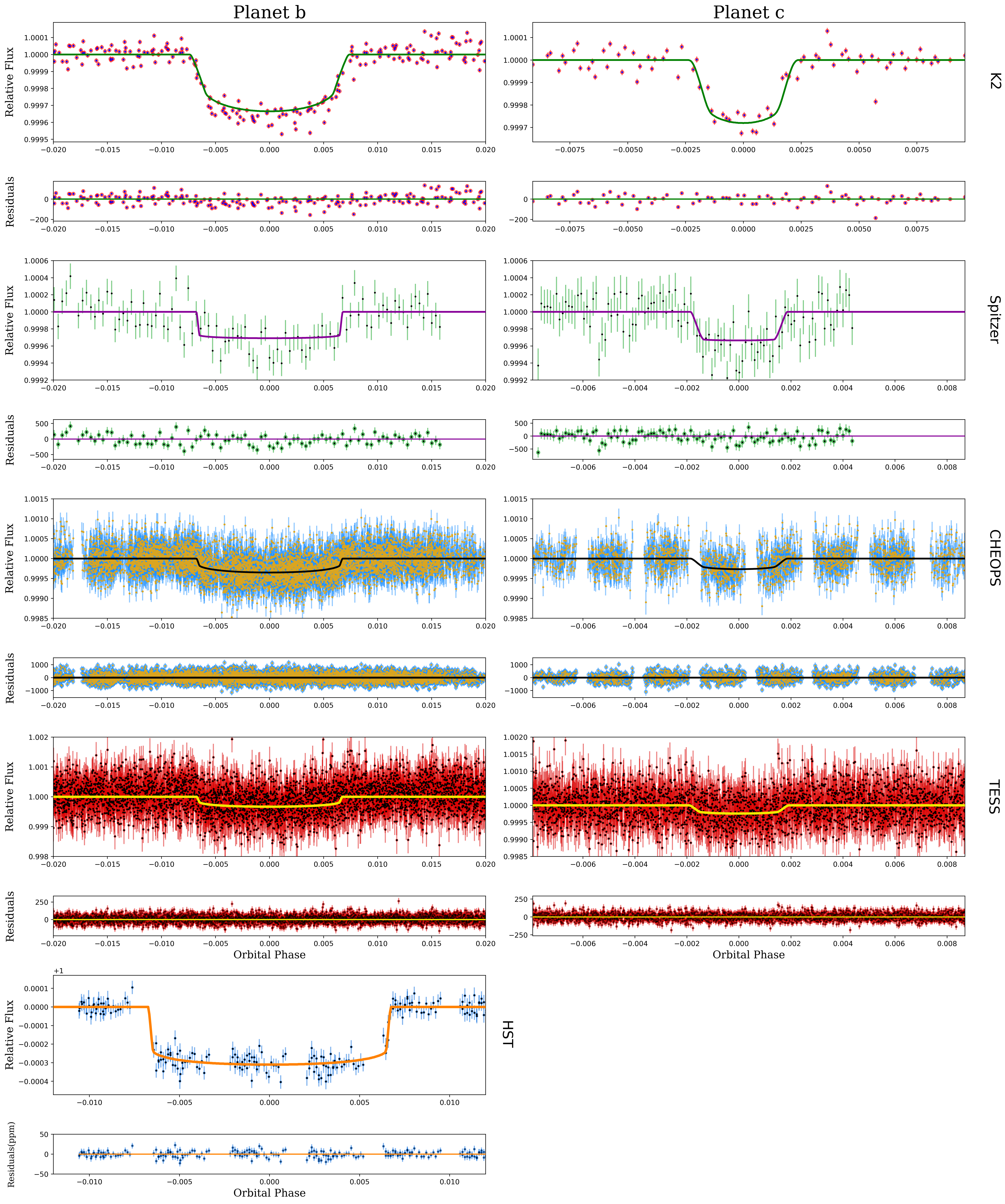}
    \caption{Phase-folded light curves of HIP\,41378\,b \&\,c, combining observations from \textit{K2}, Spitzer, HST, CHEOPS, and TESS. The oversampled best-fit transit models are overlaid on the light curves.}
    \label{Figure:LC_phase_8}
\end{figure*}

\begin{figure*}[!h]
    \centering
    \includegraphics[width=\linewidth,keepaspectratio] {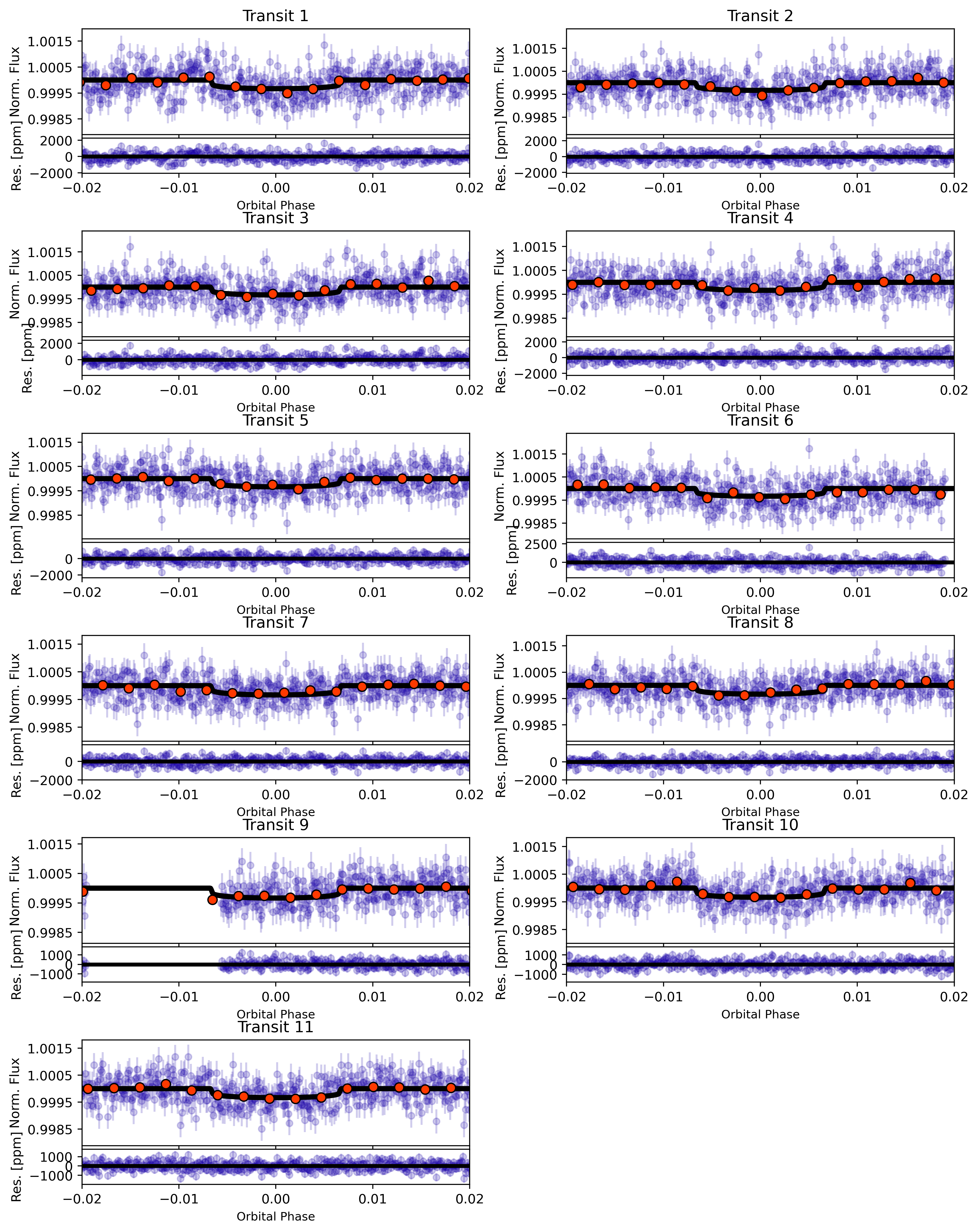}
    \caption{Individual TESS phase-folded transit light curves of HIP\,41378\,b. The binned points are showed in red. The oversampled best-fit transit models are overlaid on the light curves.}
    \label{Figure:LC_tot_TESS_b}
\end{figure*}

\begin{figure*}[!h]
    \centering
    \includegraphics[width=\linewidth,keepaspectratio] {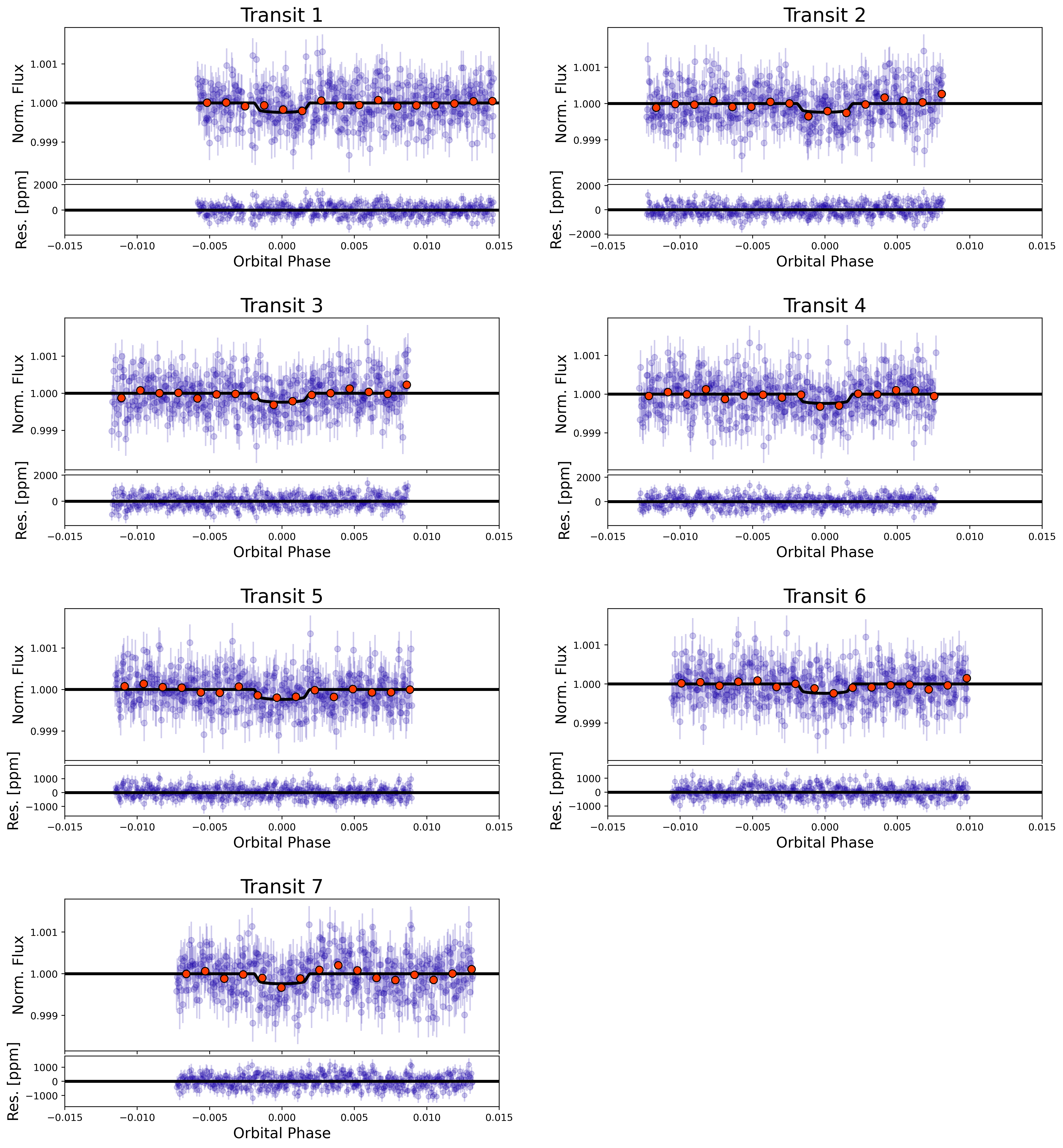}
    \caption{Individual TESS phase-folded transit light curves of HIP\,41378\,c. The binned points are showed in red. The oversampled best-fit transit models are overlaid on the light curves.}
    \label{Figure:LC_tot_TESS_c}
\end{figure*}

\begin{figure*}[!h]
    \centering
    \includegraphics[width=\linewidth,keepaspectratio] {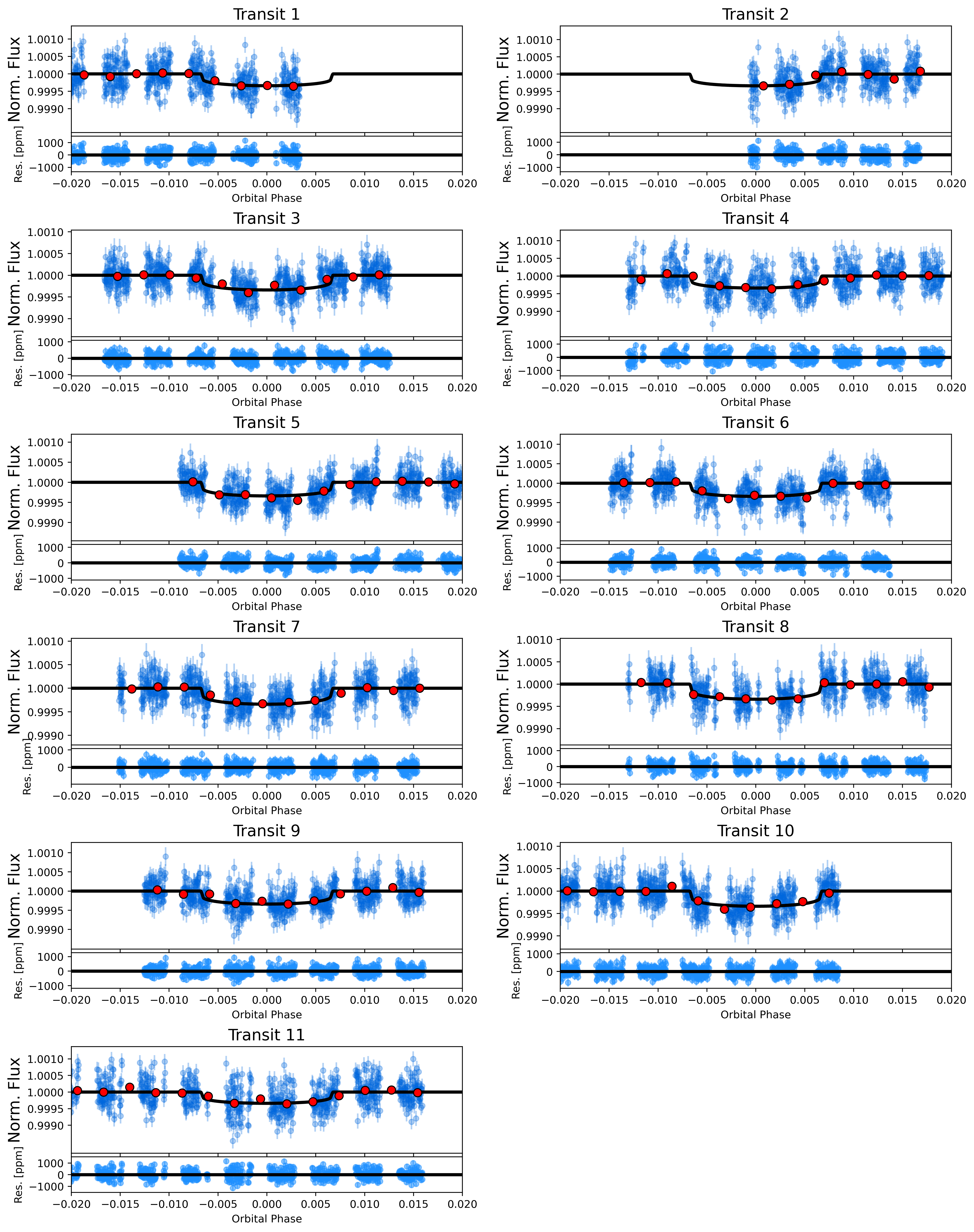}
    \caption{Individual CHEOPS phase-folded transit light curves of HIP\,41378\,b. The binned points are showed in red. The oversampled best-fit transit models are overlaid on the light curves.}
    \label{Figure:LC_tot_CHEOPS_b}
\end{figure*}

\begin{figure*}[!h]
    \centering
    \includegraphics[width=\linewidth,keepaspectratio] {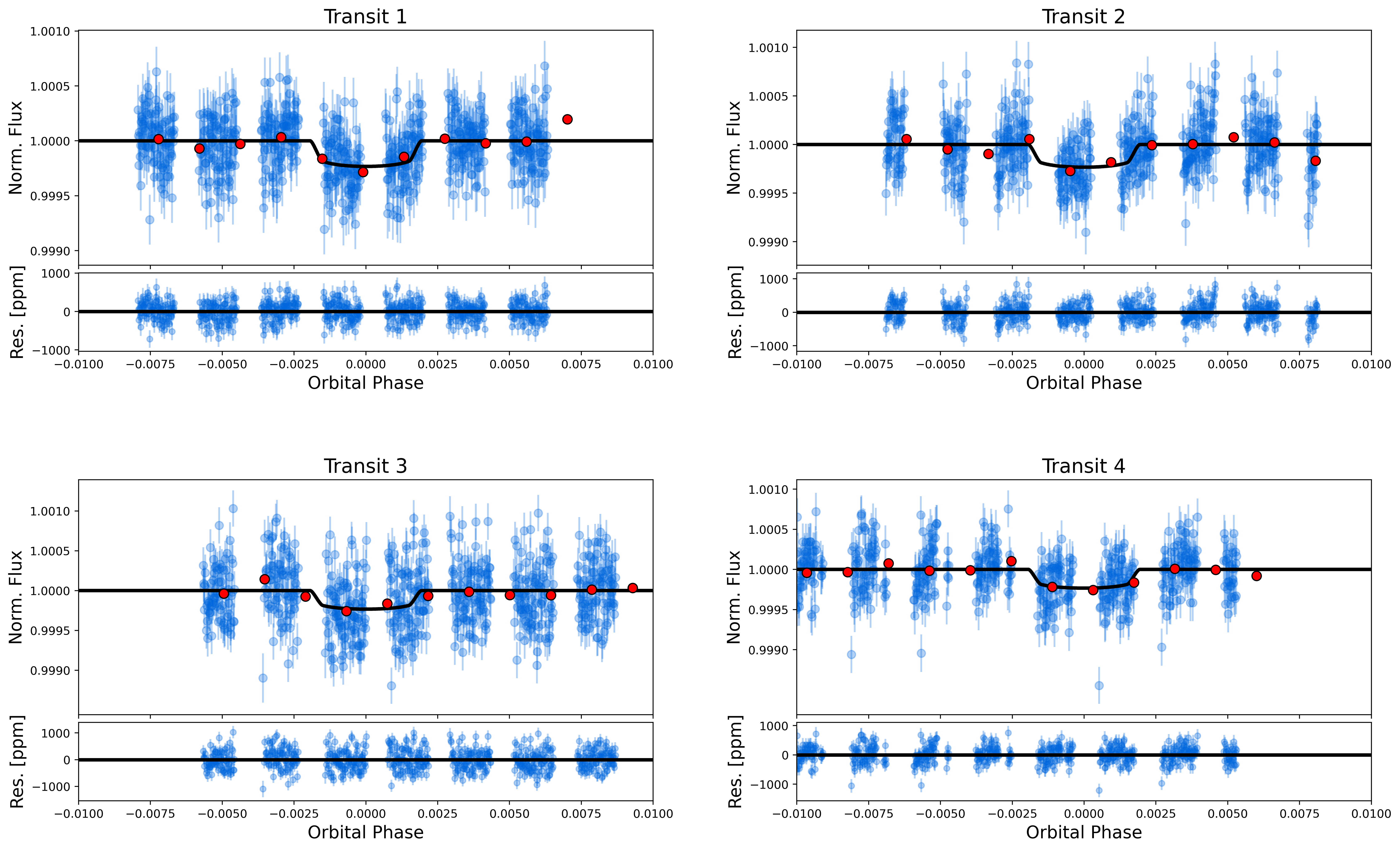}
    \caption{Individual CHEOPS phase-folded transit light-curves of HIP\,41378\,c. The binned points are showed in red. The oversampled best-fit transit models are overlaid on the light curves.}
    \label{Figure:LC_tot_CHEOPS_c}
\end{figure*}

\clearpage

\subsection{Dynamical analysis}

\begin{table*}[!h]
    \centering\small\centering\renewcommand{\arraystretch}{1.2}
    \caption{Posteriors and derived orbital parameters (MAP and HDI) for HIP\,41378\,b, c, and g obtained from the dynamical analysis with \trades.}
    \begin{tabular}{l c c r}
    \hline\hline
    Parameter & Unit & Prior & MAP (HDI$\pm 1\sigma$) \\
    \hline
    \emph{HIP\,41378\,b} \rule{0pt}{12pt} & & & \\
    &&& \\
    \textsc{Fitted Parameters} & & &\\
    $M_{\mathrm{p}}/M_{\star}$ & [$\left(\frac{M_\odot}{M_\star}\right) \times 10^{-6}$] & \unif{0.30}{954.7} & $22_{-1}^{+3}$ \\
    Orbital Period ($P$) & [days] & \unif{13.5}{17.5} & $15.571306_{-0.000323}^{+0.000011}$  \\
    $\sqrt{e} \cos \omega$ & -- & \unif{-\sqrt{0.5}}{\sqrt{0.5}} & $0.142_{-0.037}^{+0.011}$ \\
    $\sqrt{e} \sin \omega$ & -- & \unif{-\sqrt{0.5}}{\sqrt{0.5}} & $-0.033_{-0.038}^{+0.032}$ \\
    Mean Longitude ($\lambda$) & [deg] & \unif{0}{360} & $274.34_{-0.28}^{+1.09}$  \\
    \\
    \textsc{Derived Parameters} &  & & \\ 
    Mass ($m_{\mathrm{p}}$) & [$M_{\oplus}$] & -- &  $9.06_{-0.51}^{+1.41}$ \\
    Eccentricity ($e$) & -- & \halfgauss{0}{0.083}  & $0.0213_{-0.0099}^{+0.0022}$ \\
    Argument of Periastron ($\omega$) & [deg] & -- & $-13_{-17}^{+14}$ \\
    Mean Anomaly ($M_{\mathrm{A}}$) & [deg] & -- & $107_{-15}^{+16}$ \\
    \hline
    
    \emph{HIP\,41378\,c} \rule{0pt}{12pt} & & & \\
    &&& \\
    \textsc{Fitted Parameters} & & &\\
    $M_{\mathrm{p}}/M_{\star}$ & [$\left(\frac{M_\odot}{M_\star}\right) \times 10^{-6}$] & \unif{0.30}{954.7} & $16_{-1}^{+3}$ \\
    Orbital Period ($P$) & [days] & \unif{30}{34} & $31.71054_{-0.00098}^{+0.00172}$ \\
    $\sqrt{e} \cos \omega$ & -- & \unif{-\sqrt{0.5}}{\sqrt{0.5}} & $-0.242_{-0.010}^{+0.039}$ \\
    $\sqrt{e} \sin \omega$ & -- & \unif{-\sqrt{0.5}}{\sqrt{0.5}} & $0.096_{-0.015}^{+0.054}$  \\
    Mean Longitude ($\lambda$) & [deg] & \unif{0}{360} & $333_{-2}^{+8}$ \\
    Longitude of Ascending Node ($\Omega$) & [deg] & -- & $173_{-1}^{+9}$ \\
    \\
    \textsc{Derived Parameters} &  & & \\ 
    Mass ($m_{\mathrm{p}}$) & [$M_{\oplus}$] & -- & $6.53_{-0.42}^{+1.33}$ \\
    Eccentricity ($e$) & -- & \halfgauss{0}{0.083} & $0.0678_{-0.0097}^{+0.0078}$  \\
    Argument of Periastron ($\omega$) & [deg] & -- & $158_{-13}^{+4}$ \\
    Mean Anomaly ($M_{\mathrm{A}}$) & [deg] & -- &  $2_{-3}^{+13}$ \\
    \hline
    
    \emph{HIP\,41378\,g} \rule{0pt}{12pt} & & & \\
    &&& \\
    \textsc{Fitted Parameters} & & &\\
    $M_{\mathrm{p}}/M_{\star}$ & [$\left(\frac{M_\odot}{M_\star}\right) \times 10^{-6}$] & \unif{0.30}{901} & $17_{-2}^{+3}$ \\
    Orbital Period ($P$) & [days] &  \unif{32}{200} & $64.067_{-0.067}^{+0.026}$  \\
    $\sqrt{e} \cos \omega$ & -- & \unif{-\sqrt{0.5}}{\sqrt{0.5}} & $-0.091_{-0.116}^{+0.073}$ \\
    $\sqrt{e} \sin \omega$ & -- & \unif{-\sqrt{0.5}}{\sqrt{0.5}} & $-0.043_{-0.011}^{+0.231}$ \\
    Mean Longitude ($\lambda$) & [deg] & \unif{0}{360} & $350_{-22}^{+4}$ \\
    Inclination ($i$) & [deg] & -- & $95_{-10}^{+1}$  \\
    Longitude of Ascending Node ($\Omega$) & [deg] & -- & $184_{-6}^{+6}$ \\
    \\
    \textsc{Derived Parameters} &  & & \\ 
    Mass ($m_{\mathrm{p}}$) & [$M_{\oplus}$] & -- & $6.81_{-0.98}^{+1.14}$ \\
    Eccentricity ($e$) & -- & \halfgauss{0}{0.083} & $0.010_{-0.010}^{+0.031}$ \\
    Argument of Periastron ($\omega$) & [deg] & -- & $205_{-115}^{+5}$ \\
    Mean Anomaly ($M_{\mathrm{A}}$) & [deg] & -- &  $-39_{-8}^{+117}$  \\
    \hline
    
    \textsc{Additional Parameters} \rule{0pt}{12pt} & & & \\
    Radial Velocity Jitter ($\sigma_{\mathrm{jitter}}$) & [m/s] & -- & $2.36_{-0.10}^{+0.19}$  \\
    Radial Velocity offset ($\gamma_1$) & [m/s] & -- & $50711.789_{-0.241}^{+0.095}$ \\
    \hline
    \end{tabular}
    \label{table:fit_TRADES_orbital_parameters}
    \tablefoot{The symbols $\mathcal{U}$, $\mathcal{G}$, and $\mathcal{N}^{+}$ refer to uniform, Gaussian, and half-Gaussian distributions, respectively.}

\end{table*}

\clearpage

\subsection{Internal structure}

\begin{figure*}[!h]
    \centering
    \includegraphics[width=\linewidth]{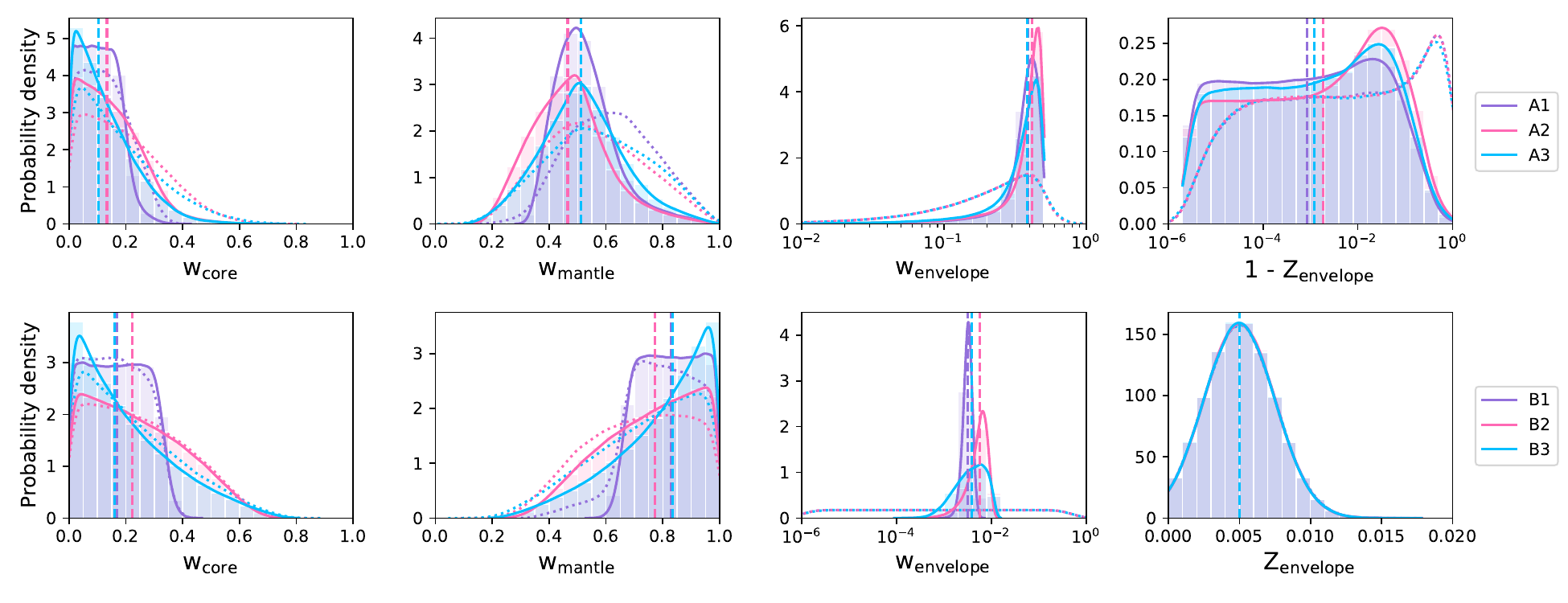}
    \caption{Posteriors for the main internal structure parameters of HIP\,41378\,b, namely the mass fractions of the planet's inner core (far left), mantle (middle left), and envelope layers (middle right), as well as the mass fraction of water in the envelope layer (far right). We show models assuming the planet's Si/Mg/Fe ratios match those of the host star exactly (purple), are Fe-enriched compared to the host star (pink), and are independent of the host star metallicity (blue). For all three options, we also use two water priors, favoring a water-rich (top row) and water-poor composition (bottom row), respectively. The vertical dashed lines show the medians of the inferred distributions and the dotted lines the chosen priors.}
    \label{fig:int_struct_b}
\end{figure*}

\begin{figure*}
    \centering
    \includegraphics[width=\linewidth]{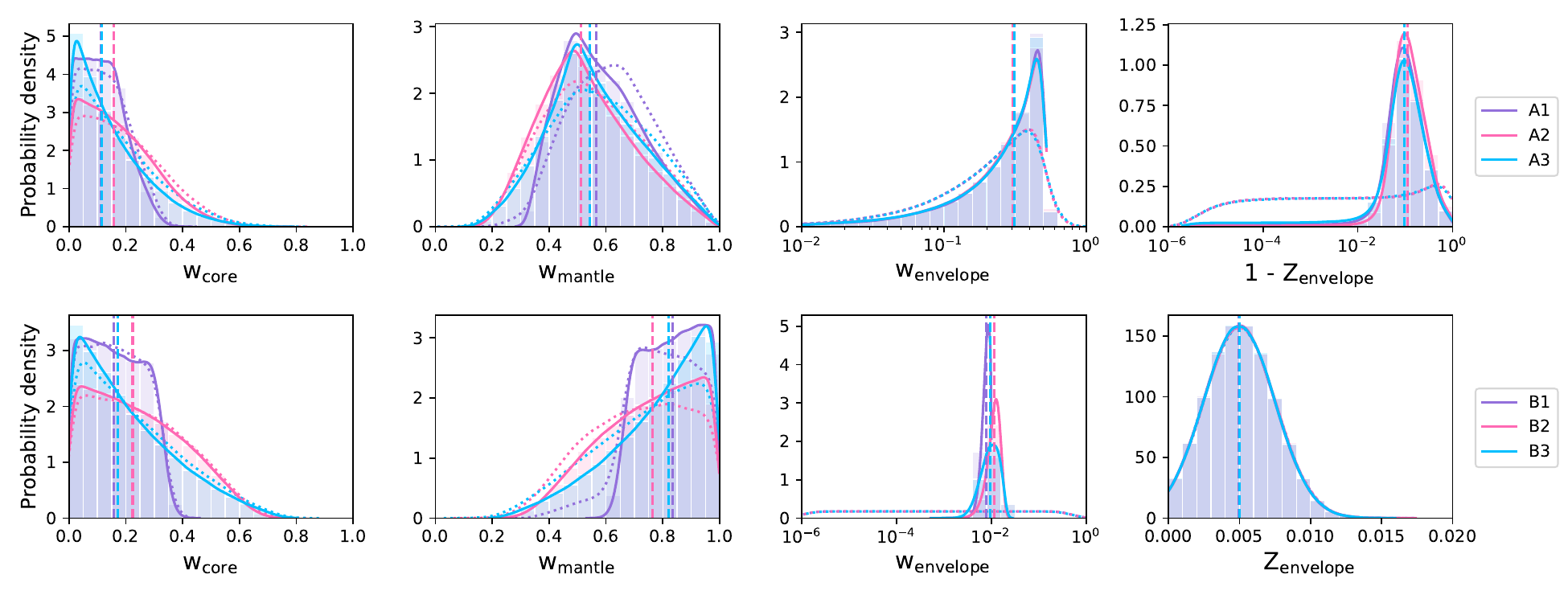}
    \caption{Same as Figure \ref{fig:int_struct_b} but for HIP\,41378\,c.}
    \label{fig:int_struct_c}
\end{figure*}

\begin{table*}[!h]
\renewcommand{\arraystretch}{1.5}
\caption{Results of the internal structure modeling for HIP\,41378 b.}
\centering
\begin{tabular}{r|ccc|ccc}
\hline \hline
Water prior &              \multicolumn{3}{c|}{Formation outside iceline (water-rich)} & \multicolumn{3}{c}{Formation inside iceline (water-poor)} \\
Si/Mg/Fe prior &           Stellar (A1) &       Iron-enriched (A2) &      Free (A3) &
                           Stellar (B1) &       Iron-enriched (B2) &      Free (B3) \\
\hline
w$_\textrm{core}$ [\%] &        $10_{-7}^{+7}$ &    $13_{-9}^{+12}$ &    $10_{-8}^{+13}$ &
                           $17_{-12}^{+12}$ &    $22_{-16}^{+21}$ &    $16_{-12}^{+22}$ \\
w$_\textrm{mantle}$ [\%] &      $51_{-9}^{+11}$ &    $47_{-13}^{+13}$ &    $51_{-14}^{+15}$ &
                           $83_{-12}^{+12}$ &    $77_{-21}^{+16}$ &    $84_{-22}^{+12}$ \\
w$_\textrm{envelope}$ [\%] &    $38.9_{-10.1}^{+6.5}$ &    $41.6_{-13.7}^{+5.8}$ &    $38.5_{-14.8}^{+7.9}$ &
                           $0.32_{-0.07}^{+0.07}$ &    $0.56_{-0.25}^{+0.23}$ &    $0.39_{-0.23}^{+0.38}$ \\
\hline
Z$_\textrm{envelope}$ [\%] &        $99.9_{-3.2}^{+0.1}$ &    $99.8_{-5.2}^{+0.2}$ &    $99.9_{-3.9}^{+0.1}$ &
                           $0.5_{-0.2}^{+0.2}$ &    $0.5_{-0.2}^{+0.2}$ &    $0.5_{-0.2}^{+0.2}$ \\
\hline
x$_\textrm{Fe,core}$ [\%] &     $90.3_{-6.4}^{+6.5}$ &    $90.3_{-6.4}^{+6.5}$ &    $90.3_{-6.3}^{+6.6}$ &
                           $90.3_{-6.4}^{+6.5}$ &    $90.4_{-6.4}^{+6.5}$ &    $90.3_{-6.4}^{+6.5}$ \\
x$_\textrm{S,core}$ [\%] &      $9.7_{-6.5}^{+6.4}$ &    $9.7_{-6.5}^{+6.4}$ &    $9.7_{-6.6}^{+6.3}$ &
                           $9.7_{-6.5}^{+6.4}$ &    $9.6_{-6.5}^{+6.4}$ &    $9.7_{-6.5}^{+6.4}$ \\
\hline
x$_\textrm{Si,mantle}$ [\%] &   $40_{-6}^{+7}$ &    $36_{-9}^{+9}$ &    $36_{-25}^{+29}$ &
                           $40_{-6}^{+7}$ &    $35_{-9}^{+10}$ &    $36_{-24}^{+29}$ \\
x$_\textrm{Mg,mantle}$ [\%] &   $42_{-7}^{+7}$ &    $37_{-10}^{+10}$ &    $38_{-26}^{+32}$ &
                           $42_{-7}^{+7}$ &    $37_{-10}^{+10}$ &    $36_{-25}^{+30}$ \\
x$_\textrm{Fe,mantle}$ [\%] &   $17_{-11}^{+9}$ &    $26_{-17}^{+19}$ &    $18_{-13}^{+22}$ &
                           $17_{-11}^{+9}$ &    $27_{-18}^{+19}$ &    $19_{-14}^{+24}$ \\
\hline
\end{tabular}
\label{tab:internal_structure_results_b}
\end{table*}
\renewcommand{\arraystretch}{1.0}

\begin{table*}
\renewcommand{\arraystretch}{1.5}
\caption{Results of the internal structure modeling for HIP\,41378 c.}
\centering
\begin{tabular}{r|ccc|ccc}
\hline \hline
Water prior &              \multicolumn{3}{c|}{Formation outside iceline (water-rich)} & \multicolumn{3}{c}{Formation inside iceline (water-poor)} \\
Si/Mg/Fe prior &           Stellar (A1) &       Iron-enriched (A2) &      Free (A3) &
                           Stellar (B1) &       Iron-enriched (B2) &      Free (B3) \\
\hline
w$_\textrm{core}$ [\%] &        $11_{-8}^{+9}$ &    $16_{-11}^{+15}$ &    $11_{-8}^{+16}$ &
                           $17_{-11}^{+12}$ &    $22_{-16}^{+20}$ &    $17_{-12}^{+22}$ \\
w$_\textrm{mantle}$ [\%] &      $57_{-13}^{+17}$ &    $51_{-14}^{+19}$ &    $54_{-14}^{+19}$ &
                           $83_{-12}^{+11}$ &    $76_{-20}^{+16}$ &    $82_{-22}^{+13}$ \\
w$_\textrm{envelope}$ [\%] &    $31.2_{-19.2}^{+14.8}$ &    $30.7_{-18.8}^{+15.0}$ &    $31.1_{-19.3}^{+14.7}$ &
                           $0.83_{-0.19}^{+0.20}$ &    $1.16_{-0.36}^{+0.35}$ &    $0.96_{-0.38}^{+0.50}$ \\
\hline
Z$_\textrm{envelope}$ [\%] &        $90.4_{-13.3}^{+5.4}$ &    $88.7_{-15.0}^{+5.8}$ &    $90.6_{-13.4}^{+6.2}$ &
                           $0.5_{-0.2}^{+0.2}$ &    $0.5_{-0.2}^{+0.2}$ &    $0.5_{-0.2}^{+0.2}$ \\
\hline
x$_\textrm{Fe,core}$ [\%] &     $90.2_{-6.3}^{+6.6}$ &    $90.3_{-6.4}^{+6.5}$ &    $90.3_{-6.3}^{+6.6}$ &
                           $90.3_{-6.4}^{+6.6}$ &    $90.4_{-6.4}^{+6.5}$ &    $90.3_{-6.4}^{+6.5}$ \\
x$_\textrm{S,core}$ [\%] &      $9.8_{-6.6}^{+6.3}$ &    $9.7_{-6.5}^{+6.4}$ &    $9.7_{-6.6}^{+6.3}$ &
                           $9.7_{-6.6}^{+6.4}$ &    $9.6_{-6.5}^{+6.4}$ &    $9.7_{-6.5}^{+6.4}$ \\
\hline
x$_\textrm{Si,mantle}$ [\%] &   $40_{-6}^{+7}$ &    $35_{-9}^{+10}$ &    $32_{-24}^{+31}$ &
                           $40_{-6}^{+7}$ &    $35_{-9}^{+10}$ &    $35_{-24}^{+29}$ \\
x$_\textrm{Mg,mantle}$ [\%] &   $42_{-7}^{+7}$ &    $37_{-10}^{+10}$ &    $40_{-27}^{+36}$ &
                           $42_{-7}^{+7}$ &    $37_{-10}^{+10}$ &    $36_{-25}^{+30}$ \\
x$_\textrm{Fe,mantle}$ [\%] &   $17_{-11}^{+9}$ &    $27_{-18}^{+19}$ &    $18_{-14}^{+24}$ &
                           $18_{-11}^{+9}$ &    $27_{-18}^{+19}$ &    $20_{-15}^{+24}$ \\
\hline
\end{tabular}
\label{tab:internal_structure_results_c}
\end{table*}
\renewcommand{\arraystretch}{1.0}

\end{appendix}

\end{document}